\documentclass[twocolumn]{aastex63}

\usepackage{natbib}
\usepackage[style=american]{csquotes}
\newcommand{\tess}{\emph{TESS}}

\usepackage{scrextend}
\let\orgautoref\autoref
\renewcommand{\autoref}
        {\def\equationautorefname{Eq.}%
         \def\figureautorefname{Fig.}%
         \def\sectionautorefname{Sect.}%
         \def\subsectionautorefname{Sect.}%
         \def\subsubsectionautorefname{Sect.}%
         \orgautoref}

\usepackage{changepage} 
\usepackage{rotating}

\revised{\today}
\accepted{----}

\usepackage[bulgarian,english]{babel}
\usepackage[T2A]{fontenc}

\shorttitle{TOI-2525 b \& c: A pair of massive warm giant planets}
\shortauthors{Trifonov et al.}

\begin{document}

\title{TOI-2525 b \& c: A pair of massive warm giant planets with a strong transit timing variations revealed by TESS\footnote{Based on observations collected at the European Organization for Astronomical Research in the Southern Hemisphere under MPG programmes 0104.A-9007. This paper includes data gathered with the 6.5 meter Magellan Telescopes located at Las Campanas Observatory, Chile.}}

\correspondingauthor{Trifon Trifonov}
\email{trifonov@mpia.de}

\author[0000-0002-0236-775X]{Trifon Trifonov}
\affiliation{Max-Planck-Institut für Astronomie,
              Königstuhl 17,
              69117 Heidelberg, Germany}
\affiliation{Department
 of Astronomy, Sofia University ``St Kliment Ohridski'', 5 James Bourchier Blvd, BG-1164 Sofia, Bulgaria}
 
\author[0000-0002-9158-7315]{Rafael Brahm}
\affil{Facultad de Ingeniera y Ciencias, Universidad Adolfo Ib\'{a}\~{n}ez, Av. Diagonal las Torres 2640, Pe\~{n}alol\'{e}n, Santiago, Chile}
\affil{Millennium Institute for Astrophysics, Chile}
\affil{Data Observatory Foundation, Chile}
 
\author[0000-0002-5389-3944]{Andr\'es Jord\'an}
\affil{Facultad de Ingeniera y Ciencias, Universidad Adolfo Ib\'{a}\~{n}ez, Av. Diagonal las Torres 2640, Pe\~{n}alol\'{e}n, Santiago, Chile}
\affil{Millennium Institute for Astrophysics, Chile}
\affil{Data Observatory Foundation, Chile}

\author{Christian Hartogh}
\affiliation{Institut f{\"u}r Astrophysik und Geophysik, Friedrich-Hund-Platz 1, 37077 G{\"o}ttingen, Germany}

\author[0000-0002-1493-300X]{Thomas Henning}
\affiliation{Max-Planck-Institut für Astronomie,
              Königstuhl 17,
              69117 Heidelberg, Germany}

\author[0000-0002-5945-7975]{Melissa J.\ Hobson}
\affiliation{Max-Planck-Institut für Astronomie,
              Königstuhl 17,
              69117 Heidelberg, Germany}
\affil{Millennium Institute for Astrophysics, Chile}

\author[0000-0001-8355-2107]{Martin Schlecker}
\affil{Department of Astronomy/Steward Observatory, The University of Arizona, 933 North Cherry Avenue, Tucson, AZ 85721, USA}
\affiliation{Max-Planck-Institut für Astronomie,
              Königstuhl 17,
              69117 Heidelberg, Germany}

\author[0000-0003-4894-7271]{Saburo Howard}
\affiliation{Universit\'e C\^ote d'Azur, Observatoire de la C\^ote d'Azur, CNRS, Laboratoire Lagrange, Bd de l'Observatoire, CS 34229, 06304 Nice cedex 4, France}

\author{Finja Reichardt}
\affil{Max-Planck-Institut f\"{u}r Astronomie, K\"{o}nigstuhl  17, 69117 Heidelberg, Germany}

\author[0000-0001-9513-1449]{Nestor Espinoza}
\affil{Space Telescope Science Institute, 3700 San Martin Drive, Baltimore, MD 21218, USA}

\author[0000-0003-1930-5683]{Man Hoi Lee}
\affil{Department of Earth Sciences, The University of Hong Kong, Pokfulam Road, Hong Kong}
\affil{Department of Physics, The University of Hong Kong, Pokfulman Road, Hong Kong}

\author{David Nesvorny}
\affil{Department of Space Studies, Southwest Research Institute, 1050 Walnut Street, Suite 300, Boulder, CO 80302, USA}

\author[0000-0003-3047-6272]{Felipe I.\ Rojas}
\affil{Instituto de Astrof\'isica, Facultad de F\'isica, Pontificia Universidad Cat\'olica de Chile, Chile}
\affil{Millennium Institute for Astrophysics, Chile}

\author[0000-0003-1464-9276]{Khalid Barkaoui}
\affil{Astrobiology Research Unit, Universit\'e de Li\`ege, All\'ee du 6 Ao\^ut 19C, B-4000 Li\`ege, Belgium}
\affil{Department of Earth, Atmospheric and Planetary Science, Massachusetts Institute of Technology, 77 Massachusetts Avenue, Cambridge, MA 02139, USA} 
\affil{Instituto de Astrof\'isica de Canarias (IAC), Calle V\'ia L\'actea s/n, 38200, La Laguna, Tenerife, Spain} 

\author[0000-0002-0436-7833]{Diana Kossakowski}
\affil{Max-Planck-Institut f\"{u}r Astronomie, K\"{o}nigstuhl  17, 69117 Heidelberg, Germany}

\author{Gavin Boyle}
\affil{El Sauce Observatory -- Obstech, Chile}
\affil{Cavendish Laboratory, JJ Thomson Avenue, Cambridge, CB3 0HE, UK}


\author[0000-0001-6187-5941]{Stefan Dreizler}
\affiliation{Institut f{\"u}r Astrophysik und Geophysik, Friedrich-Hund-Platz 1, 37077 G{\"o}ttingen, Germany}

\author{Martin\,K\"urster}
\affiliation{Max-Planck-Institut für Astronomie,
              Königstuhl 17,
              69117 Heidelberg, Germany}

\author[0000-0002-9831-0984]{Ren{\'e} Heller} 
\affiliation{Max Planck Institute for Solar System Research, Justus-von-Liebig-Weg 3, 37077 G{\"o}ttingen, Germany}

\author[0000-0002-7188-8428]{Tristan Guillot}
\affiliation{Universit\'e C\^ote d'Azur, Observatoire de la C\^ote d'Azur, CNRS, Laboratoire Lagrange, Bd de l'Observatoire, CS 34229, 06304 Nice cedex 4, France}
\author[0000-0002-5510-8751]{Amaury H.M.J. Triaud}
\affiliation{School of Physics \& Astronomy, University of Birmingham, Edgbaston, Birmingham, B15 2TT, UK}
\author{Lyu Abe}
\affiliation{Universit\'e C\^ote d'Azur, Observatoire de la C\^ote d'Azur, CNRS, Laboratoire Lagrange, Bd de l'Observatoire, CS 34229, 06304 Nice cedex 4, France}
\author{Abdelkrim Agabi}
\affiliation{Universit\'e C\^ote d'Azur, Observatoire de la C\^ote d'Azur, CNRS, Laboratoire Lagrange, Bd de l'Observatoire, CS 34229, 06304 Nice cedex 4, France}
\author{Philippe Bendjoya}
\affiliation{Universit\'e C\^ote d'Azur, Observatoire de la C\^ote d'Azur, CNRS, Laboratoire Lagrange, Bd de l'Observatoire, CS 34229, 06304 Nice cedex 4, France}
\author[0000-0001-7866-8738]{Nicolas Crouzet}
\affiliation{Leiden Observatory, Leiden University, Postbus 9513, 2300 RA Leiden, The Netherlands}
\author[0000-0002-3937-630X]{Georgina Dransfield}
\affiliation{School of Physics \& Astronomy, University of Birmingham, Edgbaston, Birmingham, B15 2TT, UK}
\author[0000-0002-7913-4866]{Thomas Gasparetto}
\affiliation{Institute of Polar Sciences - CNR, via Torino, 155 - 30172 Venice-Mestre, Italy}
\author[0000-0002-3164-9086]{Maximilian N. Günther}
\affiliation{European Space Agency (ESA), European Space Research and Technology Centre (ESTEC), Keplerlaan 1, 2201 AZ Noordwijk, The Netherlands}
\thanks{ESA Research Fellow}
\author{Wenceslas Marie-Sainte}
\affiliation{Concordia Station, IPEV/PNRA, Antarctica}
\author[0000-0001-5000-7292]{Djamel M\'ekarnia}
\affiliation{Universit\'e C\^ote d'Azur, Observatoire de la C\^ote d'Azur, CNRS, Laboratoire Lagrange, Bd de l'Observatoire, CS 34229, 06304 Nice cedex 4, France}
\author{Olga Suarez}
\affiliation{Universit\'e C\^ote d'Azur, Observatoire de la C\^ote d'Azur, CNRS, Laboratoire Lagrange, Bd de l'Observatoire, CS 34229, 06304 Nice cedex 4, France}

\author{Johanna Teske} 
\affiliation{Carnegie Institution for Science, Earth \& Planets Laboratory, 5241 Broad Branch Road NW, Washington DC 20015, USA}

 \author{R. Paul Butler}  
\affiliation{Carnegie Institution for Science, Earth \& Planets Laboratory, 5241 Broad Branch Road NW, Washington DC 20015, USA}

\author[0000-0002-5226-787X]{Jeffrey D. Crane}  
\affiliation{The Observatories of the Carnegie Institution for Science, 813 Santa Barbara Street, Pasadena, CA 91101}

\author[0000-0002-8681-6136]{Stephen Shectman}
\affiliation{The Observatories of the Carnegie Institution for Science, 813 Santa Barbara Street, Pasadena, CA 91101}




\author[0000-0003-2058-6662]{George R. Ricker} 
\affiliation{Department of Physics and Kavli Institute for Astrophysics and Space Research, Massachusetts Institute of Technology, Cambridge, MA 02139, USA}

\author[0000-0002-1836-3120]{Avi Shporer} 
\affiliation{Department of Physics and Kavli Institute for Astrophysics and Space Research, Massachusetts Institute of Technology, Cambridge, MA 02139, USA}

\author{Roland Vanderspek} 
\affiliation{Department of Physics and Kavli Institute for Astrophysics
and Space Research, Massachusetts Institute of Technology, Cambridge, MA
02139, USA}

\author[0000-0002-4715-9460]{Jon M. Jenkins} 
\affiliation{NASA Ames Research Center, Moffett Field, CA 94035, USA}
 
\author[0000-0002-5402-9613]{Bill Wohler} 
\affiliation{NASA Ames Research Center, Moffett Field, CA 94035, USA}
\affiliation{SETI Institute, Mountain View, CA 94043, USA} 
 
 \author[0000-0001-6588-9574]{Karen~A.~Collins} 
\affiliation{Center for Astrophysics \textbar \ Harvard \& Smithsonian,
60 Garden St, Cambridge, MA 02138, USA}

\author[0000-0003-2781-3207]{Kevin I.\ Collins}
\affiliation{George Mason University, 4400 University Drive, Fairfax, VA, 22030 USA}

\author[0000-0002-5741-3047]{David~ R.~Ciardi}

\affiliation{Caltech/IPAC-NASA Exoplanet Science Institute, 770 S. Wilson Avenue, Pasadena, CA 91106, USA}


\author[0000-0001-7139-2724]{Thomas~Barclay}
\affiliation{NASA Goddard Space Flight Center, 8800 Greenbelt Road, Greenbelt, MD 20771, USA}
\affiliation{University of Maryland, Baltimore County, 1000 Hilltop Circle, Baltimore, MD 21250, USA}

 \author[0000-0002-4510-2268]{Ismael~Mireles}

\affiliation{Department of Physics and Astronomy, University of New Mexico, 210 Yale Blvd NE, Albuquerque, NM 87106, USA}

%
 
\author[0000-0002-6892-6948]{Sara Seager} 
\affiliation{Department of Physics and Kavli Institute for Astrophysics
and Space Research, Massachusetts Institute of Technology, Cambridge, MA
02139, USA}
\affiliation{Department of Earth, Atmospheric and Planetary Sciences,
Massachusetts Institute of Technology, Cambridge, MA 02139, USA}
\affiliation{Department of Aeronautics and Astronautics, MIT, 77
Massachusetts Avenue, Cambridge, MA 02139, USA}

\author[0000-0002-4265-047X]{Joshua N. Winn} 
\affiliation{Department of Astrophysical Sciences, Princeton University,
NJ 08544, USA}

\begin{abstract}
TOI-2525 is a K-type star with an estimated mass of M = 0.849$_{-0.033}^{+0.024}$  M$_\odot$ and radius of 
R = 0.785$_{-0.007}^{+0.007}$ R$_\odot$ observed by the TESS mission in 22 sectors (within sectors 1 and 39). 
The TESS light curves yield significant transit events of two companions, which 
show strong transit timing variations (TTVs) with a semi-amplitude of a $\sim$6 hours.
We performed TTV dynamical, and photo-dynamical light curve analysis of the TESS data, combined with radial velocity (RV) measurements from FEROS and PFS, and we confirmed the planetary nature of these companions. 
The TOI-2525 system consists of a transiting pair of planets comparable to Neptune and Jupiter with
estimated dynamical masses of $m_{\rm b}$ = 0.088$_{-0.004}^{+0.005}$ M$_{\rm Jup.}$,
and $m_{\rm c}$ = 0.709$_{-0.033}^{+0.034}$ M$_{\rm Jup.}$, radius of 
$r_b$ = 0.88$_{-0.02}^{+0.02}$  R$_{\rm Jup.}$ and $r_c$ = 0.98$_{-0.02}^{+0.02}$ R$_{\rm Jup.}$, and with orbital periods of 
$P_{\rm b}$ =  23.288$_{-0.002}^{+0.001}$ days and $P_{\rm c}$ = 49.260$_{-0.001}^{+0.001}$
days for the inner and the outer planet, respectively. The period ratio is close to 
the 2:1 period commensurability, but the dynamical simulations of the system suggest
that it is outside the mean motion resonance (MMR) dynamical configuration. TOI-2525 b is among the lowest density Neptune-mass planets known to date, with an estimated median density of $\rho_{\rm b}$ = 0.174$_{-0.015}^{+0.016}$ g\,cm$^{-3}$.
The TOI-2525 system is very similar to the other K-dwarf systems discovered by TESS, TOI-2202 and TOI-216, which are composed of almost identical K-dwarf primary and two warm giant planets near the 2:1 MMR.


\end{abstract}



\section{Introduction}

As of September 2022, the exoplanet surveys have discovered over 5000 confirmed planets, many of which reside in multiple-planet systems. 
The current population of multiple-planet systems is a fingerprint of the planet formation and migration mechanisms. 
Many scholars are confident that planet migration must have occurred simultaneously for all planets in 
the system, but despite the large efforts to understand the interactions between planets and the proto-planetary disk \citep[][]{Goldreich1979,Lin1979,Ida2010, Kley2012,Coleman2014, Baruteau2014,Levison2015,Kanagawa2018,Bitsch2020, Schlecker2021,Matsumura2021}, the planet migration rate, direction, eccentricity excitation or damping, as a function of disk viscosity, mass, and metallicity, are still the subject of ongoing research. 
Warm massive planet pairs near the low-order, 2:1 commensurability are rare, but have the potential to reveal important details of the disk-planet interactions during the system formation stage. 

The current planet formation theories suggest that Jovian planets must have 
formed further out beyond the so called ice-line, and have migrated inwards toward warm orbits before the primordial disk dissipates. These objects are not easily understood within standard formation models that require rapid accretion of gas by a solid embryo before the stellar radiation dissipates the gas from the protoplanetary disc. This rapid, solid accretion is favored beyond the snow line. Giant planets are expected then to migrate from a couple of astronomical units to the inner regions of the system to produce the population of hot ($P<$ 10\,d) and warm (10\,d $<P<$ 300\,d) Jovian or Neptune mass planets. Typical migration mechanisms can be divided into two groups, namely: disk migration \citep[e.g.,][]{Lin1986}, and high eccentricity tidal migration \citep[e.g.,][]{Rasio1996,Fabrycky2007,Bitsch2020}. These two mechanisms predict significantly different orbital configurations for the migrating planet, and the characterization of these properties, particularly for warm Jovian planets \citep[][]{Huang2016, Petrovich2016,Santerne2016,Dong2021}, can be used to constrain migration theories.


In this context, it is fundamentally important to measure the dynamical mass and orbital eccentricity of the warm Jovian planets. For many systems, this can only be achieved by combining precise transit and RV observational data. 
NASA's {\it Transiting Exoplanet Survey Satellite} \citep[$\tess$;][]{Ricker2015}
aims to detect planets through the transit method around relatively bright stars that are suitable for precise Doppler follow-up in order to determine the planetary mass, radius, and bulk density, among other important physical parameters. $\tess$ has already led to more than 230 newly discovered planets, most of which were confirmed by Doppler spectroscopy \citep[e.g.,][among many]{Wang2019, Trifonov2019a, Dumusque2019, Luque2019,  kossakowski:2019, Teske2020, Schlecker2020, Espinoza2020}.

 \begin{figure*}[tp]
    \centering
    \includegraphics[width=4.3cm]{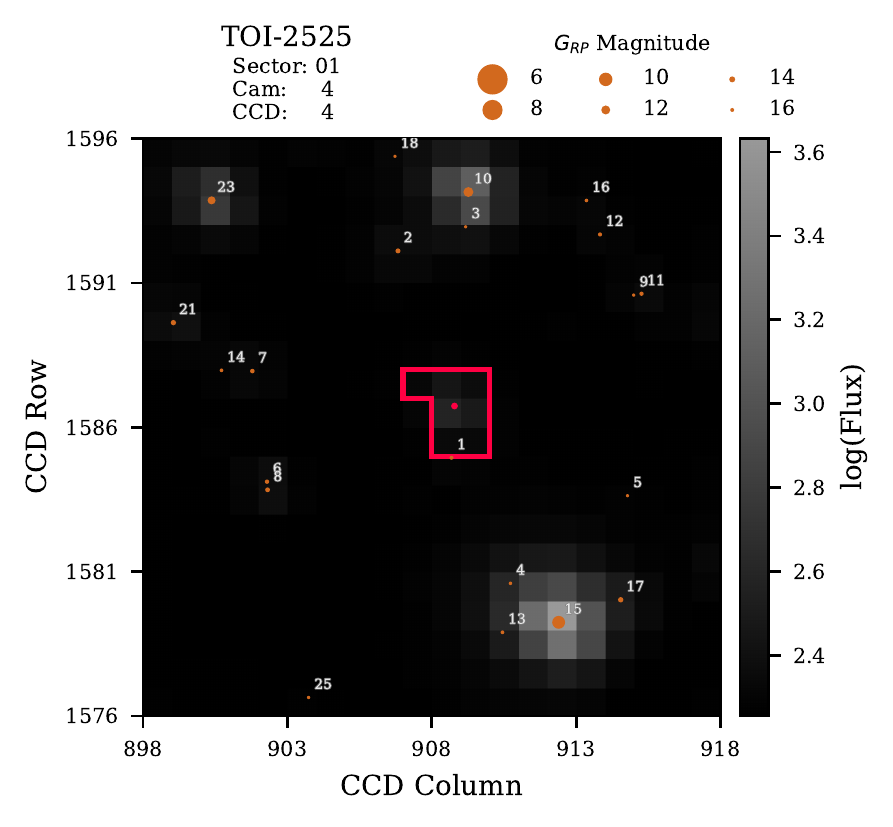} 
    \includegraphics[width=4.3cm]{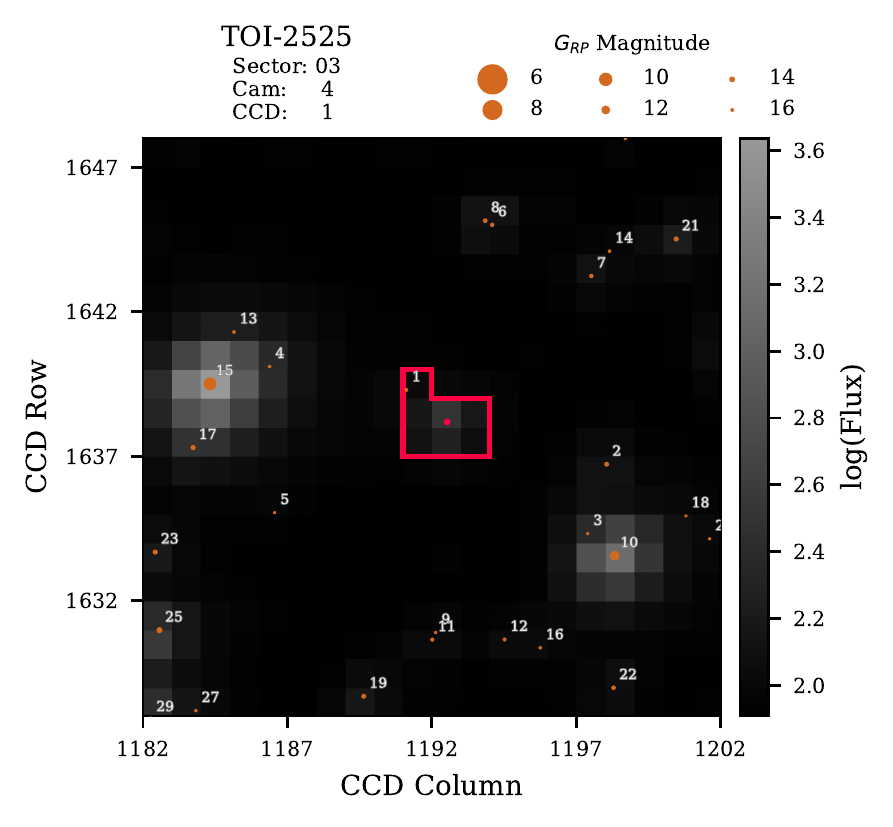} 
    \includegraphics[width=4.3cm]{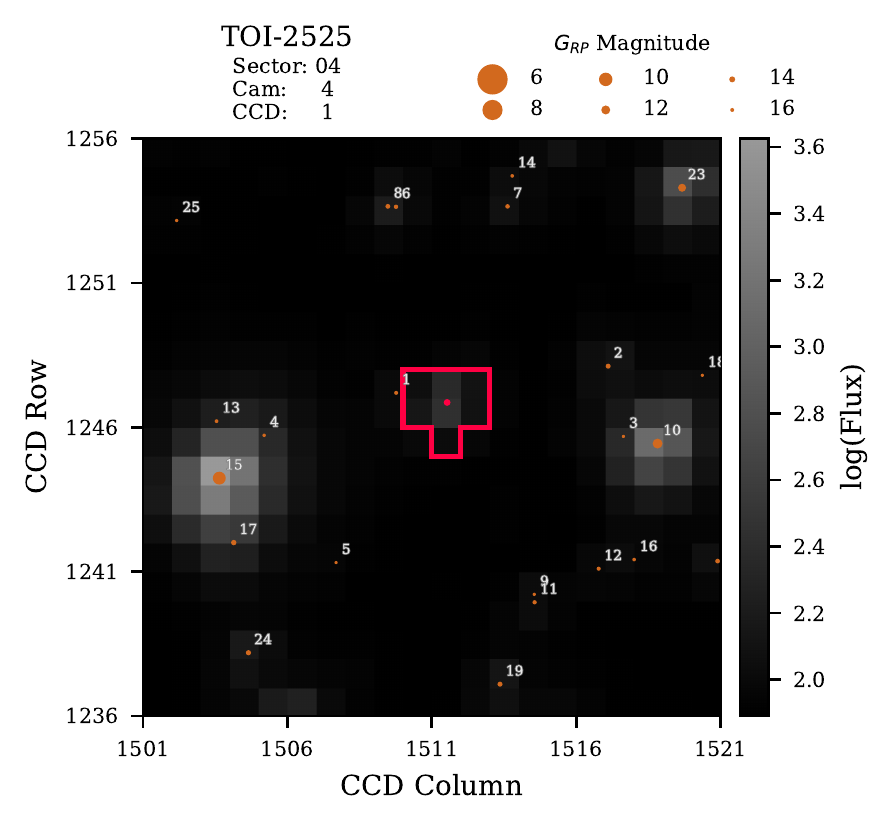} 
    \includegraphics[width=4.3cm]{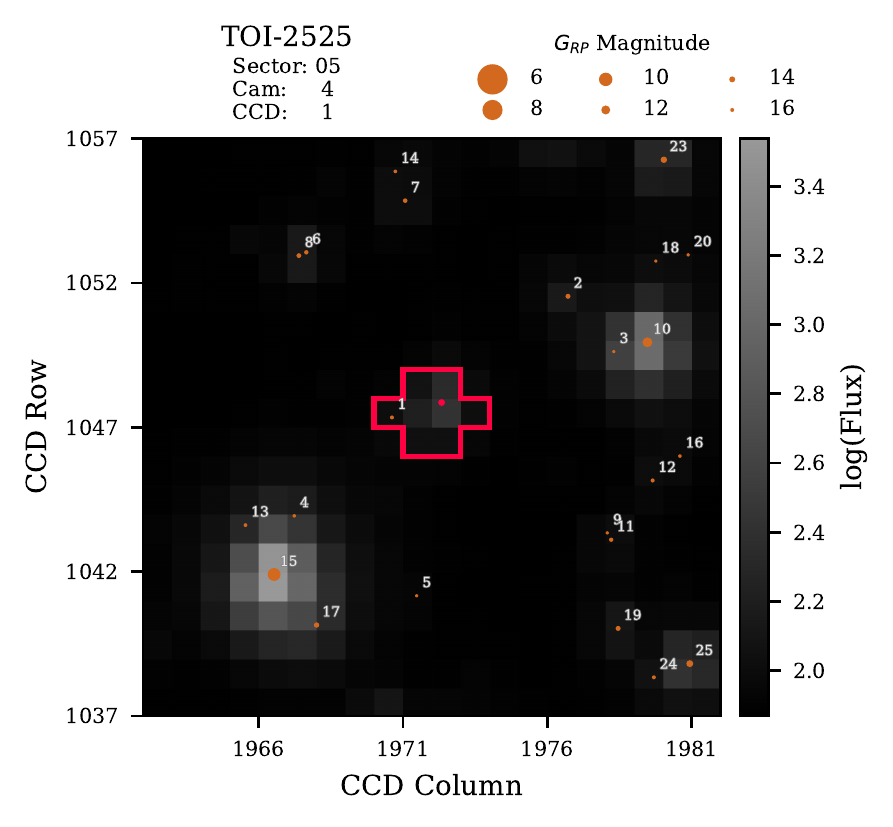} \\ 
    \includegraphics[width=4.3cm]{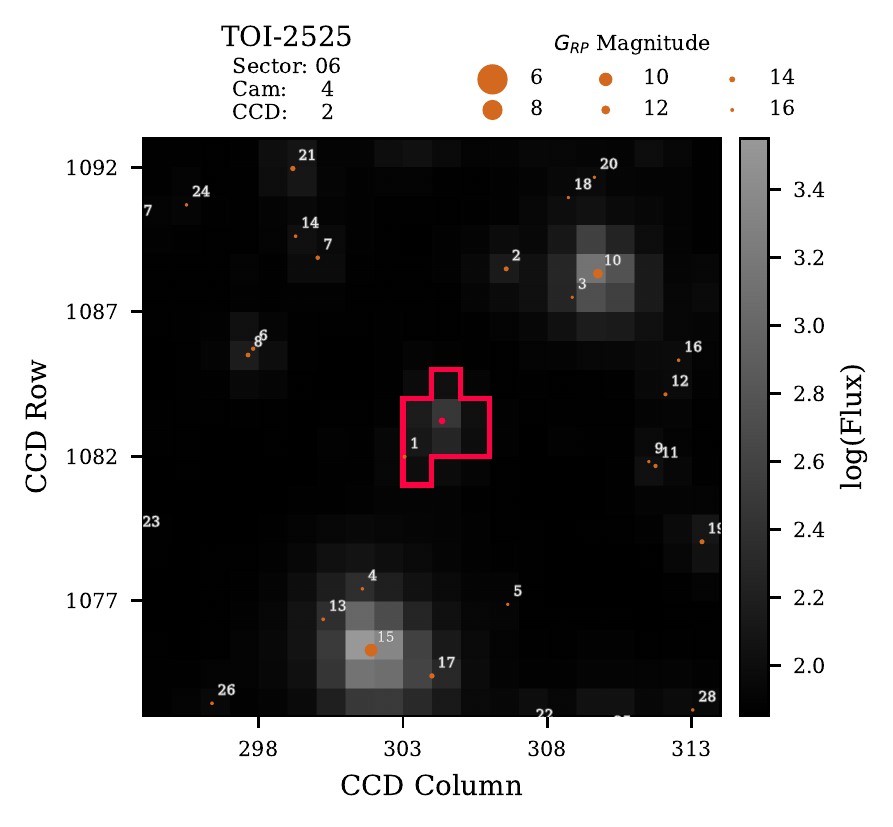} 
    \includegraphics[width=4.3cm]{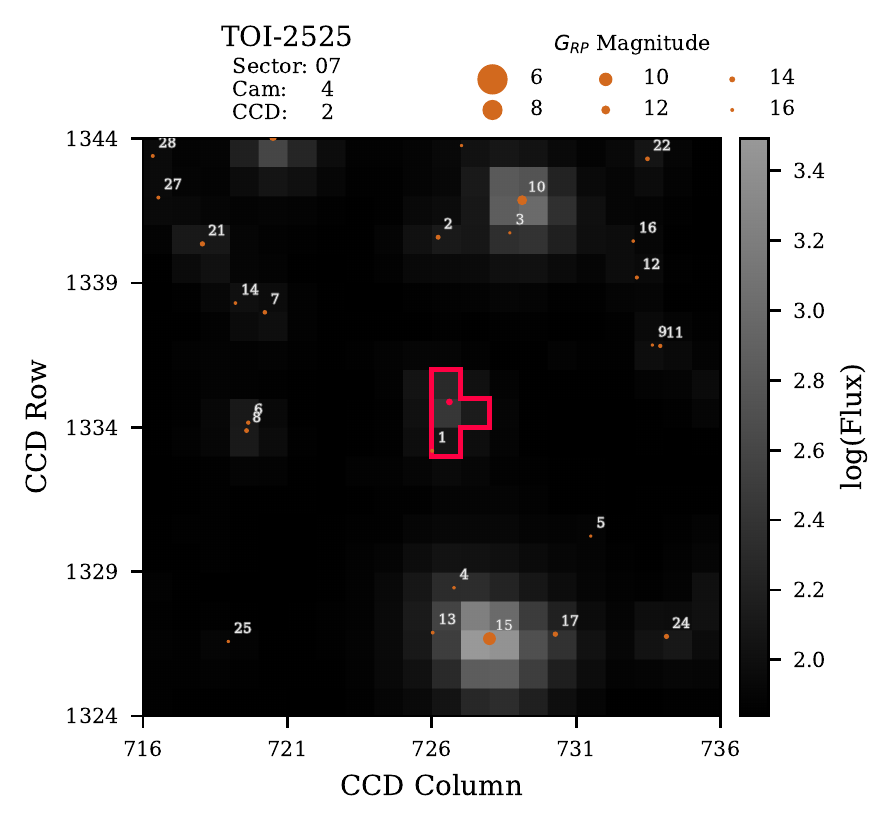} 
    \includegraphics[width=4.3cm]{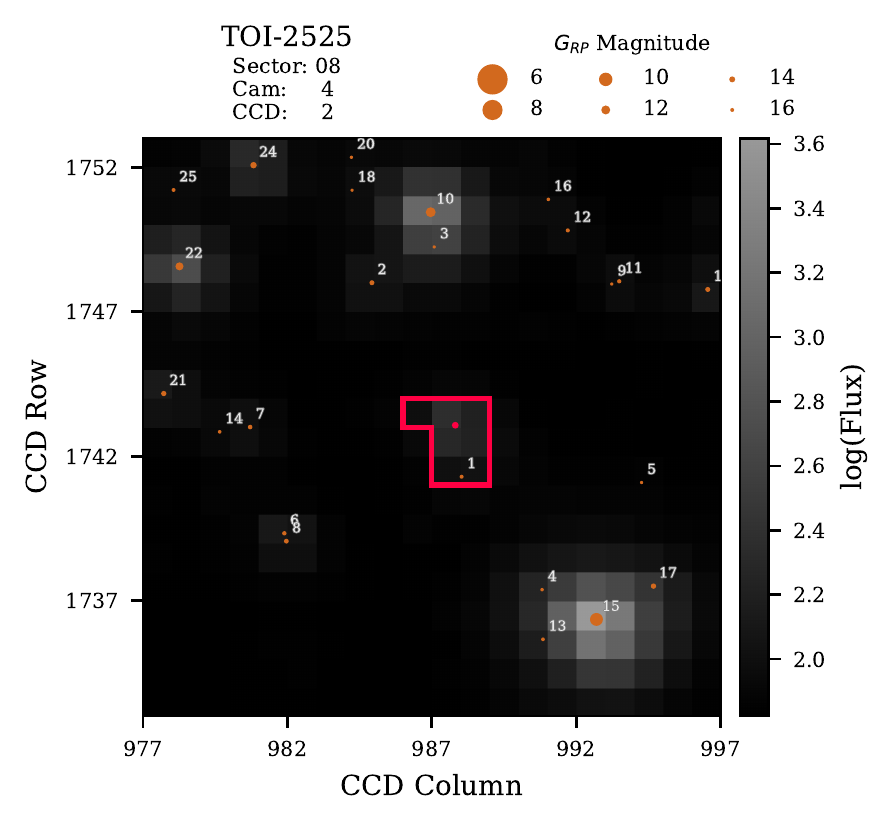} 
    \includegraphics[width=4.3cm]{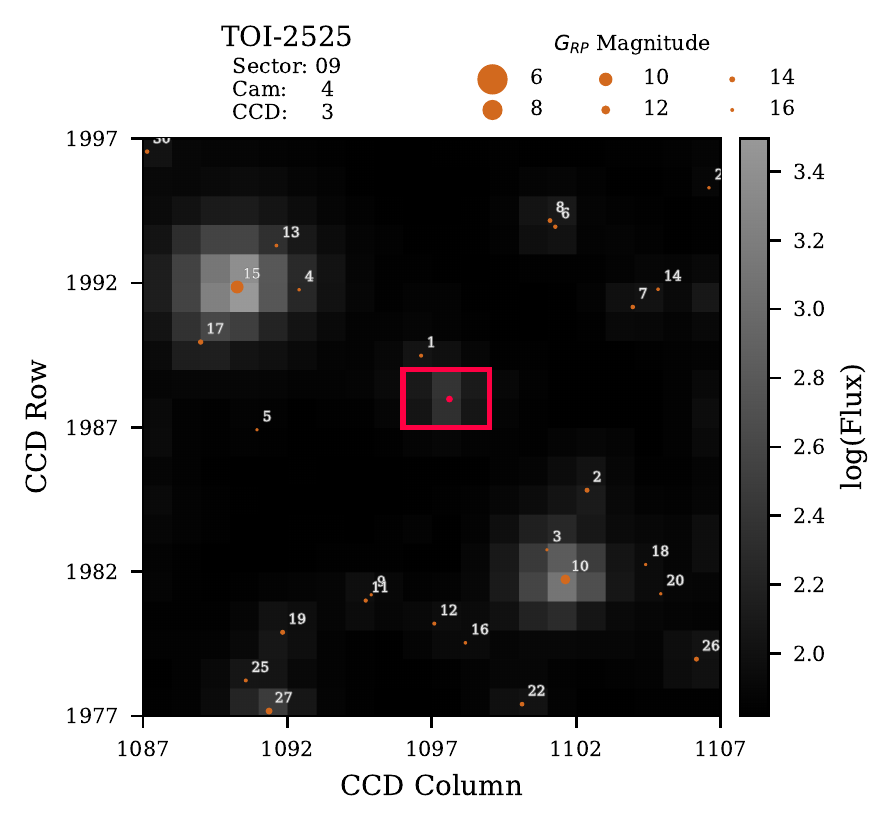} \\ 
    \includegraphics[width=4.3cm]{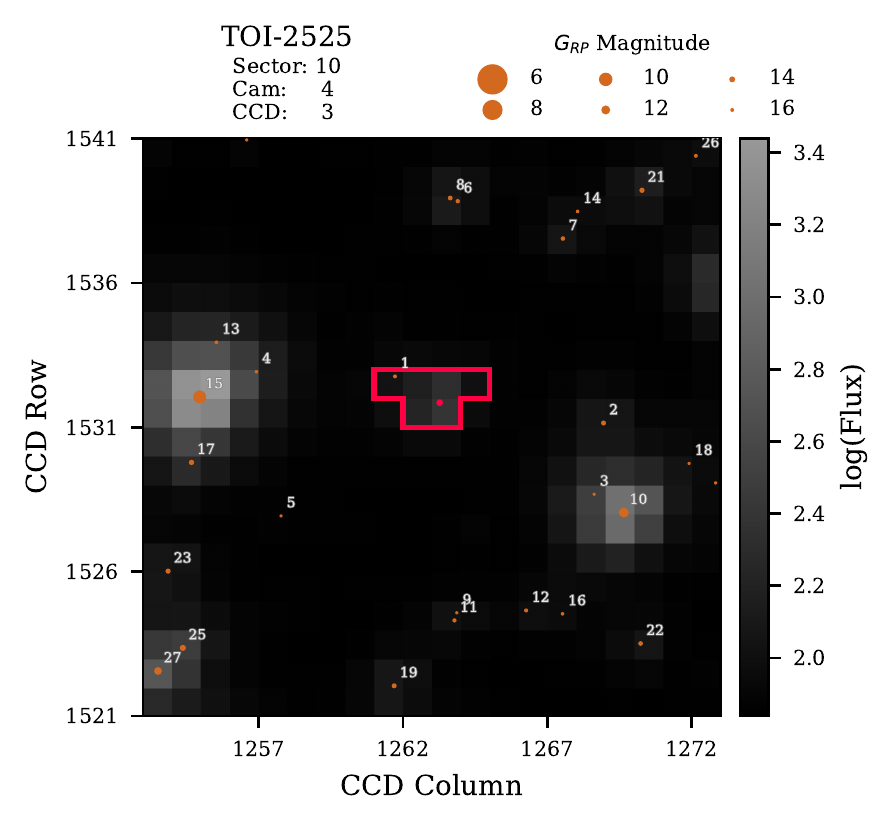} 
    \includegraphics[width=4.3cm]{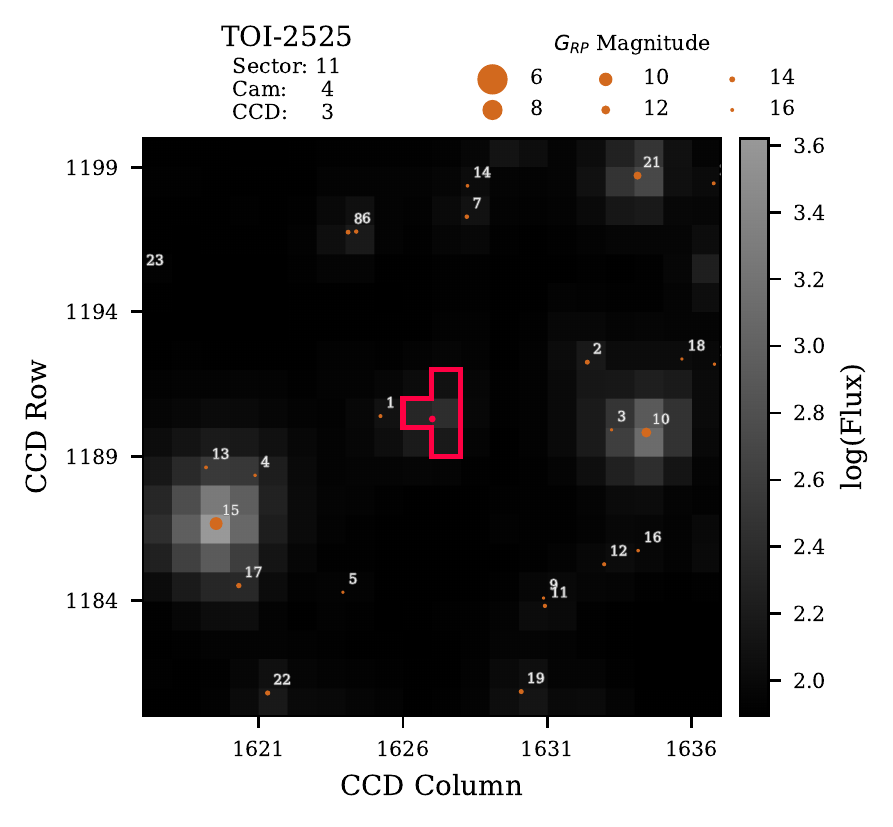} 
    \includegraphics[width=4.3cm]{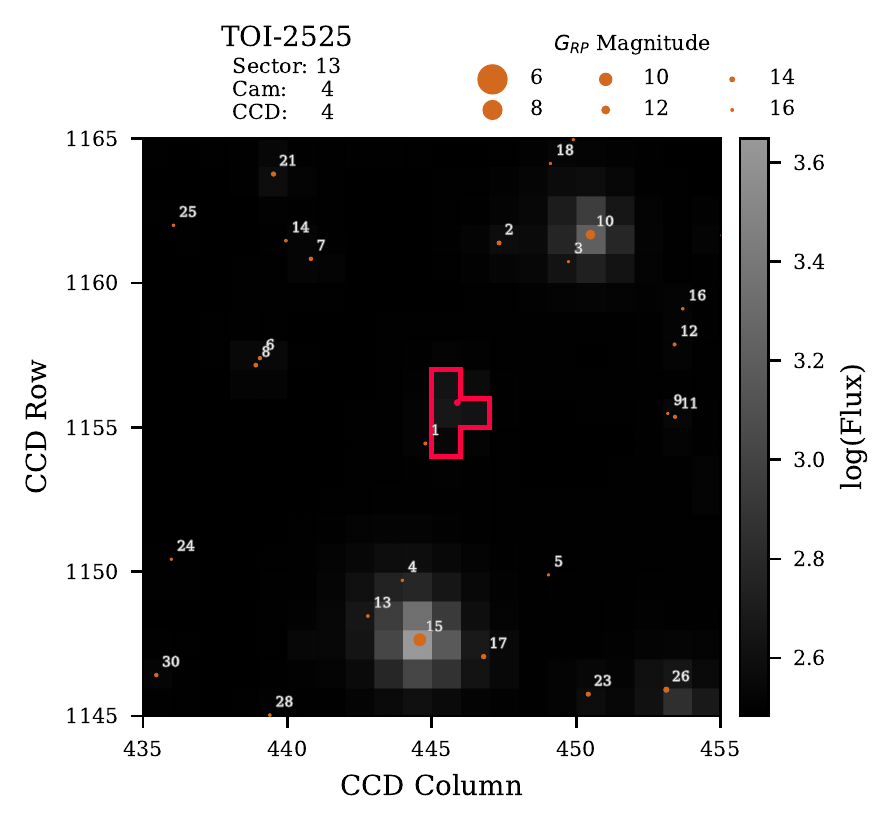} 
  
    \caption{Target pixel file (TPF) image of TOI-2525 in the FFIs of TESS Sector 1, Sectors 3 to 11, and Sector 13. The central (red)  borders in the pixel space are the ones used to construct the 
    {\sc tesseract} photometry. {\em Gaia} targets are marked with red circles, whose size has been coded by their G magnitude.
    }
    \label{tpf}
\end{figure*}

In this paper, we report the discovery of another warm massive planet pair around a K-dwarf star, which has been uncovered by $\tess$.
This work is part of our Doppler data survey and orbital analysis efforts  performed within the {\bf W}arm g{\bf I}a{\bf N}ts with t{\bf E}ss (WINE) collaboration, which focuses on the systematic characterization of $\tess$ transiting warm giant planets \citep[e.g.,][]{hd1397,jordan:2020,brahm:2020,Schlecker2020,Trifonov2021b}.
We present TOI-2525 (TIC\,149601126\footnote{Similar to TOI-2202, a.k.a. TIC\,358107516 \citep{Trifonov2021b}, TOI-2525 was known to us as TIC\,149601126. The target became an $\tess$ Object of Interest \citep[TOI,][]{Guerrero2021} while this work was in preparation. Consequently, we adopted the TOI-2525 designation for consistency with the $\tess$ survey.}), a two-planet system, which exhibits strong transit timing variations (TTVs) of the two transiting signals detected in $\tess$ multi-sector and ground-based photometry data. This strong TTV signal in the TOI -2525 system points to strongly interacting warm giant-mass planets close to the 2:1 mean motion resonance (MMR) commensurability. 

In \autoref{sec2} we present the observational data used to detect and characterize the warm pair of planets orbiting TOI-2525. In \autoref{sec3} we introduce the stellar parameter estimates of TOI-2525. In \autoref{sec4} we present our orbital analysis and results, which was performed on the extracted TTVs of the planetary signals, and a self-consistent photo-dynamical modeling scheme performed jointly with the acquired Doppler data.
In \autoref{sec4}, we also provide results from an analysis on the dynamical architecture and long-term stability of the TOI-2525 system. In \autoref{sec5}
we discuss the systems' architecture, possible formation and evolution, and interior of the planets. Finally, in \autoref{sec6} we present a brief summary and  our conclusions.


\section{Data}
\label{sec2}

Here we present the photometric light curve and Doppler data acquired for TOI-2525. The {\tess} space based photometry is used for transit event identification of TOI-2525 b \& c, whereas additional ground based photometry was used for further TTV analysis and planetary radius estimates. Precise Doppler data  were collected for further constraining the planetary masses and eccentricities, but also for the stellar parameter estimates of TOI-2525.

\subsection{TESS}
\label{Sec2.1}

TOI-2525 was observed in sectors 1, 3 $-$ 11, and 13, during the first year of the $\tess$ primary mission with a 30\,minute cadence, and in sectors 27, 28, and 30 $-$ 39 with a two-minute cadence in the third mission year.
TOI-2525 b and c were identified in the light curves extracted from the $\tess$ Full Frame Images (FFIs) using the {\sc tesseract}\footnote{\url{https://github.com/astrofelipe/tesseract}} pipeline (Rojas et al. in prep.). A brief introduction of our FFI extraction with {\sc tesseract} in the context of the WINE collaboration can be found in \citet{Schlecker2020}, \citet{gill:2020} and \citet{Trifonov2021b}.
\autoref{tpf} shows the target pixel file (TPF) image of TOI-2525\,constructed from the $\tess$ FFI image frames and {\em Gaia} DR3 data \citep{Gaia_Collaboration_2021}. \autoref{tpf} shows that there are no bright contaminators in the FFI aperture (red continuous contour), thus we concluded that the transit signals are indeed coming from TOI-2525 and not from neighboring stars. This was later confirmed by ground-based transit detections of TOI-2525\,b, and  TOI-2525\,c. However, a fainter source labeled \#1 is occasionally in the FFI aperture, which dilutes the FFI light curves. Since {\tt tesseract} does not correct contamination in the TESS apertures from nearby stars, we follow the same methodology as used in \citet{Trifonov2021b} to calculate and apply a dilution correction for the contamination of TOI-2525 on the FFI light curves.   

We retrieved the two-minute cadence light-curves from the Mikulski Archive for Space Telescopes\footnote{\url{https://mast.stsci.edu/portal/Mashup/Clients/Mast/Portal.html}}.
The Science Processing Operations Center \citep[SPOC;][]{SPOC} provides simple aperture photometry (SAP) and systematics-corrected Presearch Data Conditioning photometry \citep[PDC,][]{Smith2012, Stumpe12}. 
The PDCSAP light curves are corrected for contamination from nearby stars and instrumental systematics originating from, e.g., pointing drifts, focus changes, and thermal transients. In our work of TOI-2525, for the two-minute cadence data, we only use the corrected PDCSAP data.

\subsection{ASTEP}

The Antarctica Search for Transiting ExoPlanets \citep[ASTEP,][]{Guillot2015} instrument is a 40\,cm Newton
telescope installed in 2010 at the Concordia station located at $-$75.06$^\circ$S, 123.3$^\circ$E and
an altitude of $\sim$ 3230 meters. 
ASTEP is a robotic telescope dedicated to photometric observations of fields of stars and their exoplanets.\looseness=-5

    \begin{table}[ht]

    \centering
     \caption{FEROS and PFS RV measurements of TOI-2525.}
    
    \label{table:RVs}
 
    \begin{tabular}{p{2.0cm}  c crrrrrr}     

    \hline\hline  \noalign{\vskip 0.7mm}
    BJD \hspace{60.0 mm}& RV [$m\,s^{-1}$] & RV$_\sigma$  [$m\,s^{-1}$] & instrument \\
    \hline \noalign{\vskip 0.7mm}

2458904.623 & $-$48154.684   &   18.000  &     FEROS \\
2458923.562 & $-$48184.684   &   18.100  &     FEROS \\
2459156.792 &    320.393   &    4.235  &     PFS \\
2459157.754 &    310.269   &    6.285  &     PFS  \\
2459238.635 &    225.396   &    6.960  &     PFS \\
2459239.607 &    255.721   &    6.470  &     PFS  \\
2459501.824 &    337.951   &    5.040  &     PFS \\
2459504.843 &    345.583   &    5.150  &     PFS \\
2459505.837 &    341.083   &    5.400  &     PFS  \\
2459531.781 &    295.292   &    7.340  &     PFS\\
2459534.809 &    320.755   &    5.480  &     PFS  \\

     \hline \noalign{\vskip 0.7mm}

    \end{tabular}



    \end{table}

Due to the extremely low data transmission rate at the Concordia station, the data are processed automatically on-site using an IDL-based aperture photometry pipeline \citep{Mekarnia2016}. The raw light curves of up to 1\,000 stars of the field are transferred to Europe on a server in Roma, Italy, and are then available for deeper analysis. These data files contain each star’s flux computed through 10 fixed circular apertures radii, so that optimal calibrated light curves can be extracted.

Thanks to the accurate TTV model prediction constructed on the TESS data, we scheduled successful observations with ASTEP on TOI-2525\,b, and c.  For TOI-2525\,b, we detected a full transit event and a partial one on the nights UT\,2021-09-17 and UT\,2022-06-23, respectively, whereas for TOI-2525\,c, we observed two full transit events on the nights UT\,2021-04-15 and UT\,2022-07-02 and two partial transit events on the nights UT\,2021-06-03 and UT\,2021-09-10.

\subsection{Moana Observatoire Moana}

A partial transit of TOI-2525\,c was observed with the Siding Spring Observatory station of the  Observatorie Moana telescope network (OM-SSO). OM-SSO is an RC Optical Systems RCOS20 f8.1 telescope with a focal length of 3980 mm. OM-SSO is equipped with an FLI Microline 16803 camera with 4kx4k pixels of 9 microns with a pixel scale of 0.47\arcsec and a field of view of $30\times30\arcmin$. Observations were taken using an Astrodon Exoplanet (clear blue blocking) filter. Observations of TOI-2525\,c were performed on October 29, 2021 covering an ingress. The adopted exposure time was 147 s and the airmass ranged from 1.15 to 2. OM-SSO data was processed with a dedicated automated pipeline adapted from a version that was initially developed for obtaining differential photometry of LCOGT light curves \citep{espinoza:2019}.

\subsection{LCOGT}

We observed one partial transit of TOI-2525\,b and two partial transits of TOI-2525\,c using 
the Las Cumbres Observatory Global Telescope (LCOGT) 1.0-m network \citep{Brown2013}
nodes at Cerro Tololo Inter-American Observatory (CTIO) and South Africa Astronomical Observatory (SAAO).
The 1-m telescopes are equipped with 4096$\times$4096 SINISTRO cameras having a pixel scale 
of 0.389\arcsec/pixel, resulting in a Field-Of-View of 26'$\times$26'.
The TOI-2525\,b transit was observed from SAAO on January 11, 2022 in the Sloan-$g'$ and Sloan-$i'$ filters using a 4.3\arcsec\ target aperture.
TOI-2525\,c transits were observed twice from CTIO on December 18, 2021 and February 05, 2022 in the Sloan-$g'$ and Sloan-$i'$ filters, respectively, using 4.0\arcsec-4.7\arcsec\ target apertures. LCOGT data reduction and photometric measurements were performed using the AstroImageJ \citep[AIJ,][]{Collins2017} software package.

 \subsection{PFS}

TOI-2525 was monitored with the Planet Finder Spectrograph \citep{crane2006,crane2008,crane2010} installed at the 6.5\,m Magellan/Clay telescope at Las Campanas Observatory. TOI-2525 was observed with the iodine gas absorption cell of the instrument at four different observing runs between November 03, 2020, and November 16, 2021, adopting an exposure time of 1200 sec, and using a 3$\times$3 CCD binning mode to minimize read-noise. TOI-2525 was also observed without the iodine cell in order to generate the template for computing the RVs, which were derived following the methodology of \cite{butler1996}.
The mean uncertainty of the PFS RVs of TOI-2525 is 5.7 m\,s$^{-1}$. The PFS RVs are presented in \autoref{table:RVs}.

\begin{table}[tp]

\caption{Stellar parameters of TOI-2525 and their 1$\sigma$ uncertainties derived using 
ZASPE spectral analyses, {\em Gaia} parallax, broad band photometry and PARSEC models. 
}
\label{table:phys_param}    


\centering          
\begin{tabular}{ p{3.0cm} l r}     
\hline\hline  \noalign{\vskip 0.5mm}        
  Parameter   & TOI-2525   &  reference \\  
\hline    \noalign{\vskip 0.5mm}                   
   Spectral type                            & K8V          & [1] \\ 
   Distance  (pc)                           & 400$_{-2.3}^{+2.3}$   & [2] \\   
   Mass    ($M_{\odot}$)                    & 0.849$_{-0.033}^{+0.024}$  (0.042)   & This paper\\
   Radius    ($R_{\odot}$)                  & 0.785$_{-0.007}^{+0.007}$  (0.031)   & This paper  \\
   Luminosity    ($L{_\odot}$)             & 0.363$_{-0.016}^{+0.016}$  (0.008)   & This paper \\
   Age    (Gyr)                         & 3.99$_{-2.60}^{+4.30}$    &  This paper\\  
   A$_V$   (mag)                            & 0.287$_{-0.074}^{+0.070}$& This paper     \\
   $T_{\mathrm{eff}}$~(K)                   & 5096 $\pm$ 80 (102)   & This paper \\
   $\log g~[\mathrm{cm\cdot s}^{-2}]$       & 4.58 $\pm$ 0.20    & This paper \\   
   {}[Fe/H]                                 & 0.14 $\pm$ 0.05    & This paper  \\
   $v\cdot\sin(i)$ (km\,s$^{-1}$)         & 1.5  $\pm$ 0.3     & This paper   \\                                          
    
\hline\hline \noalign{\vskip 0.5mm}   

\end{tabular}
 

\tablecomments{\small  
[1] \citet{ESA}, 
[2] \citet{Gaia_Collaboration2016, Gaia_Collaboration2018b}.
 The values in parentheses are "floor" (i.e., more realistic, minimum) uncertainties predicted by \citet{Tayar2020} and adopted in our work.
 }

\end{table}

\subsection{FEROS}

We obtained three Doppler measurements of TOI-2525 with the FEROS spectrograph \citep{Kaufer1999} installed at the MPG 2.2\,m telescope in La Silla Observatory. These spectra were taken on BJD = 2458904.623, 2458914.612, 2458923.562, with the simultaneous  ThAr wavelength calibration technique. The exposure times were set to 1800 seconds, yielding an average signal-to-noise ratio of \textbf{25}. 
The FEROS data were reduced, extracted and analyzed with the \texttt{ceres} pipeline \citep{ceres} delivering radial velocity and bisector span measurements with a mean uncertainty of 19 m\,s$^{-1}$.
The RV datum obtained on BJD = 2458914.612, however, was a clear outlier with a poor accuracy due to bad weather conditions. Therefore, we could rely on only two FEROS spectra, which are fully consistent with the orbital fit to the TTVs and the PFS data.
However, the two FEROS RVs have no effective weight on the orbital fit, since their contribution is canceled by the two additional fitting parameters RV$_{\rm off.~FEROS}$ and RV$_{\rm jit.~FEROS}$. 
Thus, we decided to not use the FEROS RVs in our orbital analysis. The obtained FEROS radial velocities are presented in \autoref{table:RVs}.\looseness=-5

\section{Stellar parameters of TOI-2525}
\label{sec3}

TOI-2525 is an early K-type star visible in the southern hemisphere. 
The star has a distance of $400.0 \pm 2.3$ pc from the Sun and an apparent magnitude of $13.4$ mag in the $\tess$ bandpass. 
The atmospheric and physical parameters were obtained using three co-added FEROS spectra and the {\tt ZASPE} code \citep[][]{zaspe}. This code compares the stellar spectra with synthetic atmospheric models generated from the ATLAS9 model atmospheres \citep{atlas9}. The result is generated using the $\chi^2$ method at regions in the spectra which are most sensitive to changes. Parameters obtained from these regions are calculated iteratively. The errors of the values are generated with Monte Carlo simulations for the depth of the spectral lines. For TOI-2525, an effective temperature of $T_{\rm eff} = 5096 \pm 80 $ K, a metallicity of $[{\text{Fe/H}}] =  0.14\pm 0.05$ dex, with respect to the solar metallicity, and a projected rotational velocity of $v\text{sin} i = 1.5 \pm 0.3$ km\,sec$^{-1}$ were calculated.

    \begin{table}[t]

    \centering
     \caption{Planetary radii, orbital inclinations, and stellar density and mass estimates of the TOI-2525 system, derived during TTV extraction.}
    
    \label{table2}
 
    \begin{tabular}{p{5.6cm}  c crrrrrr}     

    \hline\hline  \noalign{\vskip 0.7mm}
    Parameter \hspace{60.0 mm}& median & $\sigma$ \\
    \hline \noalign{\vskip 0.7mm}

$r_b$ (R$_{\rm Jup.}$)             &	 0.88	  &  0.02 \\
$r_c$ (R$_{\rm Jup.}$)             &	 0.98	  &  0.02 \\
$i_b$ (deg)         &	 89.31	  &  0.03 \\
$i_c$  (deg)         &	 89.96	  &  0.03 \\

$\rho_\star$ (gr cm$^{-3})$ & 2.14 & 0.04\\
$M_{\star}$ ($M_{\odot}$) & 0.85 & 0.05 \\ 
     \hline \noalign{\vskip 0.7mm}

    \end{tabular}

\tablecomments{The remaining transit light curve data parameter estimates are listed in \autoref{table2a}, whereas the individual transit times for TOI-2525 b and TOI-2525 c are listed in \autoref{table:TTVdata}. \looseness=-4}

    \end{table}

The physical parameters are estimated as in \cite{brahm:2020}. We used the PARSEC stellar isochrones \citep{parsec}, which contain the absolute magnitudes of several band passes for a set of ages, masses, and metallicities. Since the latter were already calculated in the first step, they are fixed in the subsequent iteration. Using the spectroscopic temperatures, the {\em Gaia}  parallaxes, and the observed magnitudes, the age and the mass were obtained via a Markov Chain Monte Carlo (MCMC) exploration of the parameter space by using the \texttt{emcee} package \citep{emcee}. The result is an age of $3.99^{+4.3}_{-2.6}$ G\,yr, a mass of $\text{M}_\star = 0.849^{+0.024}_{-0.033}\,\text{M}_{\Sun}$ and a radius of $\text{R}_\star = 0.785 \pm 0.007\,\text{R}_{\Sun}$.
Our relatively small uncertainties in the {\tt ZASPE} stellar parameters, however, are internal and do not include possible systematic differences with respect to other stellar models. 
Therefore, we followed the prescription of \citet{Tayar2020} who suggest systematic uncertainty floor of order $\sim$5\% in mass, $\sim$4\% in radius and $\sim$2\% in temperature and luminosity, respectively \citep[see,][for more details]{Tayar2020}. We adopt 
these relative uncertainties to access more realistic stellar parameter errors through this work. 
The full set of atmospheric and physical parameters are listed in \autoref{table:phys_param}.


 \begin{figure}[tp]
    \centering
    \includegraphics[width=9cm]{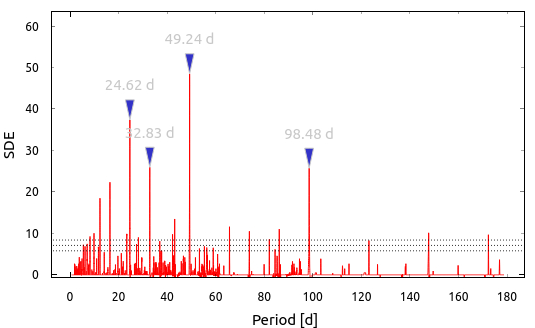}\\
    \includegraphics[width=9cm]{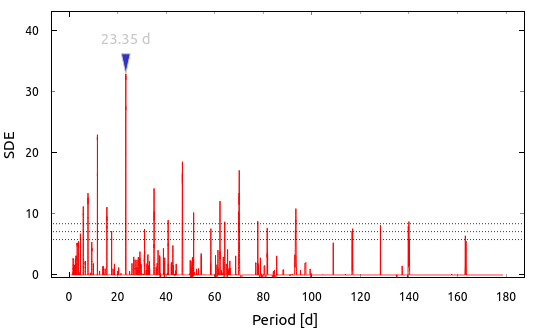}\\
 
    \caption{
    The top panel shows the TLS power spectra of the detrended $\tess$ 
    FFI light curve data of TOI-2525.    
    The planetary transit signal of the more massive planet c is detected at $P_c=49.24$\,d.
    The bottom panel shows the TLS power spectra of the residuals, which reveal the transit signal of the inner less-massive planet at $P_b=23.3$\,d.
    The remaining TLS peaks are harmonics and sub-harmonics of the transit signals.
    Horizontal dashed lines indicate the signal detection efficiency  \citep[SDE;][]{Hippke2019} power level of 5.7, 7.0, and 8.3, which correspond to the TLS false positive rate of 10\%,1\%, and 0.1\%.
    }
    \label{TLS_results} 
\end{figure}

\begin{figure*}[tp]
    \centering
    \includegraphics[width=\textwidth]{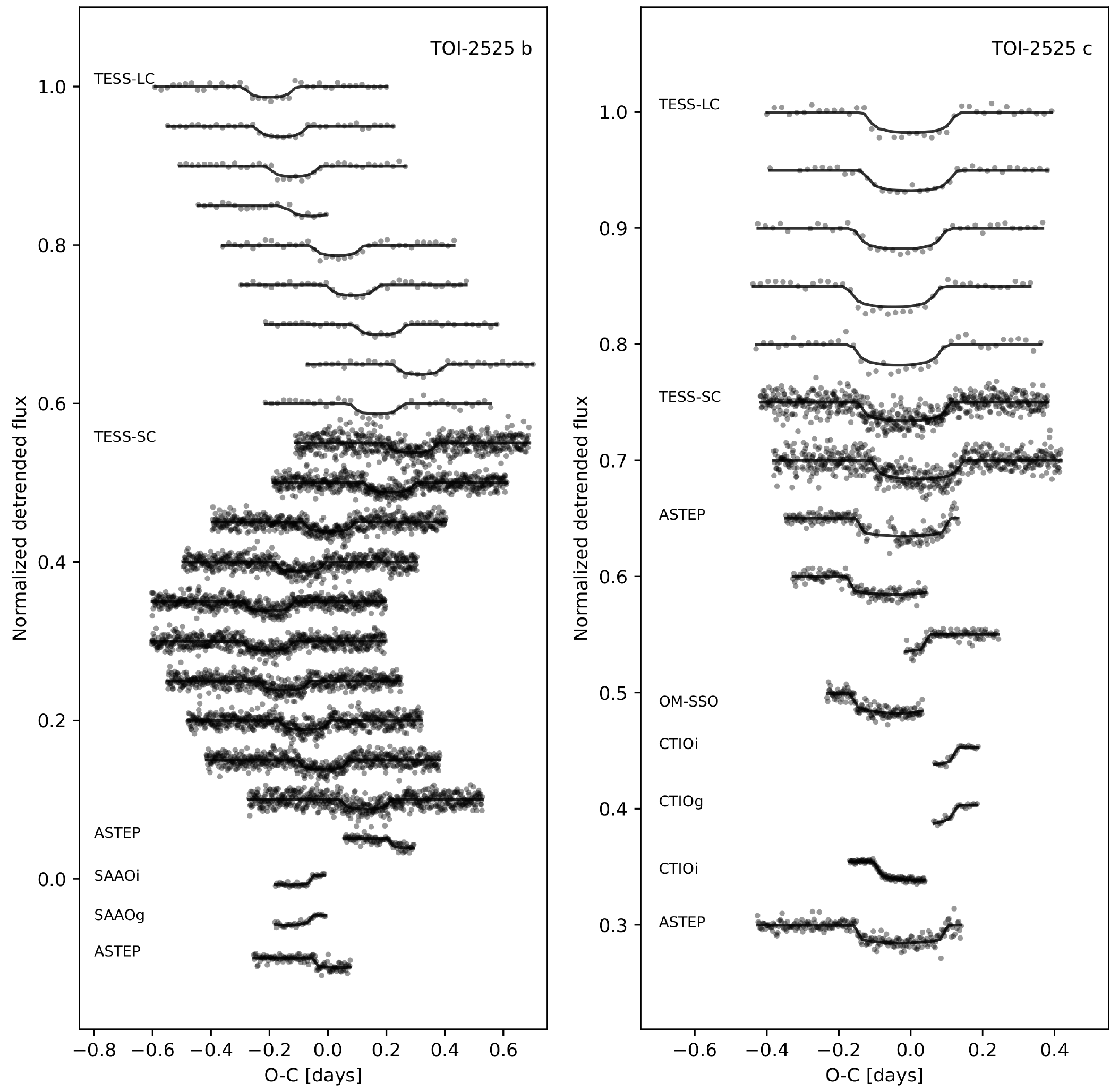} 

\caption{
    $\tess$ light-curve data and model of the TOI-2525 transit signals, plotted with arbitrary vertical offsets. 
    The left panel shows the transit signal of the inner planet TOI-2525 b
    and its strong TTV variation, whereas the right panel shows the same for the outer, more massive Jovian planet TOI-2525 c. The x-axis of the curves for TOI-2525 b is $t_{\rm n}$-1333.5070$\bmod$23.2915\,days, and for TOI-2525 c is $t_{\rm n}$-1335.4101$\bmod$49.2424\,days, respectively.
}
 
\label{new_data} 
\end{figure*}


\section{Analysis and results}
\label{sec4}

\subsection{Preliminary data vetting}


We identify the transit events of TOI-2525 b \& c in the first-year $\tess$ FFI light curves. Our preliminary transit characterization methodology includes detrending of the {\tt tesseract} FFI light curves with a robust (iterative) Mat\'ern GP kernel via the {\sc wotan} package \citep[see][]{Hippke2019} and a transit signal search with the \texttt{transitleastsquares} \cite[TLS;][]{Hippke2019b} algorithm.\looseness=-4

\autoref{TLS_results} shows our TLS results on the combined $\tess$
FFI 
light curve data of TOI-2525.     
We first detect the stronger transit signal of TOI-2525\,c with a period of $\approx$ 49.3$\,$d. We filter this signal by applying a Keplerian transit model with the obtained parameters from the TLS, and we seek additional transit signals in the model residuals. We clearly find the shallower, but more frequent, transit signal of the inner planet TOI-2525\,b at a period of $P_b=23.3$\,d.  \autoref{TLS_results} shows many other significant TLS peaks with a signal detection efficiency \citep[SED, see,][]{Hippke2019b}) of $>$ 8.3, which are nothing more than the integer sub-harmonics of the transit signals.

From the FFI data, we found that the light curve mid-transit time of TOI-2525\,b and TOI-2525\,c are not linear in time and show strong deviations in the expected time-of-transits, i.e., TTVs.\looseness=-4


The available Doppler data of TOI-2525 are too few for an independent RV validation of the two-planet system (see \autoref{table:RVs}). Since we identified the transit events in 2019, we have made many attempts to collect precise spectroscopic data from the southern hemisphere. However, TOI-2525 is faint and requires significant observational efforts and excellent sky conditions. This, in combination with the COVID-19 pandemic closures of the  ESO and Las Campanas observatories, prevented us from obtaining sufficient RV data. Nonetheless, we conclude that at this point, the PFS data alone are adequate for validation of the system when combined with the transit periods from $\tess$, and can contribute to the planetary mass estimates when combined with the strong TTVs.

\begin{figure*}[tp]
\begin{center}$
\begin{array}{ccc} 

\includegraphics[width=8.9cm]{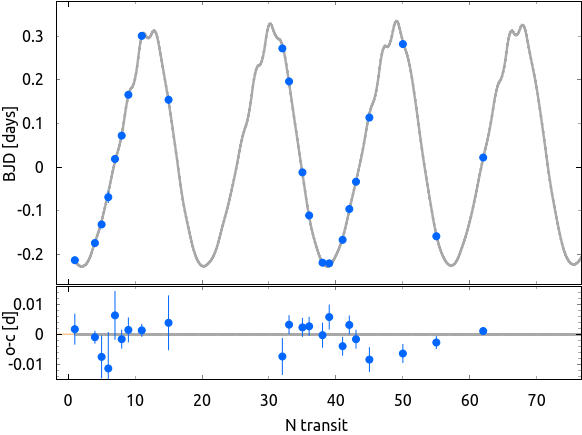} \hspace{0.2cm}
\includegraphics[width=8.9cm]{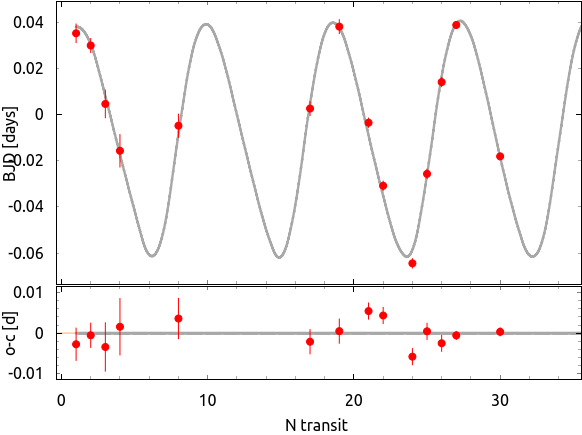} \\
 
\end{array} $
\end{center}

\caption{
    TTVs of TOI-2525 b (blue) and TOI-2525 c (red), fitted with a two-planet dynamical model jointly with the RVs from PFS.
     The top panels show the TTVs time series and best-fit model for each planet, whereas the bottom panels show the TTVs residuals.  
}
 
\label{TTV_plot1} 
\end{figure*}

\subsubsection{Extraction of TTVs}
\label{4.1.1}

Transit timing variations were estimated by fitting all the available photometric data using the {\sc exoplanet} software package \citep{exoplanet:joss}. We used the descriptive model {\sc TTVOrbit} which assumes  Keplerian orbits for each planet but allows for the central time of each transit to be a free parameter in the model. The parameters of the model are the values of $R_p$ for each planet, their impact parameters $b$, the transit times, the stellar mass $M_\star$ and density $\rho_\star$, quadratic limb darkening coefficients $u_1$ and $u_2$ for each instrument used, and parameters to describe trends and correlations in the data. Regarding the latter, we adopt a linear model in time for the ground based transits, which are often only partially observed, and a Gaussian process (GP) for each of the short and long cadence TESS datasets. The Gaussian process kernel adopted is a damped simple harmonic oscillator \citep{exoplanet:foremanmackey17} with $Q=1/3$, variance $\sigma_{\rm GP}$ and correlation length parameter $\rho_{\rm GP}$.\looseness=-4

\autoref{table2} lists the resulting light curve parameters, whereas
 \autoref{new_data} shows the resulting light curve models applied to all photometric data sets. Both TOI-2525 transit signals exhibit strong TTV libration. The values of the fitted transit times for each transit are listed in \autoref{table:TTVdata}.

\begin{figure*}[tp]
\begin{center}$
\begin{array}{ccc}

\includegraphics[width=6cm]{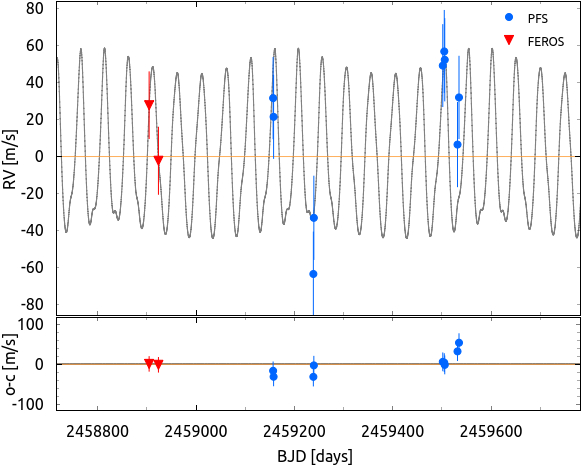}  \hspace{0.05cm} 
\includegraphics[width=6cm]{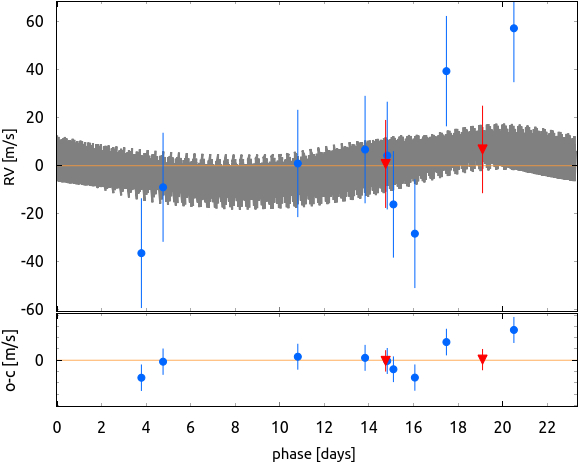}   \hspace{0.05cm} 
\includegraphics[width=6cm]{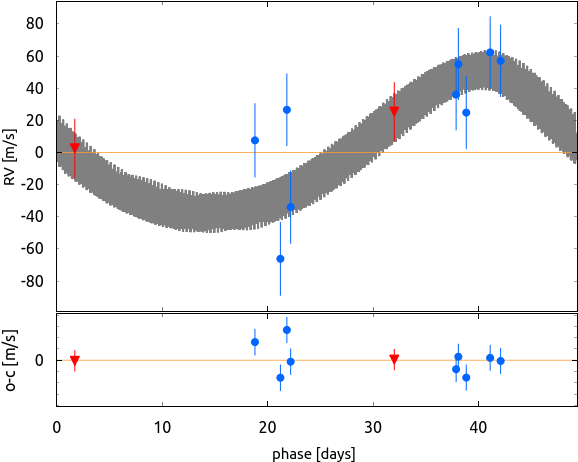}   \\

\end{array} $
\end{center}

\caption{
    RV component of the two-planet dynamical model of TOI-2525, constructed jointly with the TTVs data and with the RVs from PFS (blue circles). The two RVs from FEROS (red triangles) are overploted with an optimized RV offset to the model. The left panel shows the time series 
    and the N-body model,
    the middle and right panels show a phase-folded representation of the RV data, modeled with 
    the dynamical model (with an osculating period). The PFS data uncertainties include the estimated RV jitter, added in quadrature to the error budget, while the FEROS data are plotted with their nominal RV uncertainties. The small sub-panels show the RVs residuals.
}
 
\label{RV_plot1} 
\end{figure*}

\subsection{Orbital analysis}

\subsubsection{Joint RV and TTV analysis}
\label{4.1.2}

For the joint RV and TTV orbital analysis of the TOI-2525 system, we followed a similar route to the one we used in \citet{Trifonov2021b} for the modeling of the TOI-2022 system. We refer the reader to that paper for a more detailed description of the chosen methodology. Briefly, for TOI-2525 we performed orbital fitting with the {\tt Exo-Striker} exoplanet toolbox\footnote{\url{https://github.com/3fon3fonov/exostriker}} \citep{Trifonov2019_es}
by adopting the self-consistent dynamical model on the extracted TTVs from \autoref{4.1.1}. The TTV model can provide a relatively inexpensive orbital and dynamical solution to the system, in terms of CPU-time. However, it also comes with a severe ambiguity in eccentricity versus dynamical planetary-mass \citep[e.g.,][]{Lithwick2012,Dawson2019,Trifonov2021b}. Therefore, we fit the TTVs jointly with the RV data from PFS to better constrain the planetary dynamical masses and eccentricities. The RV model is intrinsic to {\tt Exo-Striker}, whereas the TTV model is wrapped around the \mbox{\sc TTVfast} package \citep{Deck2014}. The fitted parameters for each planet were the RV semi-amplitude $K$ (which is automatically converted to the dynamical planetary mass $m_p$), 
orbital period~$P$, eccentricity $e$, argument of periastron $\omega$, and mean anomaly $M_0$. Our TTVs+RVs modelling scheme allows a difference in the planetary inclinations ($\Delta i$ $\ge$ 0$^\circ$). Thus, we also model the orbital inclinations $i$ and the difference between the line of node $\Delta\Omega$. All these parameters are valid for
BJD = 2458333.52, which is an arbitrarily epoch, chosen slightly before the first transit event of the inner planet.
The RV data offset and  jitter\footnote{I.e., the unknown excess variance of the data, which we add in quadrature to the RV error budget while we evaluate the model's $-\ln\mathcal{L}$ \citep{Baluev2009}, and substantially build the nested sampling posteriors.} parameters of PFS added two more free parameters. 

Finally, our dynamical model, particularly the planetary masses, also depends on the stellar mass estimate of TOI-2525, for which we adopt a fixed value of 0.849 $M_\odot$. We set the time step in the dynamical model to $dt$ = 0.02 days to assure sufficient orbital resolution and accuracy.\looseness=-4

\begin{table*}[ht]

\centering   
\caption{{Nested sampling priors, posteriors, and the optimum $-\ln\mathcal{L}$ orbital parameters of the two-planet system TOI-2525 derived by joint dynamical modeling of TTVs ($\tess$, ASTEP, SSO, and LCOGT) and radial velocities (FEROS, PFS).}}
\label{NS_params}

 \begin{adjustwidth}{-3.1cm}{}
 \resizebox{0.92\textheight}{!}
 {\begin{minipage}{1.1\textwidth}

\begin{tabular}{lrrrrrrrrrrrr}     

\hline\hline  \noalign{\vskip 0.7mm}

\makebox[0.1\textwidth][l]{\hspace{45 mm} Median and $1\sigma$  \hspace{15 mm} Max. $-\ln\mathcal{L}$     \hspace{30 mm} Adopted priors  \hspace{10 mm} \hspace{1.5 mm} } \\
\cline{1-9}\noalign{\vskip 0.7mm}

Parameter &\hspace{10.0 mm} Planet b & Planet c &  & Planet b & Planet c  & & \hspace{10.0 mm}Planet b & Planet c  \\
\cline{1-9}\noalign{\vskip 0.7mm}

$K$  [m\,s$^{-1}$]            &  7.1$_{-0.3}^{+0.3}$ & 44.4$_{-1.5}^{+1.5}$ &  
                              &  7.1 & 44.4 &
                              &  $\mathcal{U}$(5.0,30.00) & $\mathcal{U}$(20.0,60.0)  &  \\ \noalign{\vskip 0.9mm}

$P$  [day]                    & 23.288$_{-0.002}^{+0.001}$ & 49.260$_{-0.001}^{+0.001}$ & 
                              & 23.289 & 49.260 &  
                              & $\mathcal{U}$(23.2,23.4) &  $\mathcal{U}$(49.1,49.4)  &  \\ \noalign{\vskip 0.9mm}
                              
$e$                           & 0.159$_{-0.007}^{+0.012}$ & 0.152$_{-0.005}^{+0.006}$ &
                              & 0.171 &  0.160 &  
                              & $\mathcal{U}$(0.0,0.4) &  $\mathcal{U}$(0.0,0.4)  &  \\ \noalign{\vskip 0.9mm}

$\omega$  [deg]               & 346.3$_{-0.7}^{+0.7}$  & 21.8$_{-0.8}^{+0.8}$ &
                              & 346.5  & 22.0 &   
                              & $\mathcal{U}$(0.0,360.0) &  $\mathcal{U}$(0.0,360.0) &  \\ \noalign{\vskip 0.9mm}

$M_{\rm 0}$  [deg]            & 120.8$_{-0.5}^{+0.6}$  & 71.2$_{-1.0}^{+1.2}$ &
                              & 120.6  & 71.6 &  
                              & $\mathcal{U}$(0.0,360.00) &  $\mathcal{U}$(0.0,360.00)  &  \\ \noalign{\vskip 0.9mm}

$\lambda$  [deg]          &  107.0$_{-0.8}^{+0.8}$  & 93.4$_{-1.0}^{+0.9}$ & 
                              & 107.1  & 93.1 &    
                              &  (derived) &   (derived)  &  \\ \noalign{\vskip 0.9mm}

$i$          [deg]            & 89.96$_{-0.07}^{+0.08}$  & 89.99$_{-0.06}^{+0.08}$ &
                              & 89.97  & 89.99& 
                              & $\mathcal{N}$(90.0,0.1) &  $\mathcal{N}$(90.0,0.1)  &  \\ \noalign{\vskip 0.9mm}  
                              
$\Omega$     [deg]            & 0.0   & 2.0$_{-1.2}^{+ 1.9}$ &
                              & 0.0  & 1.9&  
                              & (fixed) &  $\mathcal{N}$(0.0,15.0)  &  \\ \noalign{\vskip 0.9mm}                               


 


$\Delta i$  [deg]             & 2.0$_{-1.2}^{+1.9}$  & $\dots$ &  
                              &  1.9 & $\dots$ &     
                              &  (derived) &   $\dots$  &  \\ \noalign{\vskip 0.9mm}

$a$  [au]                     &  0.1511$_{-0.0025}^{+0.0025}$  & 0.2491$_{-0.0042}^{+0.0041}$  &  
                              &  0.1511 & 0.2491&     
                              &  (derived) &   (derived)  &  \\ \noalign{\vskip 0.9mm}

$m$  [$M_{\rm jup}$]          & 0.088$_{-0.004}^{+0.005}$  & 0.709$_{-0.034}^{+0.034}$ & 
                              & 0.089 & 0.710 &     
                              &  (derived) &   (derived)  &  \\ \noalign{\vskip 0.9mm}

 $\rho$ [g\,cm$^{-3}$]       & 0.174$_{-0.015}^{+0.016}$ & 1.014$_{-0.076}^{+0.084}$ & 
                              & 0.174 & 1.014 &     
                              &  (derived) &   (derived)  &  \\ \noalign{\vskip 0.9mm}





RV$_{\rm off.}$ PFS  [m\,s$^{-1}$]               &       $-$28.5$_{-7.4}^{+7.1}$  &   $\dots$ &   &    $-$32.2     &  $\dots$   & &$\mathcal{U}$(-300.00,100.0)&  $\dots$ \\ \noalign{\vskip 0.9mm}

RV$_{\rm jit.}$ PFS  [m\,s$^{-1}$]                &       26.8$_{-5.4}^{+6.8}$  &  $\dots$ &    & 21.5 &    $\dots$   &          & $\mathcal{J}$(0.0,50.0)&  $\dots$    \\ \noalign{\vskip 0.9mm}


\\
\hline \noalign{\vskip 0.7mm}

\end{tabular}

\end{minipage}}
\end{adjustwidth}
\tablecomments{The orbital elements are in the Jacobi frame and are valid for epoch BJD = 2458333.52. The adopted priors are listed in the right-most columns and their meanings are $\mathcal{U}$ -- Uniform, $\mathcal{N}$ -- Gaussian, 
    and $\mathcal{J}$ -- Jeffrey's (log-uniform) priors. The derived planetary posterior parameters of $a$, and $m$ are calculated taking into account the stellar parameter uncertainties (see Note in \autoref{table:phys_param}.)
}
\end{table*}

We ran a nested sampling (NS) scheme \citep{Skilling2004}, which allowed us to efficiently explore the complex parameter space of osculating orbital elements and study the parameter posteriors and overall dynamics. Our NS run was performed with the {\sc dynesty} sampler \citep{Speagle2020}, which is integrated into {\sc Exo-Striker}. We ran 100 "live-points" per fitted parameter using the "Dynamic" NS scheme, focused on 100\% posterior convergence instead of log-evidence \citep[see,][for details]{Speagle2020}
Parameter priors were estimated by running several experimental NS runs 
and adopting a wide range of uniform parameter priors. After a few consecutive NS runs, we narrowed the adequate parameter space to be explored. We note that we adopted very narrow priors on the orbital inclinations to assure the TTV+RV model parameter space is consistent with the inclination estimates extracted from the TTVs. Thus, we eliminate configurations that could explain the RV data but would not lead to transit events (i.e., impact parameters $b_{\rm b}$ and $b_{\rm c}$, which are inconsistent with the light curve signal). The final adopted parameter priors are listed in \autoref{NS_params}.

\autoref{TTV_plot1} and \autoref{RV_plot1} show the TTVs and RVs data together with the best-fit joint dynamical model of TOI-2525. 
The left and the right panels of \autoref{TTV_plot1} show the TTVs of TOI-2525\,b \& c, respectively, fitted with the best-fit TTV model. 
\autoref{RV_plot1} shows the RV component of the best-fit model applied to the PSF RVs. In \autoref{RV_plot1} also shows the two FEROS RVs fitted independently to the best-fit with an optimized offset.
The middle and the right panel of  \autoref{RV_plot1} show a phase-folded representation of the RV signals of TOI-2525\,b \& c, respectively, modeled with an osculating period. The signal of the inner planet TOI-2525\,b is strongly overshadowed by the dominating signal of the outer one due to the large difference in K amplitudes (see \autoref{NS_params}).


The final posterior probability distributions with an RV linear trend are shown in \autoref{Nest_samp_ttv}.
Our final estimates for TOI-2525\,b \& c lead to planetary orbital periods of 
P$_b$ = 23.288$_{-0.002}^{+0.001}$ days, and 
P$_c$ = 49.260$_{-0.001}^{+0.001}$ days, 
eccentricities of  
$e_b$ = 0.159$_{-0.007}^{+0.012}$ and 
$e_c$ = 0.152$_{-0.005}^{+0.006}$, and dynamical masses of
$m_b$ = 0.088$_{-0.004}^{+0.005}$ $M_{\rm jup}$ and 
$m_c$ = 0.709$_{-0.034}^{+0.034}$ $M_{\rm jup}$. 
The mutual inclination is constrained to 
$\Delta i$ = 2.0$_{-1.2}^{+1.9}$ deg.
Given the planetary radii obtained during the TTV extraction in \autoref{4.1.1}, we derive a remarkably low density of $\rho_b$=0.174$_{-0.015}^{+0.016}$\,g\,cm$^{-3}$ for the inner planet, and $\rho_c$=1.014$_{-0.076}^{+0.084}$\,g\,cm$^{-3}$, for the outer one, respectively. 
The full list of posterior and maximum $-\ln\mathcal{L}$ (i.e., best-fit) estimates was derived from the joint TTVs+RVs model
and listed in \autoref{NS_params}.

\begin{figure*}[tp]
\begin{center}$
\begin{array}{ccc}

\includegraphics[width=16.0cm]{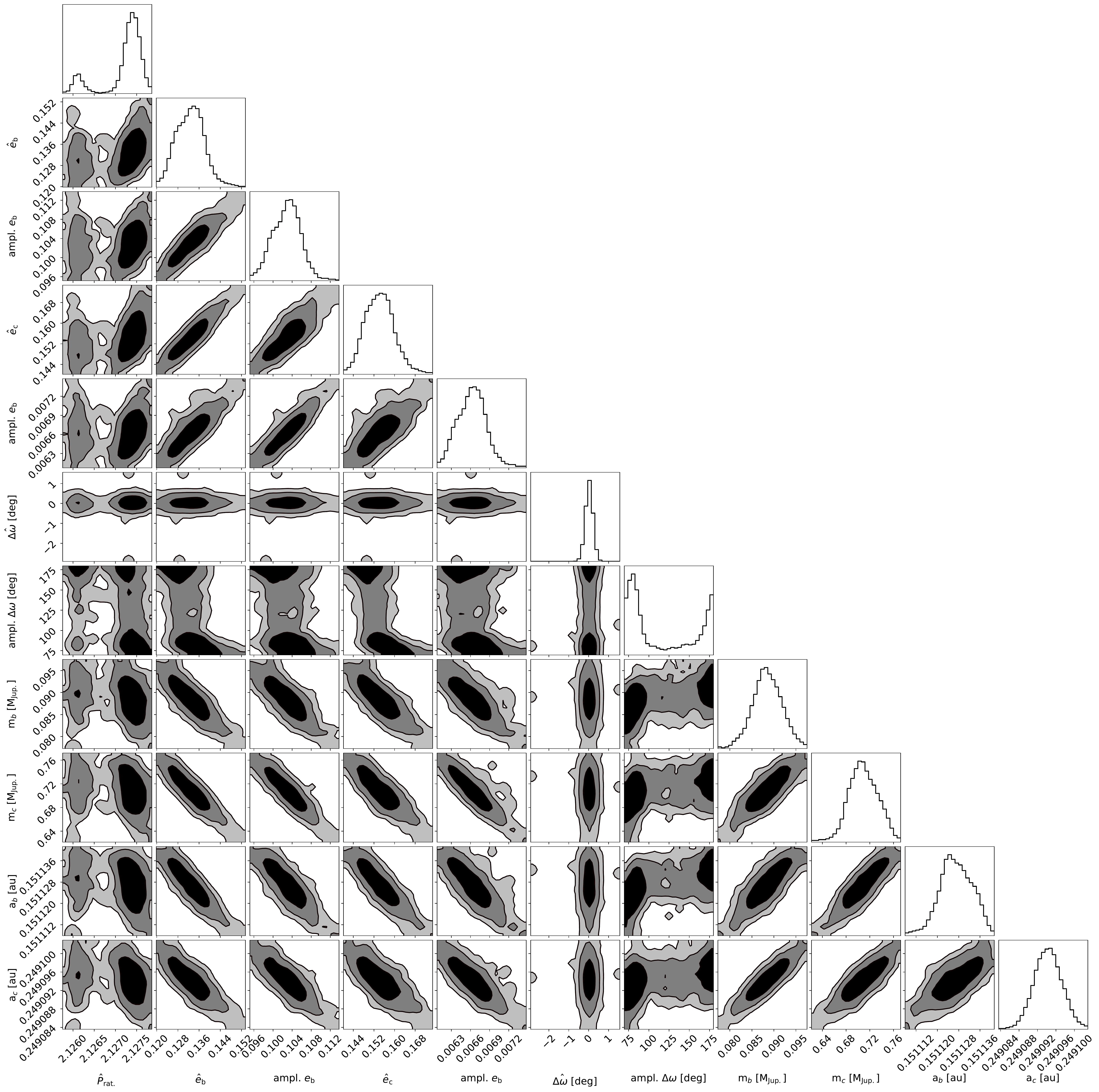}

 \end{array} $
\end{center}

\caption{Posteriors of the dynamical properties at the 2:1 period ratio commensurability of the two-planet system TOI-2525. The posteriors are generated by randomly drawing 10\,000 samples from the TTV+RV dynamical model. 
Each sample is tested for stability and the overall dynamical properties are evaluated at the 2:1 period ratio commensurability. The derived dynamical parameters are: mean period ratio $\hat{P}_{\rm rat.}$, mean eccentricities $\hat{e}_{\rm b}$, $\hat{e}_{\rm c}$, mean orbital aligment $\hat{\Delta\omega}$, and their peak-to-peak amplitudes Ampl. $e_{\rm b}$, Ampl. $e_{\rm c}$, Ampl. $\Delta\omega$, and the planetary dynamical masses and semi-major axes. Note that the posterior distribution of $\theta_1$, and $\theta_2$ are not shown, since these exhibit circulation between 0$^\circ$ and 360$^\circ$. 
The black contours on the 2D panels represent the 1, 2 and 3$\sigma$ confidence level of the overall posterior samples.
}
 
\label{dyn_samp} 
\end{figure*}




\begin{figure*}
\includegraphics[width=6cm]{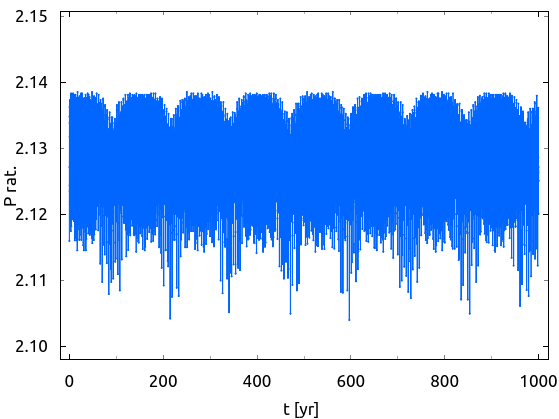}
\includegraphics[width=6cm]{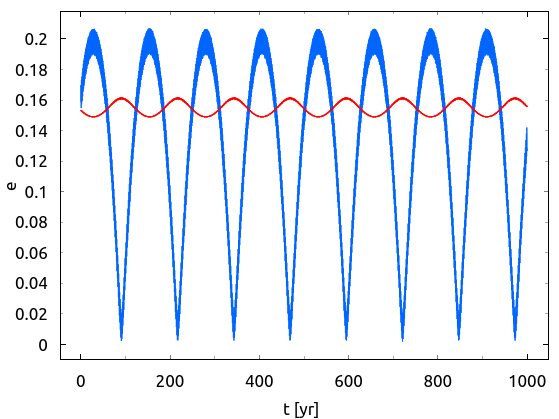} 
\includegraphics[width=6cm]{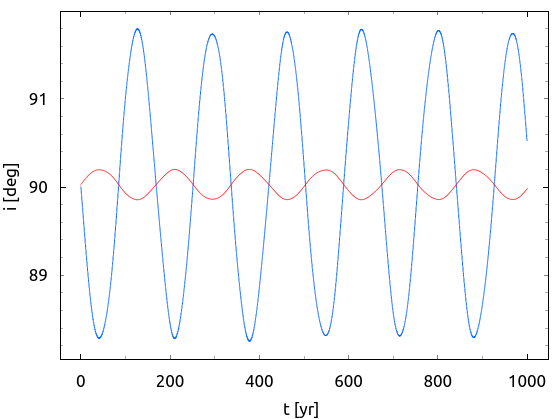} 

\caption{Orbital evolution of the TOI-2525 system for a 1\,000 yr long N-body integration using the Wisdom-Holman scheme. The left panel shows the evolution of the planetary periods, the middle panel shows the eccentricities $e_{\rm b}$ (blue) and $e_{\rm c}$ (red) of the best-fit N-body model, while the right panel shows the evolution of the line-of-sight inclinations $i_{\rm b}$  (blue) and $i_{\rm c}$ (red).}
\label{evol_plot} 
\end{figure*}

\subsubsection{Joint RV and photodynamical analysis}
\label{4.1.3}

A complementary analysis of the light curves and RVs of TOI-2525 was performed using the  \texttt{flexi-fit}\footnote{\url{https://gitlab.gwdg.de/sdreizl/exoplanet-flexi-fit}} python package. 
For modeling the photometric data, \texttt{flexi-fit} employs analytic transit light curves \citep{Mandel2002} with a quadratic limb-darkening. Dynamical effects are included via the \texttt{rebound} $N$-body package \citep{Rein2012}. Our orbital analysis with \texttt{flexi-fit} is performed in Jacobi-coordinates using the  \texttt{ias15} integrator \citep{Everhart1985}, which automatically determines the numerical time step $dt$ down to machine precision. 
The fitting with \texttt{flexi-fit} relies  on a Markov Chain Monte Carlo (MCMC) parameter sampling provided by the \texttt{emcee} sampler \citep{emcee}, which converges to the posterior probability distribution.

The \texttt{flexi-fit} model provides orbital parameters valid for the epoch BJD = 2458333.52, which is the same as the one chosen for joint TTVs + RVs analysis (see \autoref{4.1.1}).
We chose uniform parameter priors  
 which define an equal probability of occurrence between predefined borders. Most of the priors were generously defined, covering a large section of parameter space. 
The only priors worth mentioning are for the periods $P_b\in\mathcal{U}$(23.1,23.4), $P_c\in\mathcal{U}$(49.1,49.4) days which is sufficient considering the mid-transit times and the inclination $i_b\in\mathcal{U}$(81,99), $i_c\in\mathcal{U}$(81,99) degrees, which can be restricted to these intervals for geometrical reasons, because we do observe transits. 
The remaining priors of the photodynamical model are listed in \autoref{photodynparameters}. 
Having the stellar parameters (\autoref{table:phys_param}) and quadratic limb-darkening coefficients given, our free orbital parameters for each planet are the orbital period $P$, the mass $m$, the eccentricity $e$, the longitude of periastron $\omega$, the time of first conjunction with respect to the start of integration $t_{conj}$, the inclination $i$ and the planet to star radius ratio $R_p/R_s$. The longitude of the ascending node $\Omega$ was fixed $\Omega_b=0$ for the inner planet and free for the outer planet. 
For the instrumental and observational errors from the radial velocity data from PFS, a jitter parameter and an offset were determined. 

The photometric data were separated into 8 distinct data sets: TESS FFIs (year 1), TESS PDC (year 3), ASTEP 1-6, CTIO, SSO and SAAO 1-2.
The photometric data from the TESS-FFIs in the first TESS-year were de-trended sector-wise using the \texttt{rspline} in the \texttt{wotan} package \citep{Hippke2019} package. For the third year of the TESS observations, the 2-min cadence data from the PDCSAP pipeline, as well as the ground based photometry data from ASTEP, CTIO, SSO and SAAO were included into the model without further de-trending. 
For each data set, an offset was fitted. In total, this adds up to 30 dimensions of fitting, all the other parameters including the RV semi-amplitude $K$ are derived.

\autoref{Transits2525b} and \autoref{Transits2525c} show an impression of the light curves for TESS FFI, TESS PDC, ASTEP, and SSO, fitted jointly with \texttt{flexi-fit}. The posterior distributions of the MCMC can be seen in \autoref{photodyposterior}. 
Both masses $m_b=0.084_{-0.005}^{+0.005}\mathrm{M_{Jup}}$ and $m_c=0.657_{-0.032}^{+0.031}\mathrm{M_{Jup}}$ were estimated to an error of less than 10\%. Combined with the planetary radii of $R_b=0.774_{-0.010}^{+0.010}R_{Jup}$ and $R_c=0.904_{-0.010}^{+0.010}R_{Jup}$ we got a very low density of $\rho_b=0.226_{-0.014}^{+0.015}$ g\,cm$^{-3}$ for the inner planet. The density for the outer planet is $\rho_c=1.11_{-0.07}^{+0.07}$ g\,cm$^{-3}$. 
The planetary densities are slightly larger than our results from the TTVS+RV analysis. The orbital parameters, offsets, and jitter parameters from the photodynamical orbital analysis are listed in \autoref{photodynparameters}.

Here we note that applying a photodynamical model on a system having RV data and large photometric data sets (of the order 200\,000 data points) for more than $1000~\mathrm{d}$ is computationally expensive. On a standard CPU, an iteration of the MCMC takes approximately $2.5~\mathrm{s}$, which keeps the length of the chain limited. Since the analysis was very time-consuming, the analysis was stopped after 100\,000 MCMC iterations. The convergence criterion of 20 \% mean acceptance fraction was not reached \citep[see,][]{emcee}, but the physical and dynamical parameters correspond to those from the light curve and RV+TTV analysis presented in \autoref{4.1.1} and \autoref{4.1.2} and give us confidence in our orbital solution.
\looseness=-4

\begin{figure*}
    \includegraphics[width=9cm]{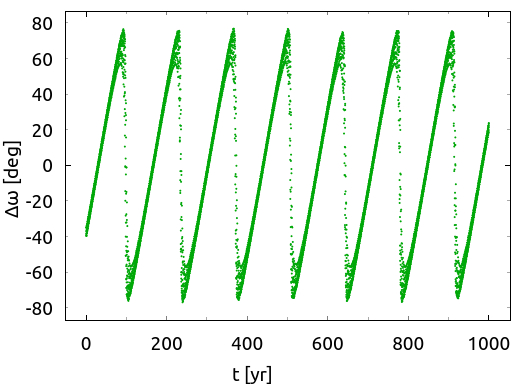} \put(-215,165){a)} 
    \includegraphics[width=9cm]{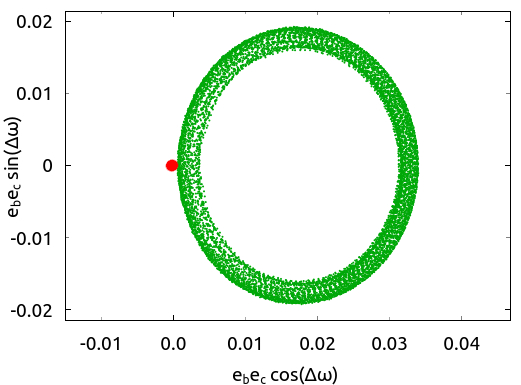} \put(-215,165){b)} \\ 
    \includegraphics[width=9cm]{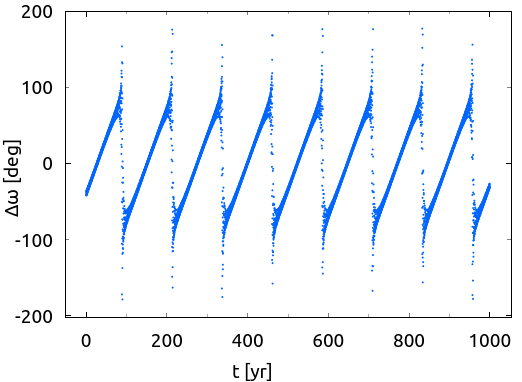}  \put(-215,165){c)} 
    \includegraphics[width=9cm]{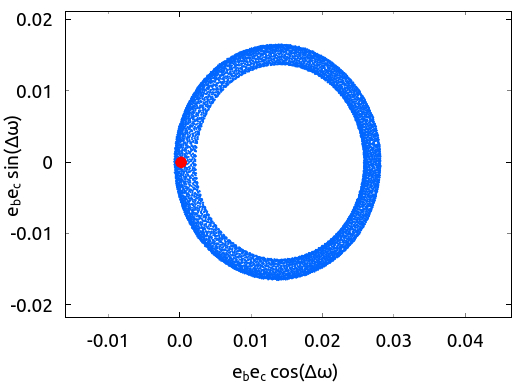}  \put(-215,165){d)} 
    
    \caption{
 The top panels; a) and b) show the same as in \autoref{evol_plot}, but for the evolution of the apsidal angle $\Delta\omega$ = $\omega_{\rm b}$ - $\omega_{\rm c}$, and the trajectory evolution as a function $e_b e_c$ times sine/cosine of  $\Delta\omega$, respectively.
The trajectory just misses the origin (red dot, centred at [0,0]), which leads to libration of $\Delta\omega$ with a semi-amplitude of $\sim$ 80$^\circ$.
The bottom panels c) and d) show the same as in a) and b), but for a random posterior fit which exhibits circulation of $\Delta\omega$ (see \autoref{dyn_samp}). The trajectory evolution in this case passes through the origin, which leads to brief episodes of circulation of $\Delta\omega$.
    }
    \label{oct_evol_traj}
\end{figure*}

\subsection{Dynamics and long-term stability}
\label{sec4.3}

Following \citet{Trifonov2021b}, we inspected the Hill \citep[see,][]{Gladman1993} and 
Angular Momentum Deficit \citep[AMD, see,][]{Laskar2017} stability criteria of the TOI-2525 system.
In terms of the classical Hill stability criterion, the TOI-2525 planetary system is predicted to be stable. We estimate a mutual Hill distance of $\sim$ 7.4 R$_{\rm Hill,m}$, which is above the widely accepted distance of $\sim$ 3.5 R$_{\rm Hill,m}$ for the system to be considered Hill-stable.
However, accounting for the estimated orbital eccentricities, semi-major axes, mutual orbital inclination, and planetary masses of TOI-2525 b \& c, the AMD criterion suggests that the TOI-2525 planetary system is unstable. The AMD criterion is very sensitive to eccentricities. Thus, the moderately eccentric orbits of the pair are the reason for the negative AMD result. 

As discussed in \citet{Trifonov2021b}, the AMD and Hill stability criteria
are only a proxy for long-term stability and do not account for the system's dynamics near mean motion resonances.
Therefore, to test the long-term dynamics and possible MMR in the system, we adopt exactly the same $N$-body numerical setup used in our recent analysis of TOI-2202 \citep{Trifonov2021b}. This is adequate because TOI-2202 and TOI-2525 share somewhat similar physical and orbital characteristics. 
We refer the reader to Section 5 in \citet{Trifonov2021b} for more details about the stability test performed here for TOI-2525. Briefly, we performed numerical integrations using a custom version of the Wisdom-Holman $N$-body algorithm \citep[][]{Wisdom1991}, which directly adopts and integrates the Jacobi orbital elements from the posterior orbital analysis.   
We test the stability of the TOI-2525 system up to 1\,Myr with a small time-step of 0.02 \,d for 10\,000 randomly chosen samples from the achieved orbital parameter posteriors from the TTV+RV dynamical modelling scheme. 
We automatically monitored the evolution of the planetary semi-major axes, eccentricities, secular apsidal angle $\Delta\omega$ = $\omega_{\rm b}$ - $\omega_{\rm c}$, and first-order 2:1 MMR angles $\theta_1 =  \lambda_{\rm b} - 2\lambda_{\rm c} + \omega_{\rm b},~~\theta_2 =  \lambda_{\rm b} - 2\lambda_{\rm c} + \omega_{\rm c}$
\citep[where $\lambda_{\rm b,c}$ = $M_{\rm b,c}$ + $\omega_{\rm b,c}$ is the mean longitude of planet b and c, respectively; see][]{Lee2004}.
These angles are important for libration in secular apsidal alignment or mean motion resonance.

We found that all examined 10\,000 samples are stable for 1\,Myr.~\autoref{dyn_samp} presents the resulted posterior probability distribution of some of the important dynamical properties of the system, such as mean period ratio $P_{\rm rat.}$, mean eccentricities $\hat e_{\rm b}$, $\hat e_{\rm c}$, their peak-to-peak amplitudes Ampl. $e_{\rm b}$, Ampl. $e_{\rm c}$, dynamical masses of the planets $m_{\rm b}$, $m_{\rm c}$, and the orbital semi-major axes $a_{\rm b}$, $a_{\rm c}$. \autoref{dyn_samp} shows that the period ratio evolution is bimodal, oscillating either around $\sim$ 2.127, or more plausibly around $\sim$ 2.126, which are too far from the exact 2:1 period ratio for the system to librate in the 2:1 MMR. This is confirmed by the circulation of $\theta_1$ and $\theta_2$, which distributions are not shown in \autoref{dyn_samp}.
On the other hand, we observe a bi-modality of the evolution of $\Delta\omega$, which is intriguing. About half of the sampled initial conditions lead to libration of $\Delta\omega$ about $0^\circ$ with a mean semi-amplitude around 80$^\circ$, whereas the rest of the sampled initial conditions seem to exhibit a mean semi-amplitude around $180^\circ$, i.e., circulation (see \autoref{sec5.1} for more explanation on this dynamical behavior).

\autoref{evol_plot} shows an example of a 1000\,yr extent of the orbital evolution simulation started from the best-fit (i.e., maximum $-\ln\mathcal{L}$, see \autoref{NS_params}). We show the evolution of the eccentricities, mutual period ratio $P_{\rm rat.}$ the eccentricities $e_{\rm b}$ and $e_{\rm c}$, and the orbital inclinations $i_{\rm b}$ and $i_{\rm c}$. The TOI-2525 system is consistent with moderate eccentricity evolution, and appears to osculate outside of the low-order eccentricity-type 2:1 MMR. It is interesting that the secular evolution time scales are rather short, of the order of $\sim$ 120\,yr. Therefore, future observations might be sensitive to transit depth variations. Furthermore, TOI-2525\,b may soon become a non-transiting planet (different timescales are possible within the mutually inclined posteriors).

\section{Discussion}
\label{sec5}

\subsection{Dynamical state of the system}
\label{sec5.1}

Our numerical orbital analysis of the system's configuration revealed that the TOI-2525 pair of planets are outside of the 2:1 MMR.
However, the posterior of the apsidal alignment angle $\Delta\omega$ shows bi-modality with approximately equal fraction exhibiting libration and circulation. Thus, we took a closer look at the dynamical evolution of these two populations. The upper and lower panels of \autoref{oct_evol_traj} show 
the characteristic evolution of $\Delta\omega$ for the same stable configuration as in \autoref{evol_plot} (i.e., our best TTV+RV fit)
and for a random posterior fit with $\Delta\omega$ in circulation, respectively.
The left panels of \autoref{oct_evol_traj} show
$\Delta\omega$ as a function of time, whereas the right panels show the trajectory in the polar plot of $e_b e_c \cos\Delta\omega$ vs $e_b e_c \sin\Delta\omega$. In both cases, the polar plot shows that the system circulates around a point along the positive $e_b e_c \cos\Delta\omega$ axis, i.e., an aligned configuration. For the best fit shown in the upper panels, the trajectory in the polar plot is just small enough to miss the origin, and $\Delta\omega$ exhibits large amplitude libration about $0^\circ$. For the example shown in the lower panels, the trajectory in the polar plot is just large enough to touch the origin, which leads to brief episodes of circulation of $\Delta\omega$. Thus, the bi-modality of the apsidal alignment angle is simply due to slightly different sizes of the trajectories in the secular polar variables.

\begin{figure} 
\begin{center}$
\begin{array}{ccc}

\includegraphics[width=8.7cm]{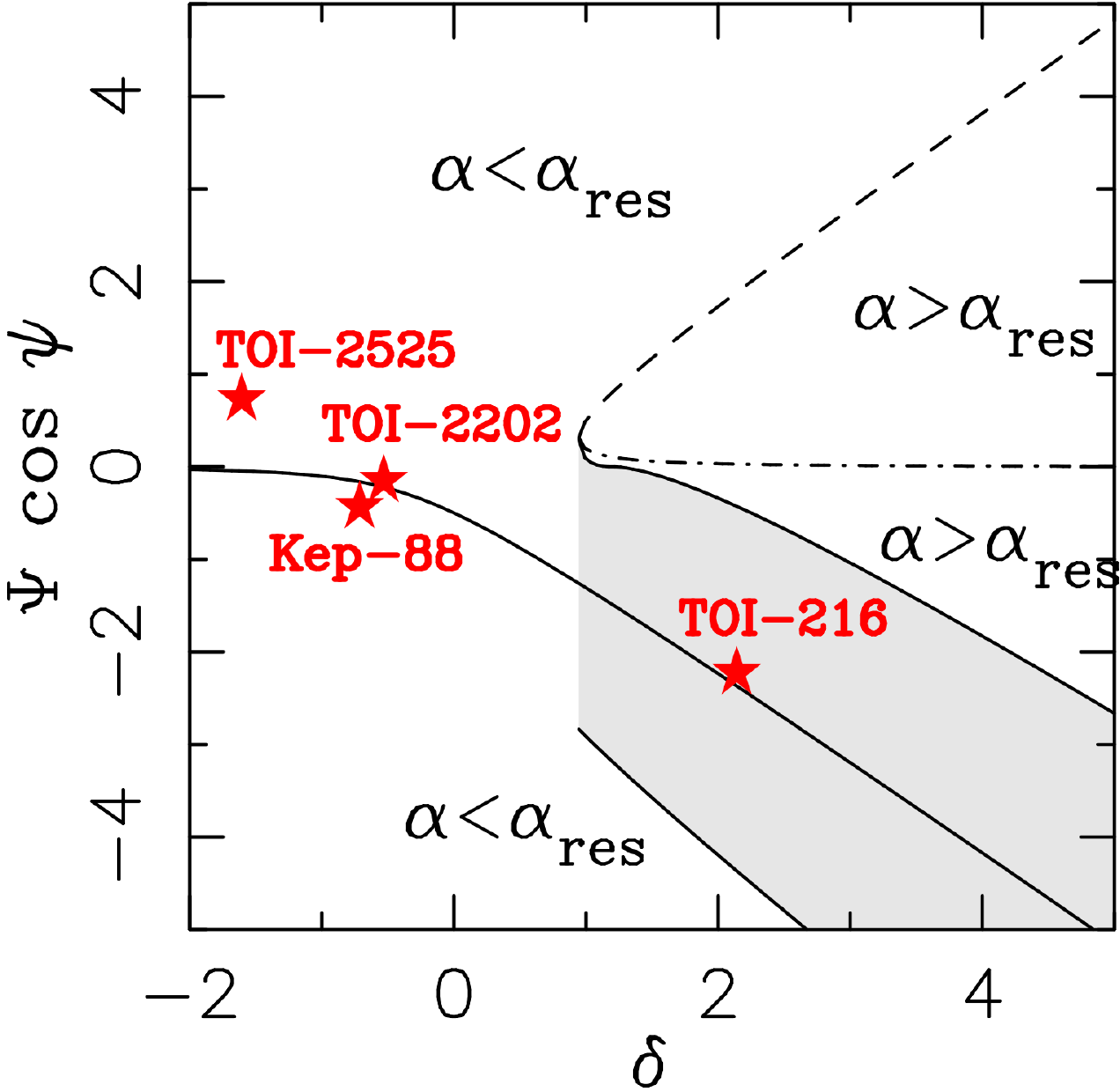}\\

\end{array} $
\end{center}

\caption{The 2:1 MMR structure diagram following \citet{Nesvorny2016}. Four planetary systems are plotted: TOI-2525 (this work), TOI-2202 \citep{Trifonov2021b}, TOI-216 
\citep{Dawson2021} and Kepler-88 \citep{Nesvorny2013}. 
The separatrices and stable point are solid. The dotted line is an approximation of the stable point inside the resonance, as explained in  \citet{Nesvorny2016}. The dashed line is the unstable point. Dot-dashed is the stable point. See \citet{Nesvorny2016} for more explanations. 
}
\label{evol_plot3} 
\end{figure}

\begin{figure*} 
\begin{center}$
\begin{array}{ccc}

\includegraphics[width=9cm,height=6.5cm]{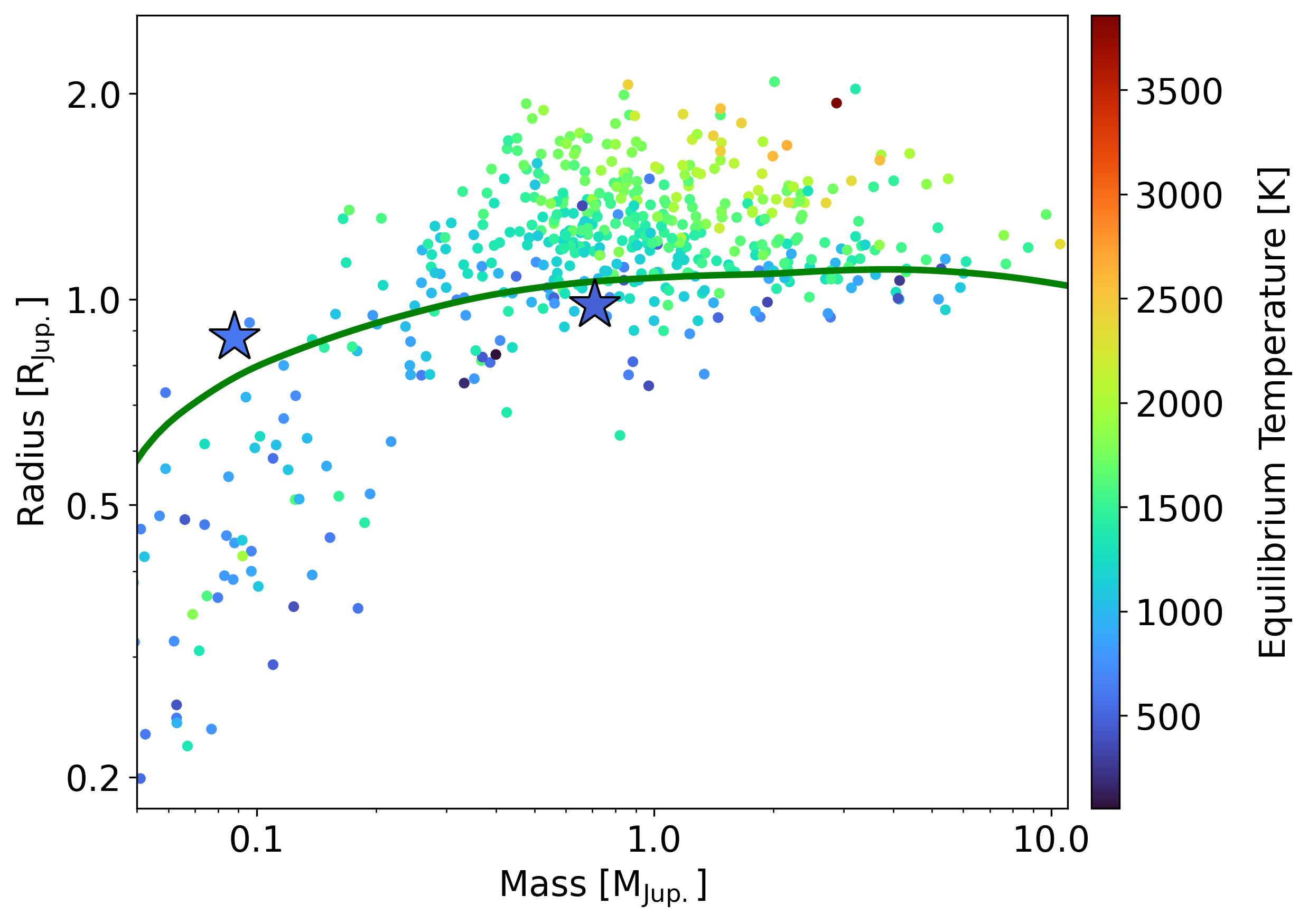} \put(-215,165){a)} 
\includegraphics[width=9cm,height=6.5cm]{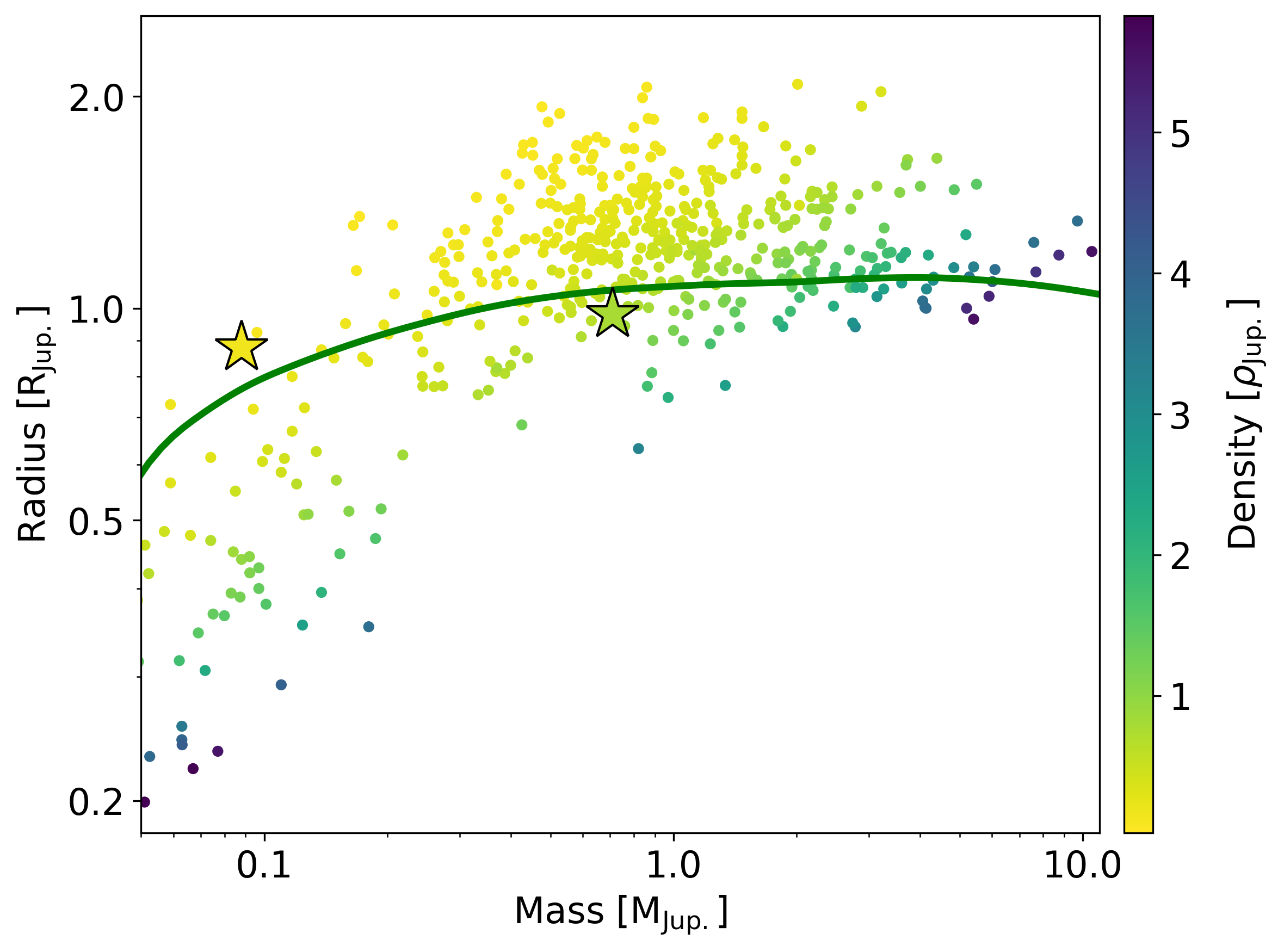}\put(-215,165){b)} \\

\end{array} $
\end{center}

\caption{Radius-mass demographics for all giant transiting planets with measured
masses by TTVs or RVs. The position of TOI-2525\,b and c are plotted with "star" symbols. Panel a) shows the mass-radius relation color-coded by T$_{\rm eq.}$. Panel b) shows the same as panel a), but is color-coded with the planetary density.
The solid curve corresponds to the predicted radius using the models of \citet{Fortney2007} given the estimated luminosity of TOI-2525, a semi-major axis=0.2 AU, and an age of 3.1 G\,yr.\looseness=-5
}
 
\label{hhe_fraction} 
\end{figure*}

\begin{figure} 
\includegraphics[width=8.7cm]{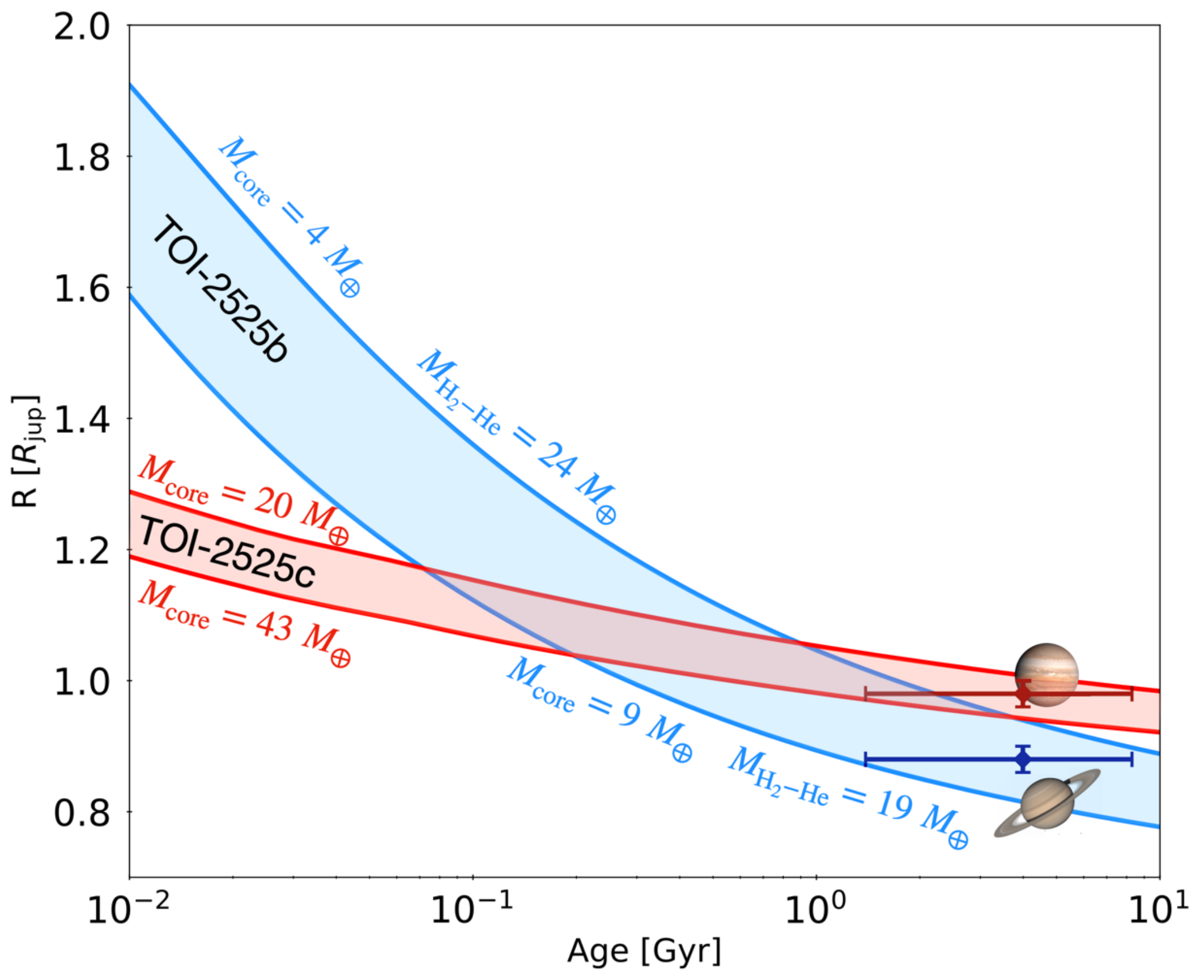}\\
\caption{Evolution models of TOI-2525\,b and TOI-2525\,c. All models assume a central ice-rock core surrounded by a hydrogen-helium envelope of solar composition. For both planets, the range of core masses is shown. For TOI-2525\,b, the range of envelope masses is also shown. The errorbars correspond to observational constraints on their age and radius, compared to the ones of Jupiter and Saturn.
}
\label{interiors} 
\end{figure}

We ruled out a 2:1 MMR librating configuration of the system based on the orbital posterior probability distribution constructed from the available transit and RV data. Similarly to the TOI-2202 system, the osculating period ratio of the TOI-2525 pair of planets is well above two, consistent with the prominent peak of period ratios of planet pairs observed by the Kepler mission \citep{Lissauer2011, Fabrycky2014}. \autoref{evol_plot3} shows an analytical analysis of the resonant and near-resonant dynamics in the 2:1  commensurability, following \citet{Nesvorny2016}. 
The constant $\delta$ is an orbital 
invariant that defines the position of the system relative to the 2:1  period ratio, $\psi$ is a combination of the resonant angles $\theta_1$ and $\theta_2$, whereas the variable $\Psi$ is a combination of planetary masses, semi-major axes and eccentricities \citep[see,][for details]{Nesvorny2016}. 
\autoref{evol_plot3} is an updated version of 
Figure 12 in \citet{Trifonov2021b}, which now includes the position of TOI-2525, in addition to TOI-2202 \citet{Trifonov2021b}, TOI-216 \citep{Dawson2021} and Kepler-88 \citep{Nesvorny2013}.  
The median posterior probability values of TOI-2525, listed in \autoref{NS_params}, lead to $\delta \simeq$ -1.60, therefore, firmly 
outside the libration region together with Kepler-88 and TOI-2525 systems.
The only system that is in a 2:1 MMRs is TOI-216 \citep[see,][for details]{Dawson2021}.

\subsection{Possible formation mechanisms}

An interesting question is how frequently state-of-the-art planet formation models produce systems like TOI-2525, TOI-2202, or TOI-216 and what drives their formation.
We explored the abundance and origins of systems with multiple giant planets in synthetic planet populations from the Generation~III Bern model~\citep{Emsenhuber2021,Emsenhuber2021b,Schlecker2021,Schlecker2021b,Burn2021,Mishra2021}.
The model includes the mechanisms relevant for the dynamical evolution of multi-planet systems, in particular type~I and type~II planet migration~\citep{Paardekooper2011,Dittkrist2014}, eccentricity and inclination damping through planet-disk interaction~\citep{Cresswell2008}, and dynamical evolution modeled via an N-body integrator~\citep{Chambers1999}.
The planet population with 0.7\,$M_\odot$ host star mass introduced in~\citet{Burn2021} is most suitable for the comparison with the K~dwarf system presented here.
Out of its 999 systems, 92 contain planets more massive than  30\,$M_\oplus$.
We find 51 systems with more than one such planet within orbital periods of 1000\,$\,$d in the population, typically around host stars with enhanced metallicity.
Only a single system includes a pair of warm giant planets in an MMR-like configuration: It contains three giant planets with periods of $48.0\,\mathrm{d}$, $96.5\,\mathrm{d}$, and $479.8\,\mathrm{d}$.



The system emerged from a numerical disk with a large solid material content, which is reflected in a high dust-to-gas ratio of 0.024 as compared to the population median of 0.014. 
This led to efficient core growth and runaway gas accretion of three protoplanets.
Through simultaneous inward migration and gravitational interaction, the two warm gas giants were eventually captured in an MMR, a configuration that persisted until the end of the N-body integration at 20\,Myr.
This comparison proposes that the TOI-2525 system was formed in a similar metal-enriched disk. 
TOI-2525 is thus in a configuration whose realization through core accretion is rare but possible, according to the simulations.

We note that the literature contains other examples of pairs of warm Jovian planets that have been extensively studied. For instance, the Kepler-9 system, consisting of a G-dwarf star orbited by a mini-Saturn planet pair in a 2:1 MMR \citep{Holman2010}. Other examples include the HD\,82943 \citet{Tan2013} and TIC\,279401253 (Bozhilov et al., submitted) systems, both of which are G-dwarf stars with almost identical 2:1 MMR Jovian-mass planet pairs. Although rare, M-dwarfs have also been found to host 2:1 MMR warm massive systems, with the GJ\,876 multiple planet system \citep{Rivera2010, Millholland2018, Trifonov2018a}, being widely considered as a benchmark for planetary dynamics and planet formation. The formation of such systems may not depend on the stellar type, but its occurrence rate remains an important observable to be further studied.

\subsection{TOI-2525\,b; a very low-density planet }

TOI-2525\,b and c are two warm low-density giant planets, a category of planetary systems whose frequency is steadily growing in the literature. The estimated mean density of TOI-2525\,c of $\rho_{\rm c}$ = 1.014$_{-0.076}^{+0.084}$ g\,cm$^{-3}$ is lower than that of Jupiter ($\rho_{\rm Jup.}$ = 1.33 g\,cm$^{-3}$), but is higher than that of Saturn ($\rho_{\rm Sat.}$ = 0.69 g\,cm$^{-3}$). Therefore, the density of the Jovian mass planet TOI-2525\,c is not surprising. The Neptune-mass planet TOI-2525\,b, however, has a mean density of $\rho_{\rm b}$ = 0.174$_{-0.015}^{+0.016}$ g\,cm$^{-3}$, which makes it among the lowest density Neptune-mass planets known to date. The density of TOI-2525\,b is a bit larger than the Kepler-51\,b, c, and d ``super puffs'' \citep{Steffen2013, Masuda2014}, and is comparable to low-density planets like WASP-107\,b \citep{Rubenzahl2021}, WASP-131\,b, WASP-139\,b \citep{Hellier2017}, WASP-21\,b \citep{Barros2011, Ciceri2013}, HATS-46b\,b \citep{Brahm2018}, among others. We use the mass-radius models by \citet{Fortney2007} to infer the core mass of the TOI-2525\,b and c planets given the estimated masses, radii, and stellar parameters. 
We calculate core masses of 6.9 $\pm$ 1.4 $M_{\oplus}$ and 36.6 $\pm$ 9.3 $M_{\oplus}$, for TOI-2525\,b and TOI-2525\,c, respectively.
The ratio of core-mass to the total mass $Zp$, is therefore; $Zp_b$ = 0.25 $\pm$ 0.05, and $Zp_c$ = 0.17 $\pm$ 0.04.

\autoref{hhe_fraction} shows a mass-radius plot for all known transiting planets with measured masses validated by TTVs or RVs. \autoref{hhe_fraction} is limited only to the giant planets in the range of 0.05 to 11 $M_{\rm Jup.}$, and 0.15 to 3.0 $R_{\rm Jup.}$. Panel a) of \autoref{hhe_fraction} is color-coded with the planetary equilibrium temperature (T$_{\rm eq.}$), whereas panel b) is color-coded with the planetary density. In both panels, we plot an interpolated model for the mass-radius relationship assuming the estimated luminosity of TOI-2525, semi-major axis=0.2 AU, and age of 3.1 G\,yr from \citet{Fortney2007}.
From \autoref{hhe_fraction} it is clear that the higher T$_{\rm eq.}$ of the planet, the larger the radius is. Except for the massive planets with few Jupiter masses and more, the larger radius correlates with low planetary density. 
TOI-2525 c is consistent with the mass-radius model from \citet{Fortney2007}. TOI-2525\,b is consistent with a large radius, and its mean density is among the lowest for the Neptune-mass planets.

We also model the evolution of both planets in the system, using CEPAM \citep{Guillot1995,Guillot2006} and a non-grey atmosphere \citep{Parmentier2015}, to provide constraints on their interior. We assume simple structures consisting of a central dense core surrounded by a hydrogen and helium envelope of solar composition. The core is assumed made with 50\% of ices and 50\% of rocks. \autoref{interiors} shows the resulting evolution models. The core mass of TOI-2525\,c is found to be between 20 and 43\, $M_{\oplus}$. This indicates that the enrichment in heavy elements of TOI-2525\,c could be comparable to Jupiter's, which is between 8 and 46\,$M_{\oplus}$ \citep{Guillot2022}. With a radius almost similar to Saturn's (1.08 times larger) but 3.4 times less massive than Saturn, TOI-2525\,b is an uncommon example of very low-density and inflated planet with an equilibrium temperature close to 500\,K. The H-He envelope of TOI-2525\,b is found to be between 19 and 24\,$M_{\oplus}$. The case of TOI-2525\,b is challenging for the traditional core-accretion formation scenario. With our simple modeling of TOI-2525\,b, a such small envelope hints that the accretion of H-He have potentially been hindered. Characterizing the atmospheres of both planets of the system would be very useful to understand their structure and formation.

Not many inflated Neptune-mass planets are known, making TOI-2525\,b a useful addition to the sample of transiting planets with a measured mass. The low-density, and thus, large scale-height of TOI-2525\,b makes it a good target for a future atmospheric investigation with transmission spectroscopy.


\section{Summary and Conclusions}
\label{sec6}

We report the discovery of a warm pair of giant planets around a K-dwarf star, uncovered by $\tess$ with multi-sector light curve photometry. The TOI-2525 light curve shows recurrent transit events consistent with two gravitationally interacting giant planets, resulting in a robust TTVs signal with a semi-amplitude of $\sim$6 hours for the inner planet. We obtained precise spectroscopic Doppler follow-up with the FEROS and the PFS spectrographs to estimate the stellar parameters and constrain the planetary masses. Using high signal-to-noise FEROS spectra of TOI-2525, we estimate a stellar mass of M$_\star$ = 0.849$_{-0.033}^{+0.024}$ M$_\odot$ and a stellar radius of R$_\star$ = 0.785$_{-0.007}^{+0.007}$ R$_\odot$. Using these stellar mass and radius, we conducted an extensive orbital analysis of the $\tess$ TTVs and RVs using self-consistent N-body models. This analysis allowed us to construct an accurate orbital model from which we predicted future transit events, confirmed by follow-up photometry observations by ASTEP, OM-SSO, and LCOGT.

The complete collection of RVs and transit light curves allowed us to perform more extensive joint TTV+RV analyses, as well as light curve photodynamical+RV N-body orbital modeling. 
We found that TOI-2525\,b is a massive-Neptune with a dynamical mass of m$_{\rm b}$ = 0.088$_{-0.005}^{+0.005}$ $M_{\rm jup}$, and radius of $R_{\rm b}$ = 0.88$_{-0.02}^{+0.02}$ $R_{\rm jup}$. Thus, the estimated density of TOI-2525\,b is $\rho_{\rm b}$ = 0.174$_{-0.015}^{+0.016}$ g\,cm$^{-3}$, which makes it among the lowest density Neptune-mass planets known to date, similar to the Kepler 51 planets.
The outer transiting planet TOI-2525\,c is a Jovian-mass planet 
with $m_{\rm c}$ = 0.709$_{-0.034}^{+0.034}$ $M_{\rm jup}$, and planetary radius $R_{\rm c}$ =0.98$_{-0.02}^{+0.02}$ $R_{\rm jup}$, and therefore, with a relatively low mean density of $\rho_{\rm c}$ = 1.014$_{-0.076}^{+0.084}$ g\,cm$^{-3}$.\looseness=-4

The warm pair of massive planets is near the 2:1 period ratio commensurability with orbital periods of 
$P_{\rm b}$ =  23.288$_{-0.002}^{+0.001}$\,d and $P_{\rm c}$ = 49.260$_{-0.001}^{+0.001}$\,d, but the dynamics of the system clearly suggest that it is outside the mean motion resonance (MMR) dynamical configuration. \looseness=-4

The TOI-2525 system is very similar to other K-dwarf TESS systems; TOI-2202 and TOI-216 are composed of almost identical K dwarf primary and two warm giant planets near the 2:1 MMR. 
These three systems will be a useful sample for studying the formation and composition of warm giant pairs around K-dwarf stars.

\acknowledgements

This research has made use of the Exoplanet Follow-up Observation Program website, which is operated by the California Institute of Technology, under contract with the National Aeronautics and Space Administration under the Exoplanet Exploration Program.
Funding for the $\tess$ mission is provided by NASA's Science Mission directorate.
This paper includes data collected by the $\tess$ mission, which are publicly available from the Mikulski Archive for Space Telescopes (MAST).
Some of the data presented in this paper were obtained from the Mikulski. The specific observations analyzed can be accessed via {\bf \dataset[10.17909/fwdt-2x66]{https://doi.org/10.17909/fwdt-2x66}}.
This research made use of \textsf{exoplanet} \citep{exoplanet:joss,
exoplanet:zenodo} and its dependencies \citep{exoplanet:foremanmackey17,
exoplanet:foremanmackey18, exoplanet:arviz, exoplanet:astropy13,
exoplanet:astropy18, exoplanet:kipping13, exoplanet:pymc3, exoplanet:theano}.
%
%
Resources supporting this work were provided by the NASA High-End Computing (HEC) Program through the NASA Advanced Supercomputing (NAS) Division at Ames Research Center for the production of the SPOC data products.
This work makes use of observations from the LCOGT network. Part of the LCOGT telescope time was granted by NOIRLab through the Mid-Scale Innovations Program (MSIP). MSIP is funded by NSF.
This work makes use of observations from the ASTEP telescope.
ASTEP benefited from the support of the French and Italian polar
agencies IPEV and PNRA in the framework of the Concordia station
program, from INSU, ESA and the University of Birmingham.
T.T. acknowledges support by the DFG Research Unit FOR 2544 "Blue Planets around Red Stars" project No. KU 3625/2-1.
T.T. further acknowledges support by the BNSF program "VIHREN-2021" project No. КП-06-ДВ/5.
M.H.L. was supported in part by Hong Kong RGC grant HKU 17305618.
AJ, RB, MH and FR acknowledge support from ANID -- Millennium  Science  Initiative -- ICN12\_009. AJ acknowledges additional support from FONDECYT project 1210718. RB acknowledges support from FONDECYT project 11200751.
MNG acknowledges support from the European Space Agency (ESA) as an ESA Research Fellow.
The results reported herein benefited from collaborations and/or information exchange within the program “Alien Earths” (supported by the National Aeronautics and Space Administration under agreement No. 80NSSC21K0593) for NASA’s Nexus for Exoplanet System Science (NExSS) research coordination network sponsored by NASA’s Science Mission Directorate.
This research is in part funded by the European Union's Horizon 2020 research and innovation programme (grants agreements n$^{\circ}$ 803193/BEBOP), and from the Science and Technology Facilities Council (STFC; grant n$^\circ$ ST/S00193X/1).

\facilities{$\tess$, MPG-2.2m/FEROS, Magellan-6.5m/PFS, LCOGT}

\software{
          Exo-Striker~\citep{Trifonov2019_es},
          exoplanet~\citep{exoplanet:joss},
          CERES~\citep{ceres},
          ZASPE~\citep{zaspe},
          tesseract~(Rojas, in prep.),
          TESSCut~\citep{TESSCut},
          lightkurve~\citep{lightkurve},
          emcee~\citep{emcee},
          corner.py~\citep{corner},
		  dynesty~\citep{Speagle2020},
          batman~\citep{Kreidberg2015},
          celerite~\citep{celerite},
          TTVfast \citep{Deck2014}
          wotan~\citep{Hippke2019},
           transitleastsquares~\citep{Hippke2019b},
           pyExoRaMa~\citep{Francesco2022},
           AstroImageJ \citep{Collins2017}
          }

\newpage 

\bibliographystyle{aasjournal}
\bibliography{bibliography}

\begin{appendix} 

\label{appendix}

In this Appendix, we show in \autoref{table:TTVdata} we list the estimated mid-transit time estimates of TOI-2525\,b and c (i.e., TTVs),  in \autoref{photodynparameters} we list the estimated posterior probability distribution of our joint Doppler and transit light curve photodynamical model with {\tt flexifit}, in \autoref{Nest_samp_ttv} shows the posterior probability distribution of the joint Doppler and TTV modeling with {\tt Exo-Striker}, in \autoref{Transits2525b} and \autoref{Transits2525c} we show the {\tt flexifit} model applied to the available transit lightcurves, and finally, in \autoref{photodyposterior} we show the MCMC posterior probability distribution of the {\tt flexifit} joint photodynamical analysis.

 \setcounter{table}{0}
\renewcommand{\thetable}{A\arabic{table}}

\setcounter{figure}{0}
\renewcommand{\thefigure}{A\arabic{figure}}

\begin{figure*}[tp]
\begin{center}$
\begin{array}{ccc}

\includegraphics[width=18cm]{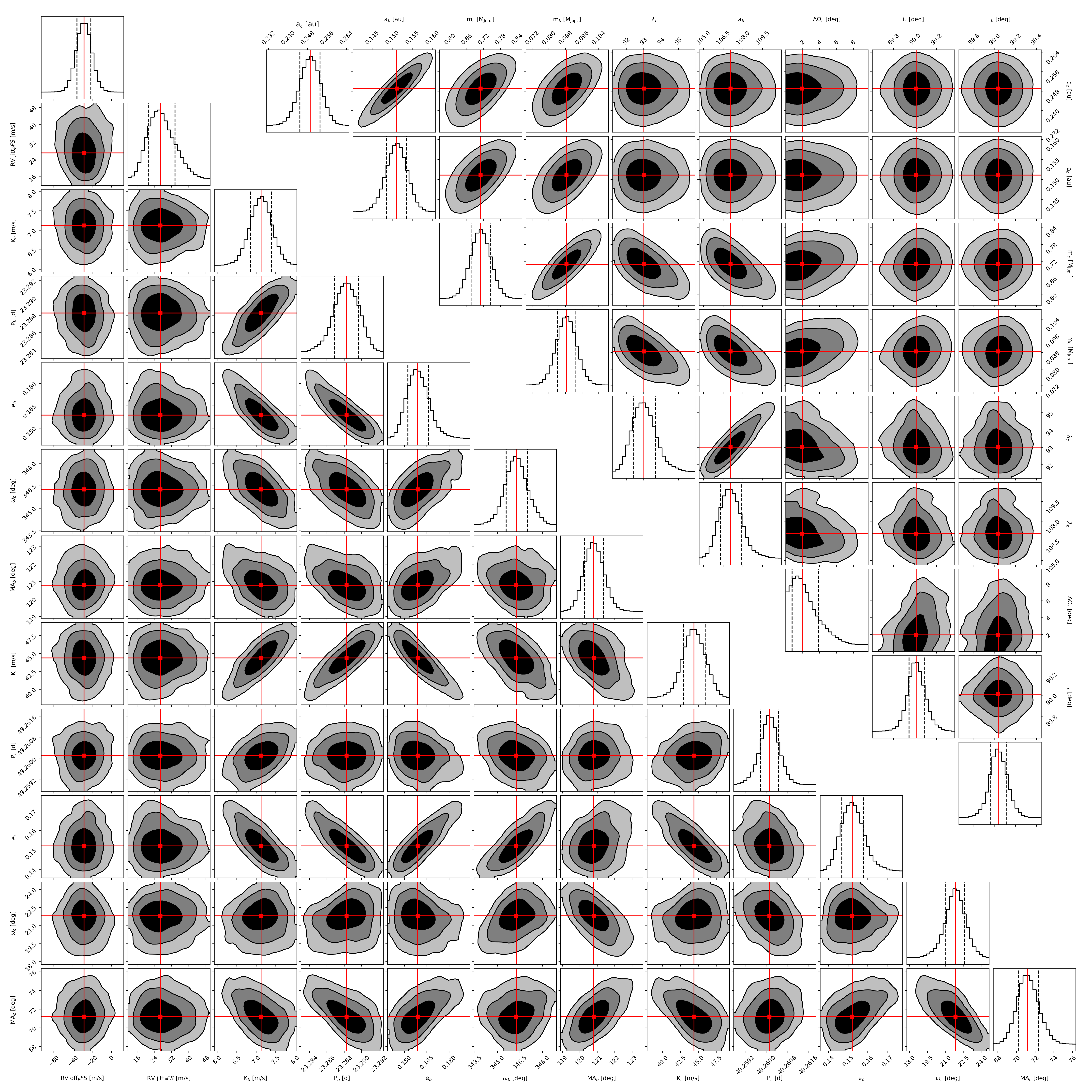} 
 \end{array} $
\end{center}

\caption{{\sc Exo-Striker} global parameters search results of the $\tess$, the ASTEP, the OM-SSO, and the LCOGT TTVs, of TOI-2525\,b and  TOI-2525\,c, and the Doppler radial velocities from FEROS and PFS. 
The posterior probability distribution is acheaved  with a nested sampling scheme employing a two-planet self-consistent dynamical model. 
The black contours on the 2D panels represent the 1, 2, and 3$\sigma$ confidence level of the overall posteriors. Red crosses point to the median position of the posterior parameters.
}
 
\label{Nest_samp_ttv} 
\end{figure*}

    \begin{table}[ht]

    \centering
     \caption{Parameters of the two-planet system TOI-2525, its host star and the light curves derived during TTV extraction (continues from \autoref{table2}).}
    
    \label{table2a}
 
    \begin{tabular}{p{5.6cm}  c crrrrrr}     

    \hline\hline  \noalign{\vskip 0.7mm}
    Parameter \hspace{60.0 mm}& median & $\sigma$ \\
    \hline \noalign{\vskip 0.7mm}
$\mu_{\rm TESS-LC}$       &  $-$0.0003 & 0.0004 \\
$\log(\rho)_{\rm TESS-LC}$  &   0.830 & 0.093 \\
$\log(\sigma)_{\rm TESS-LC}$ &  $-$5.628 & 0.056 \\
$\mu_{\rm TESS-SC}$            &   0.0000 & 0.0001 \\
$\log(\rho)_{\rm TESS-SC}$   &   0.526 & 0.156 \\
$\log(\sigma)_{\rm TESS-SC}$ &  $-$7.246 & 0.058 \\
$\mu_{\rm ASTEP1}   $          &   0.0002 & 0.0001 \\
$\beta_{\rm ASTEP1}$           &   0.0054 & 0.002 \\
$\mu_{\rm ASTEP2}$             &   0.0000 & 0.0001 \\
$\beta_{\rm ASTEP2}$       &   -0.001 & 0.002 \\
$\mu_{\rm ASTEP3}$             &   0.0000 & 0.0002 \\
$\beta_{\rm ASTEP3}$        &  $-$0.005 & 0.005 \\
$\mu_{\rm ASTEP4}$          &   0.0006 & 0.0002 \\
$\beta_{\rm ASTEP4}$      &	   0.0185 & 0.005 \\
$\mu_{\rm ASTEP5}$             &   $-$0.0001 & 0.0001 \\
$\beta_{\rm ASTEP5}$        &  $-$0.004 & 0.002 \\
$\mu_{\rm ASTEP6}$          &  $-$0.0001 & 0.0001 \\
$\beta_{\rm ASTEP6}$      &	   $-$0.004& 0.0008 \\
$\mu_{\rm SSO}$          &	 $-$0.0007 & 0.0003 \\
$\beta_{\rm SSO}$       &	 $-$0.008 & 0 006 \\
$\mu_{\rm CTIOi}$   &   0.0030 &  0.0002 \\
$\beta_{\rm CTIOi}$    & $-$0.0061 &  0.0081 \\
$\mu_{\rm CTIOg}$    &   0.0031 &  0.0002  \\
$\beta_{\rm CTIOg}$  &  $-$0.0404 &  0.0105  \\
$\mu_{\rm CTIOi2}$  &   0.0047 &  0.0001  \\
$\beta_{\rm CTIOi2}$ & $-$0.0024 &  0.0039 \\

$u_{0, \rm TESS}$	   &	 0.502	  &  0.150 \\
$u_{1, \rm TESS}$	   &	 0.422	  &  0.189 \\
$u_{0, \rm ASTEP}$	   &	 0.409	  &  0.072 \\
$u_{1, \rm ASTEP}$	   &	$-$0.159	  &  0.069 \\
$u_{0, \rm SSO}$	   &	 0.528	  &  0.220 \\
$u_{1, \rm SSO}$	   &	 0.053	  &  0.299 \\
$u_{0, \rm SAAO,g}$	   &	 0.688	  &  0.330 \\
$u_{1, \rm SAAO,g}$	   &	$-$0.055	  &  0.348 \\
$u_{0, \rm SAAO,i}$	   &	 0.403	  &  0.143 \\
$u_{1, \rm SAAO,i}$	   &	 0.057	  &  0.220 \\
 
     \hline \noalign{\vskip 0.7mm}

    \end{tabular}

\tablecomments{TESS-LC/SC means long (30-min) and short (2-min) cadance data, respectivly.  The $\mu$ values correspond to the fitted mean of each light curve set, while the $\beta$ values correspond to the coefficient of a linear trend in units of day$^{-1}$. The $\{u_0,u_1\}$ coefficients correspond to the coefficients of a quadratic limb darkening law as derived for each instrument/band as indicated in the subscript for each coefficient. \looseness=-4}

    \end{table}

\begin{table}[ht]

\centering   
\caption{{Individual mid-transit time estimates of TOI-2525\,b extracted from $\tess$, ASTEP, SSO and LCOGT used for TTV analysis.}} 
\label{table:TTVdata}

\begin{tabular}{p{2.0cm} ccccrrrr}     

\hline\hline  \noalign{\vskip 0.7mm}
N Transit \hspace{0.0 mm}& t$_{\rm 0}$ [BJD] & $\sigma$ t$_{\rm 0}$ [BJD] & Instrument \\
\hline \noalign{\vskip 0.9mm}

 & Planet b  & \\
      \hline \noalign{\vskip 0.7mm}
 1   &       2458333.527  &0.005  &$\tess$ \\
 4   &       2458403.441  &0.002  &$\tess$ \\
 5   &       2458426.775  &0.007  &$\tess$ \\
 6   &       2458450.130  &0.011  &$\tess$ \\
 7   &       2458473.508  &0.008  &$\tess$ \\
 8   &       2458496.853  &0.003  &$\tess$ \\
 9   &       2458520.237  &0.004  &$\tess$ \\
 11  &       2458566.956  &0.002  &$\tess$ \\
 15  &       2458659.976  &0.009  &$\tess$ \\
 32  &       2459056.051  &0.006  &$\tess$ \\
 33  &       2459079.266  &0.003  &$\tess$ \\
 35  &       2459125.641  &0.003  &$\tess$ \\
 36  &       2459148.834  &0.003  &$\tess$ \\
 38  &       2459195.308  &0.004  &$\tess$ \\
 39  &       2459218.599  &0.004  &$\tess$ \\
 41  &       2459265.237  &0.003  &$\tess$ \\
 42  &       2459288.598  &0.003  &$\tess$ \\
 43  &       2459311.952  &0.003  &$\tess$ \\
 45  &       2459358.682  &0.004  &$\tess$ \\
 50  &       2459475.305  &0.004  &ASTEP \\
 55  &       2459591.328  &0.002 &LCOGT-SAAO\\
       \hline \noalign{\vskip 0.9mm}
& Planet c & & \\
     \hline \noalign{\vskip 0.7mm}
 
 1   &    2458335.410   &0.004 &$\tess$ \\
 2   &    2458384.648   &0.002 &$\tess$ \\
 3   &    2458433.866   &0.006 &$\tess$ \\
 4   &    2458483.090   &0.007 &$\tess$ \\
 8   &    2458680.073   &0.005 &$\tess$ \\
 17  &    2459123.270   &0.003 &$\tess$ \\
 19  &    2459221.791   &0.003 &$\tess$ \\
 21  &    2459320.238   &0.002 &ASTEP\\
 22  &    2459369.454   &0.003 &ASTEP\\
 24  &    2459467.906   &0.003 &ASTEP\\
 25  &    2459517.190   &0.003 &SSO \\

\hline \noalign{\vskip 0.7mm}

\end{tabular}

\end{table}



\begin{figure}[]
\centering
  \begin{tabular}{@{}cccc@{}}
    \includegraphics[width=.23\textwidth]{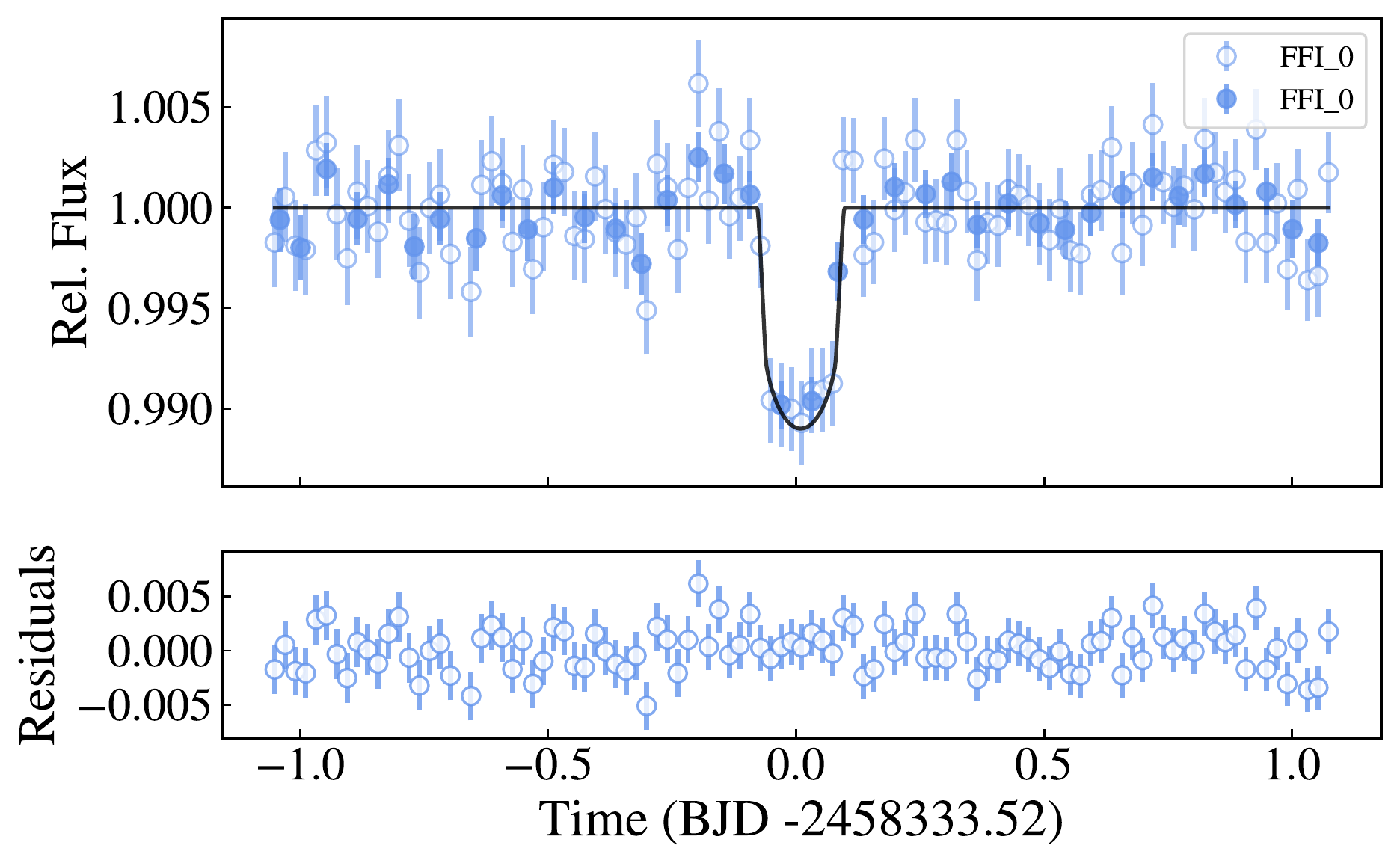} &
    \includegraphics[width=.23\textwidth]{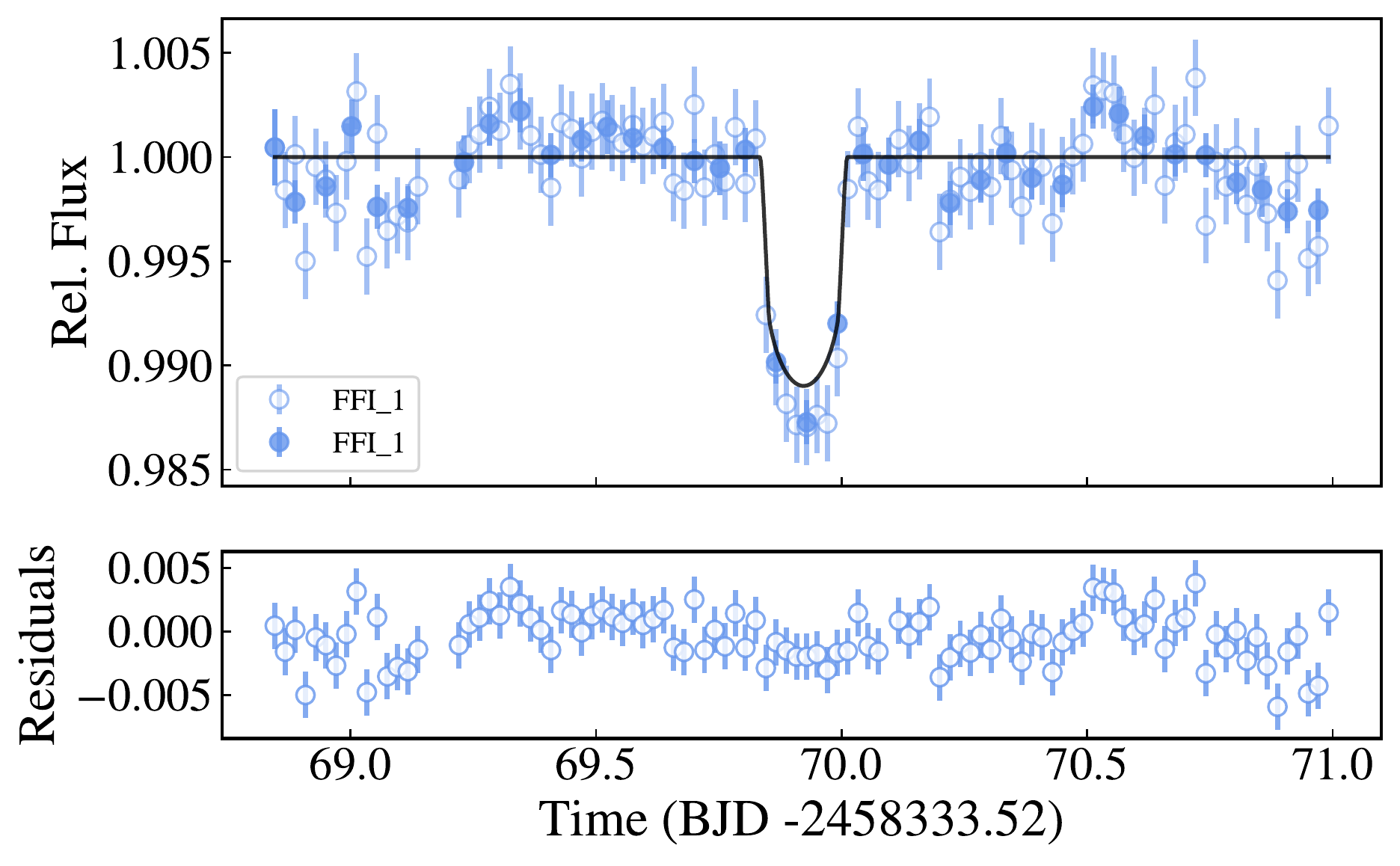} &
    \includegraphics[width=.23\textwidth]{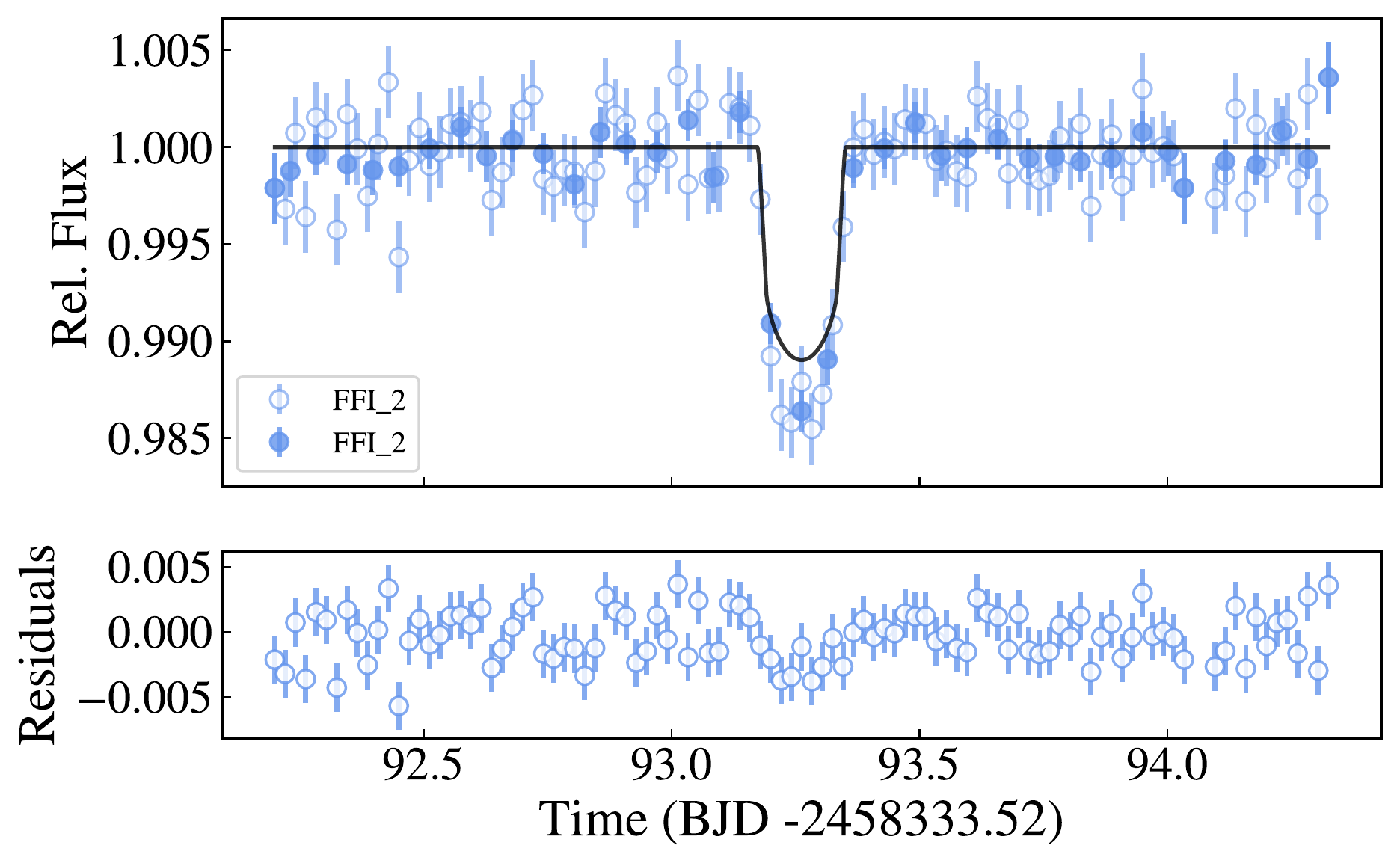} &
    \includegraphics[width=.23\textwidth]{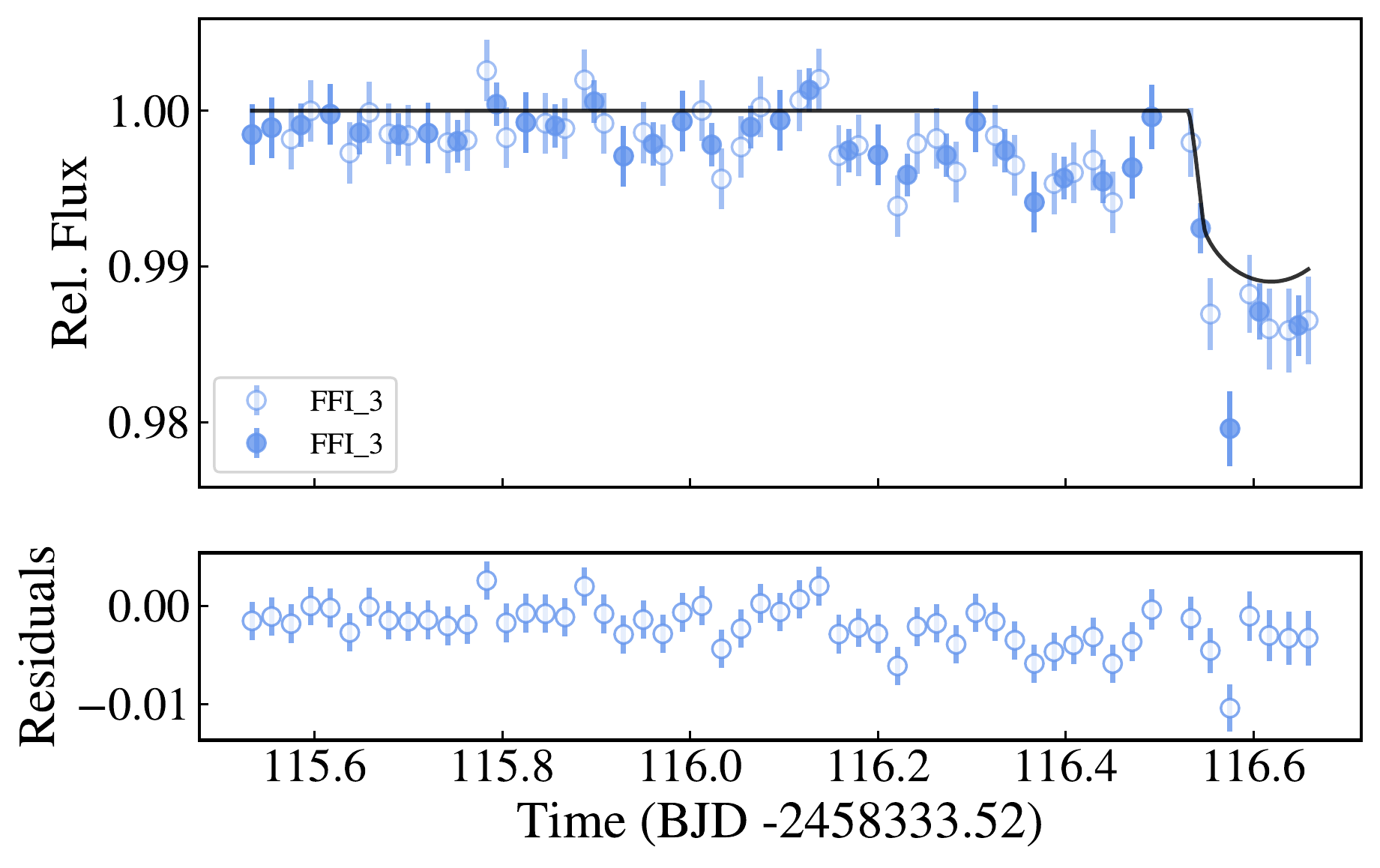} \\
    \includegraphics[width=.23\textwidth]{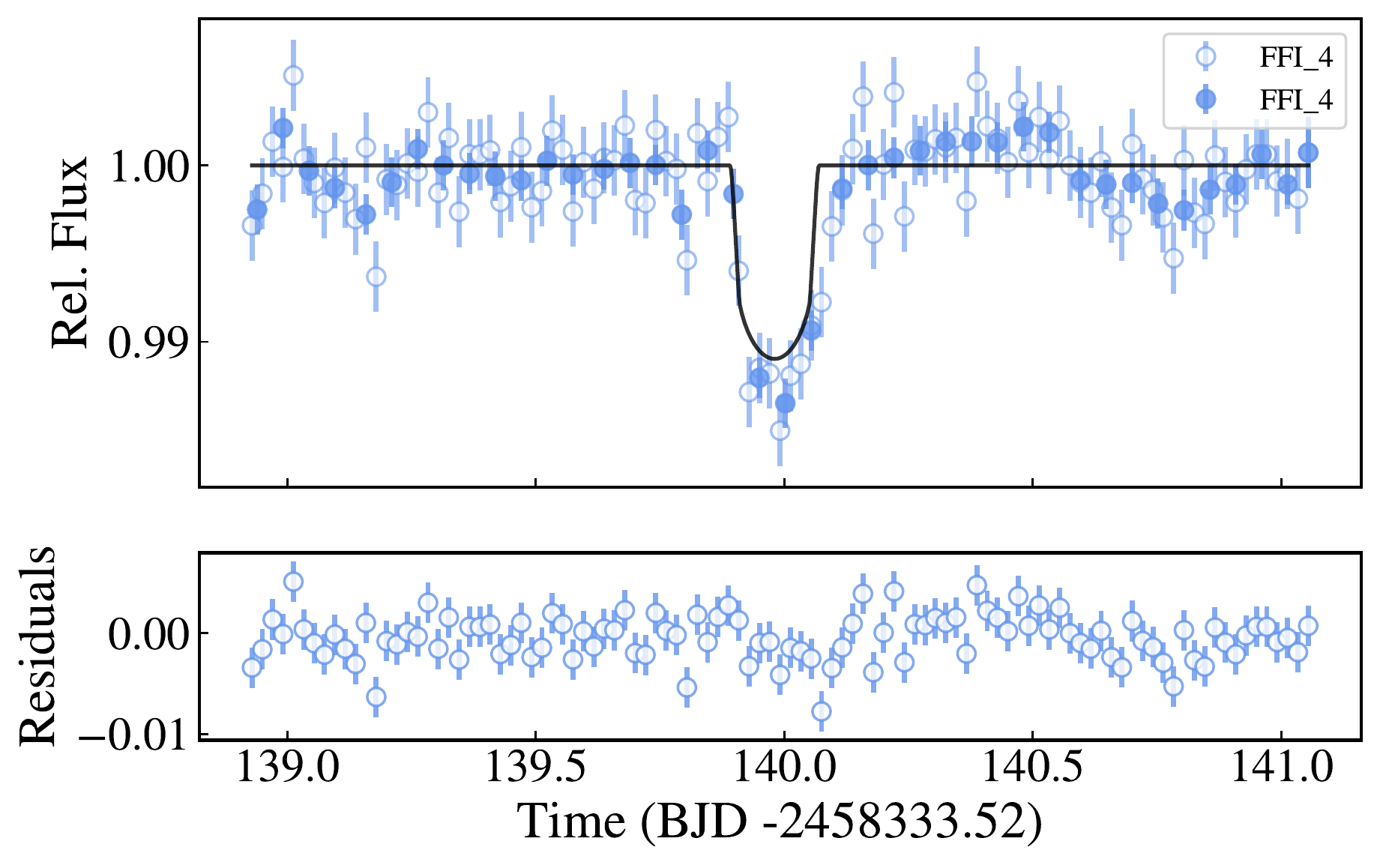} &
    \includegraphics[width=.23\textwidth]{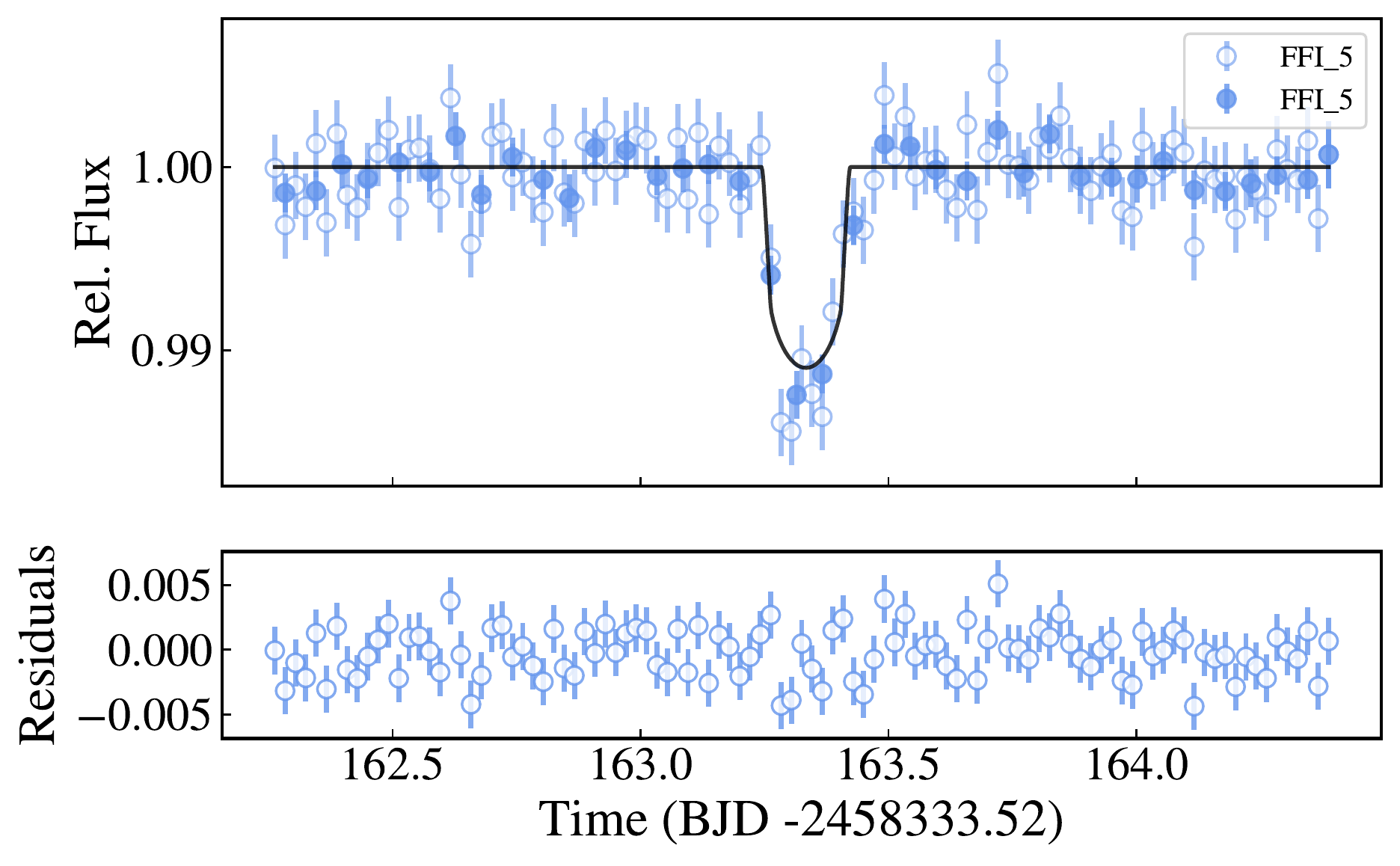} &
    \includegraphics[width=.23\textwidth]{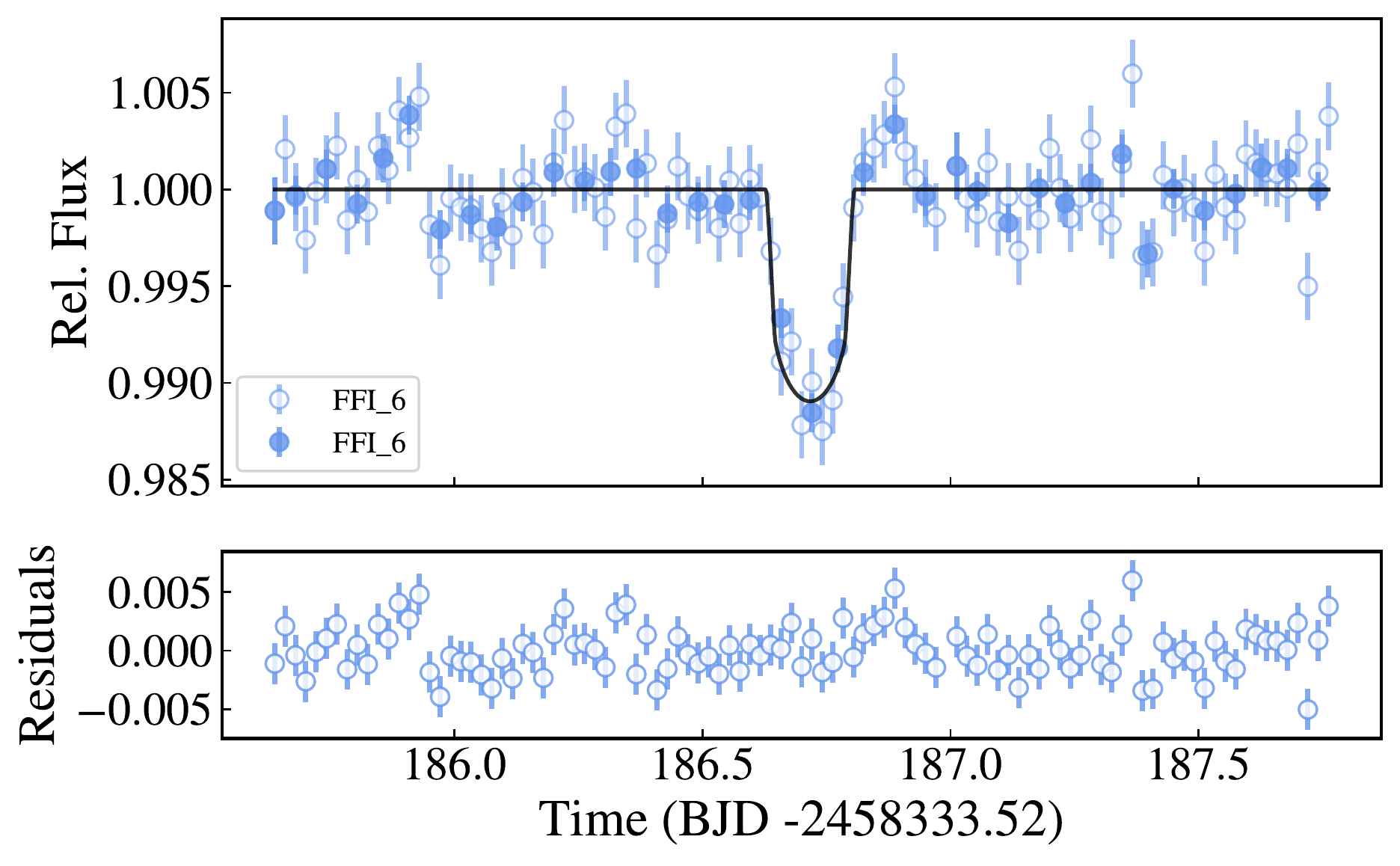} &
    \includegraphics[width=.23\textwidth]{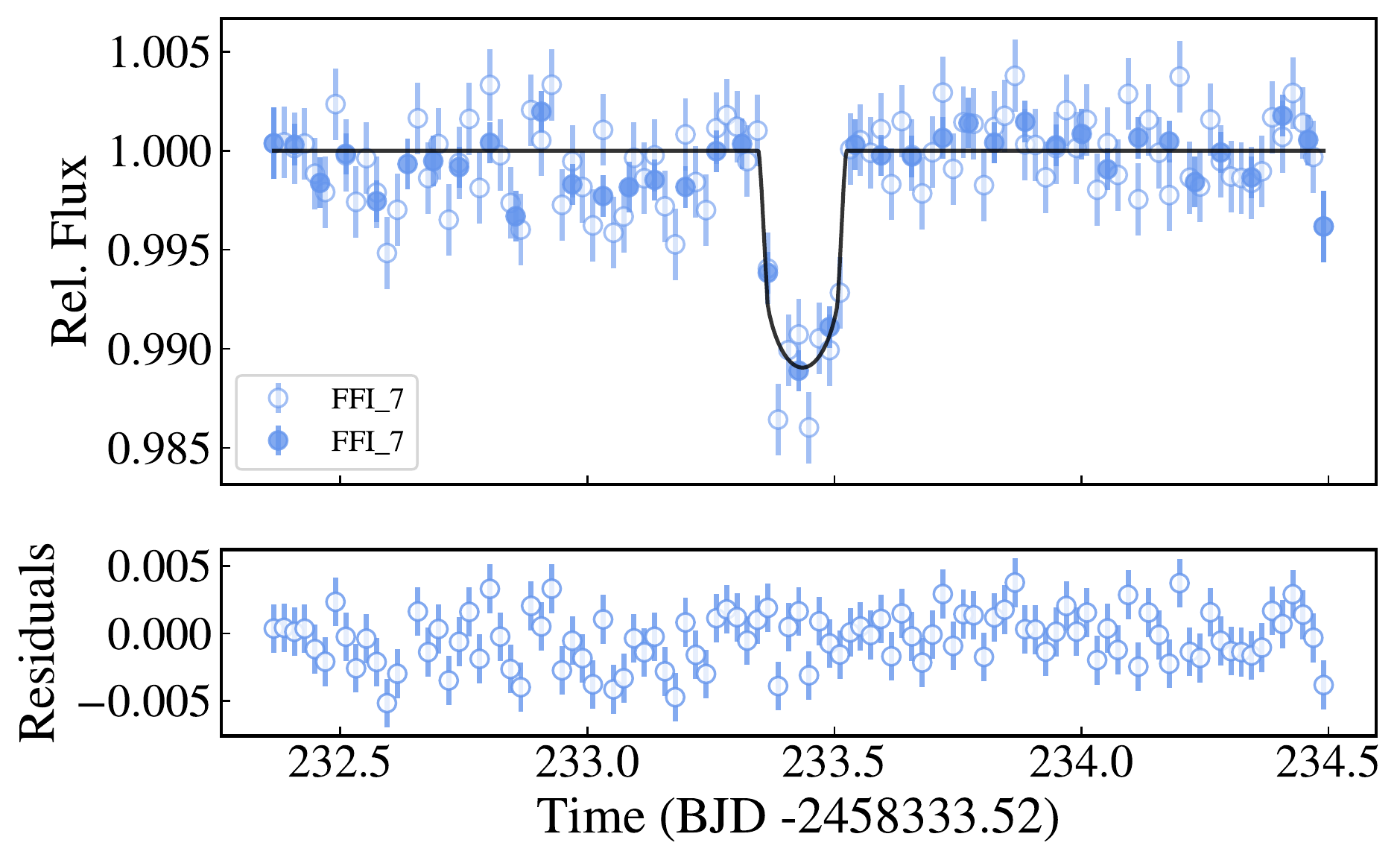} \\
    \includegraphics[width=.23\textwidth]{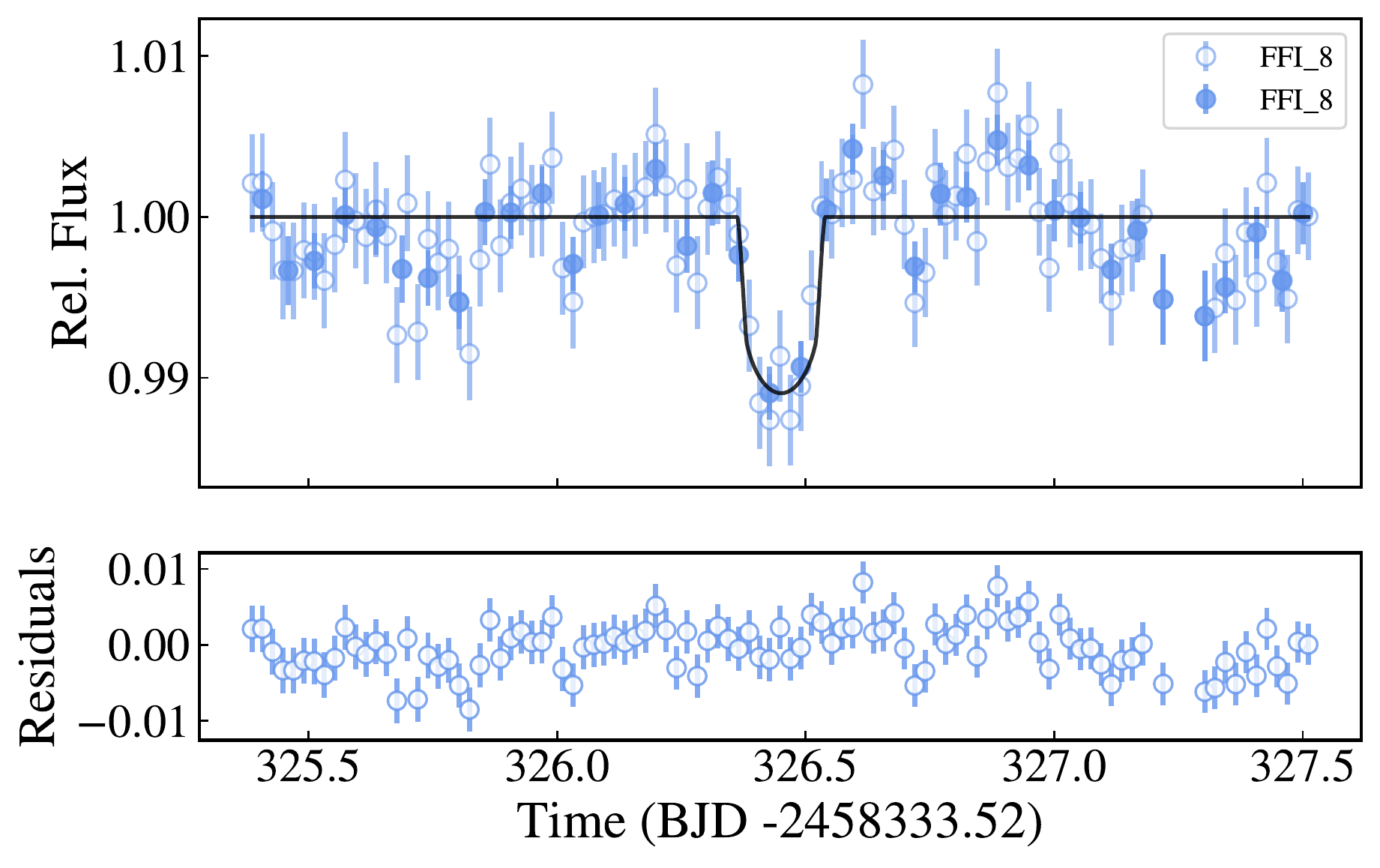}  &
    \includegraphics[width=.23\textwidth]{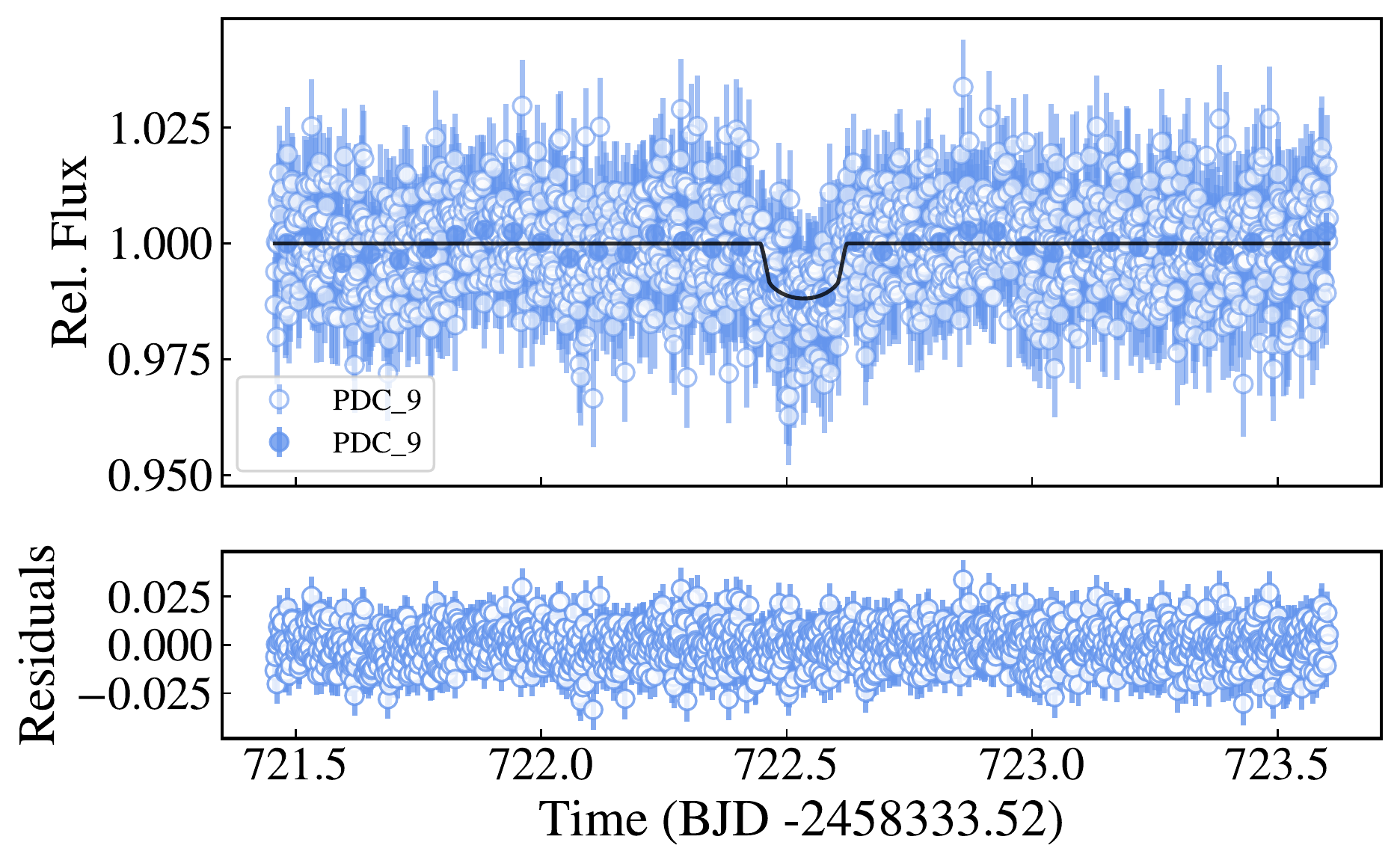} &
    \includegraphics[width=.23\textwidth]{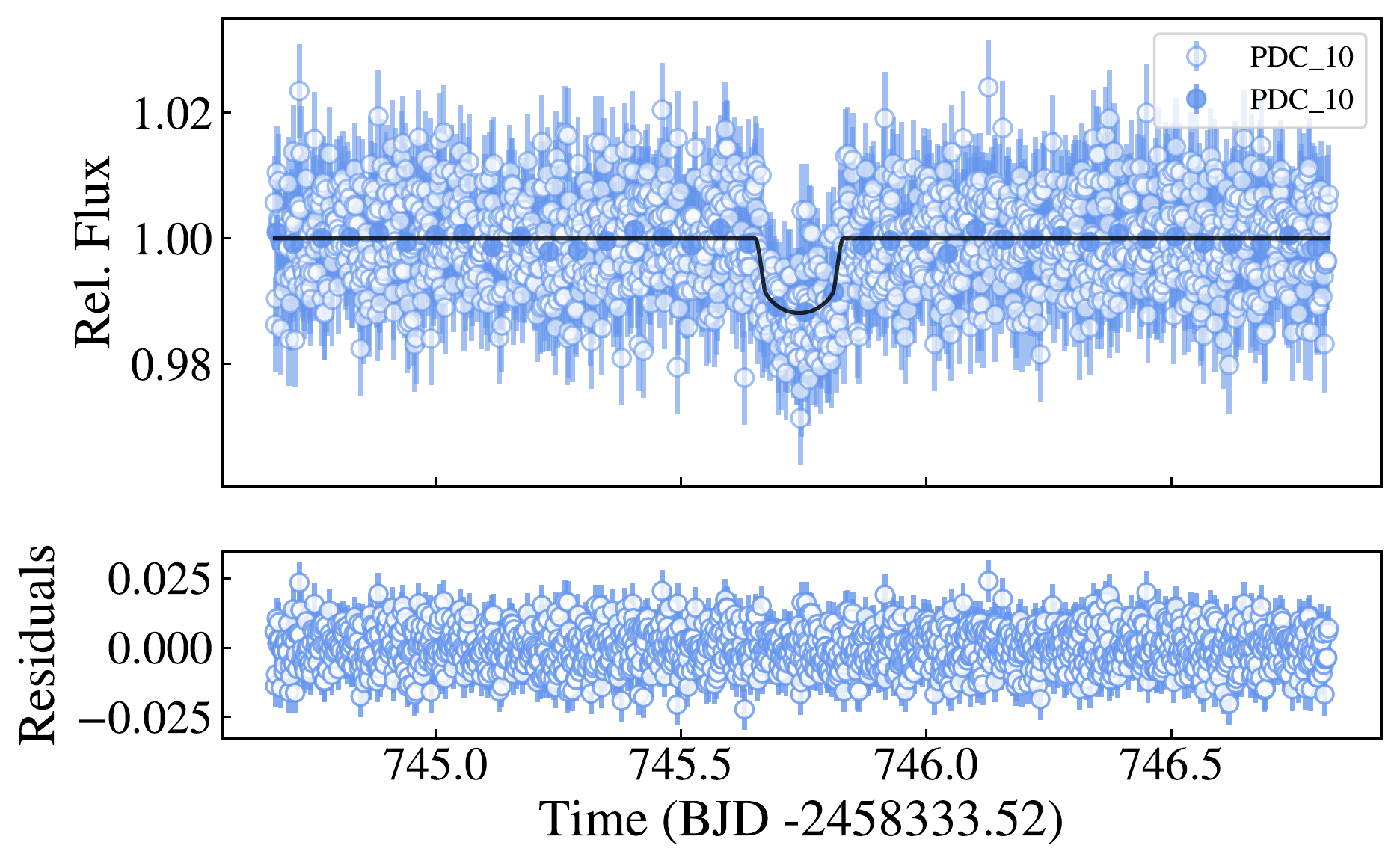} &
    \includegraphics[width=.23\textwidth]{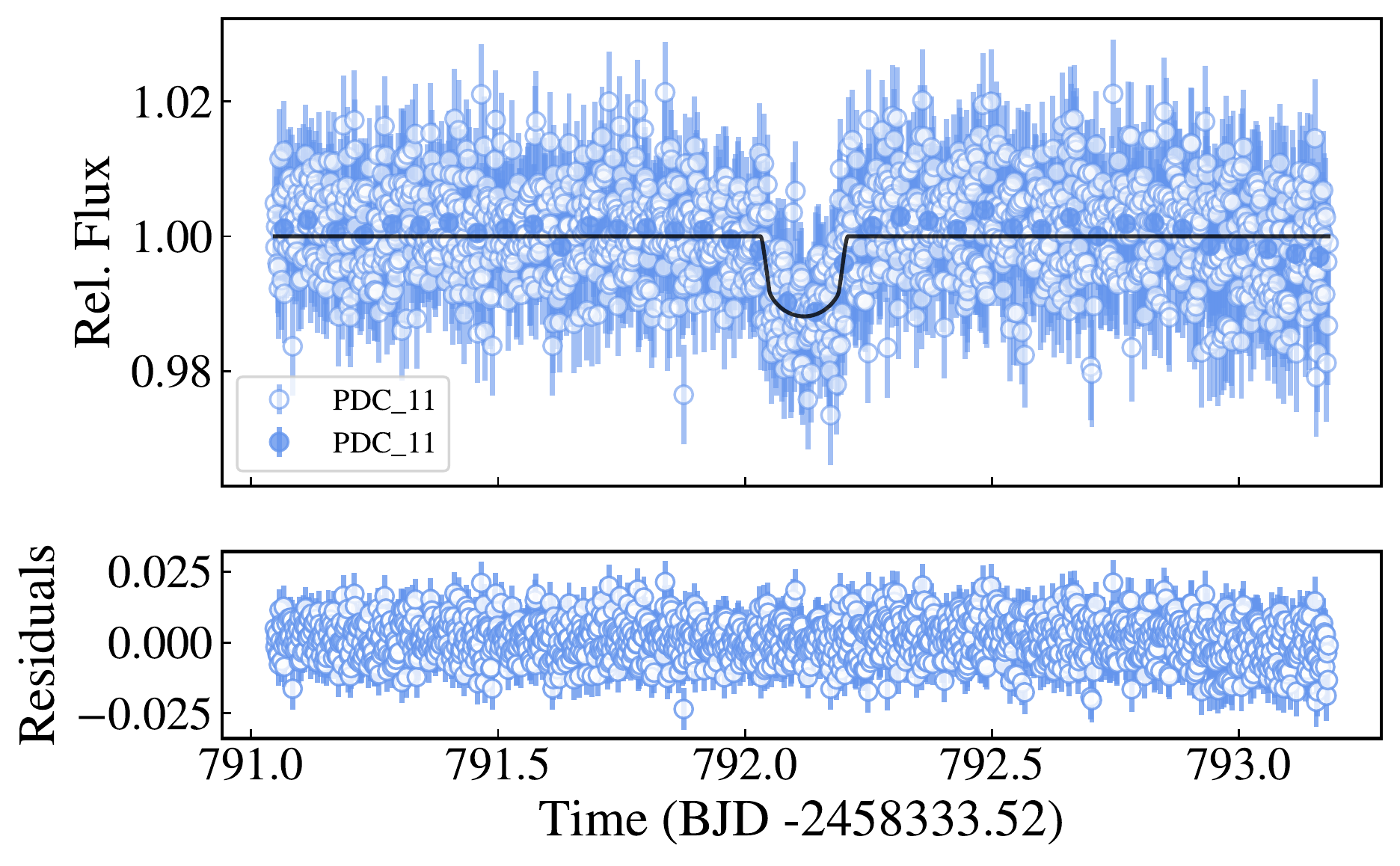}   \\
    \includegraphics[width=.23\textwidth]{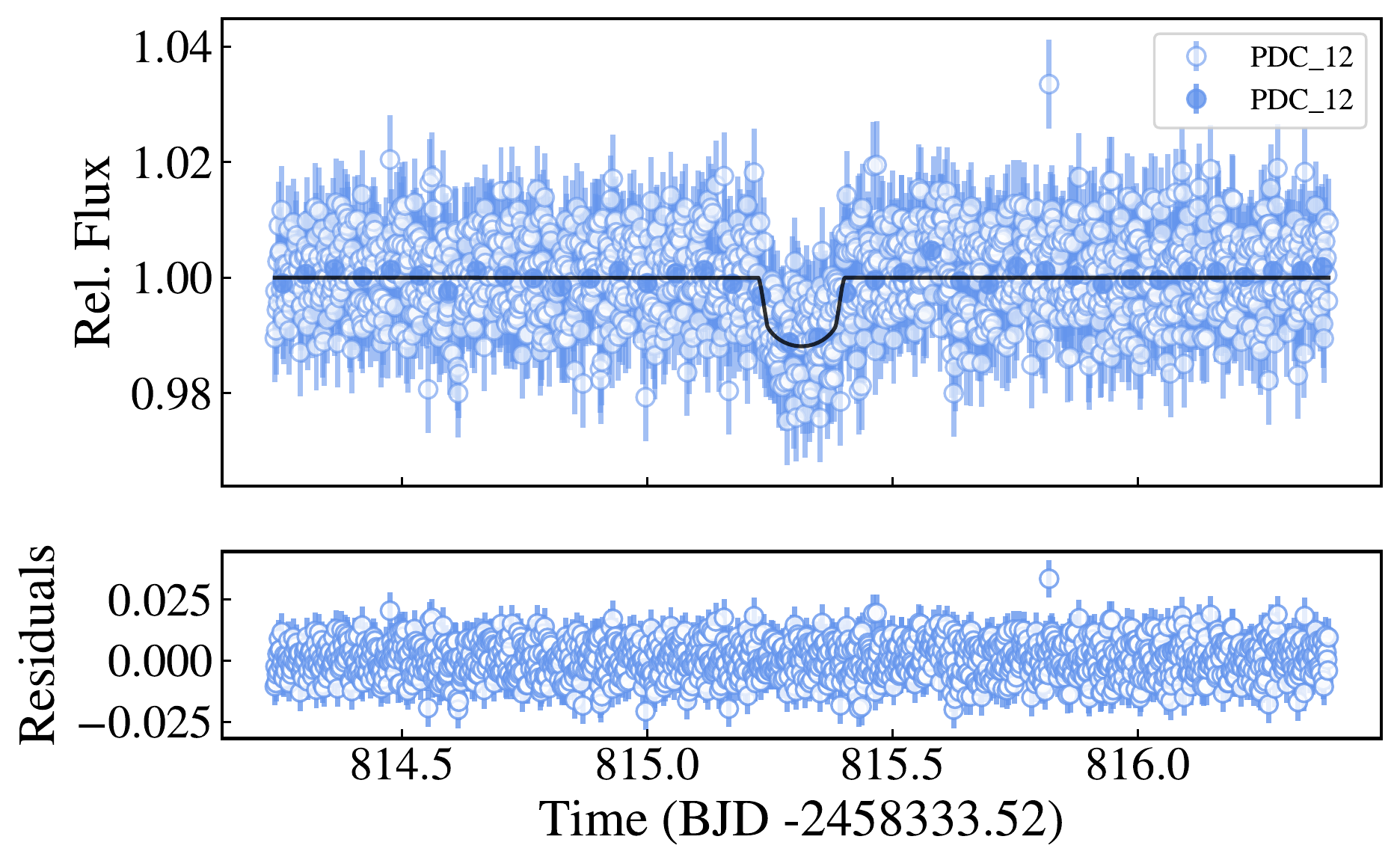} &
    \includegraphics[width=.23\textwidth]{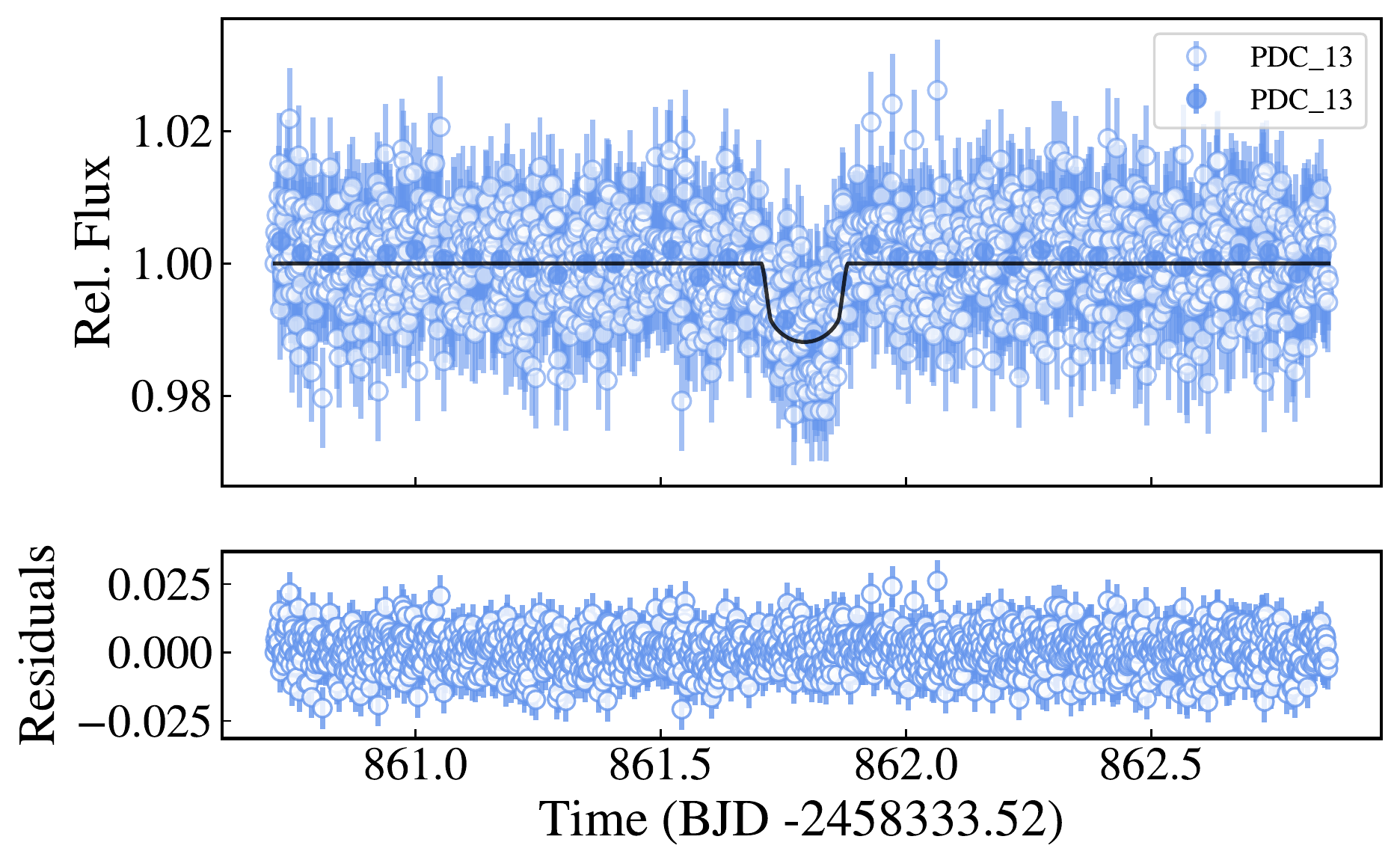} &
    \includegraphics[width=.23\textwidth]{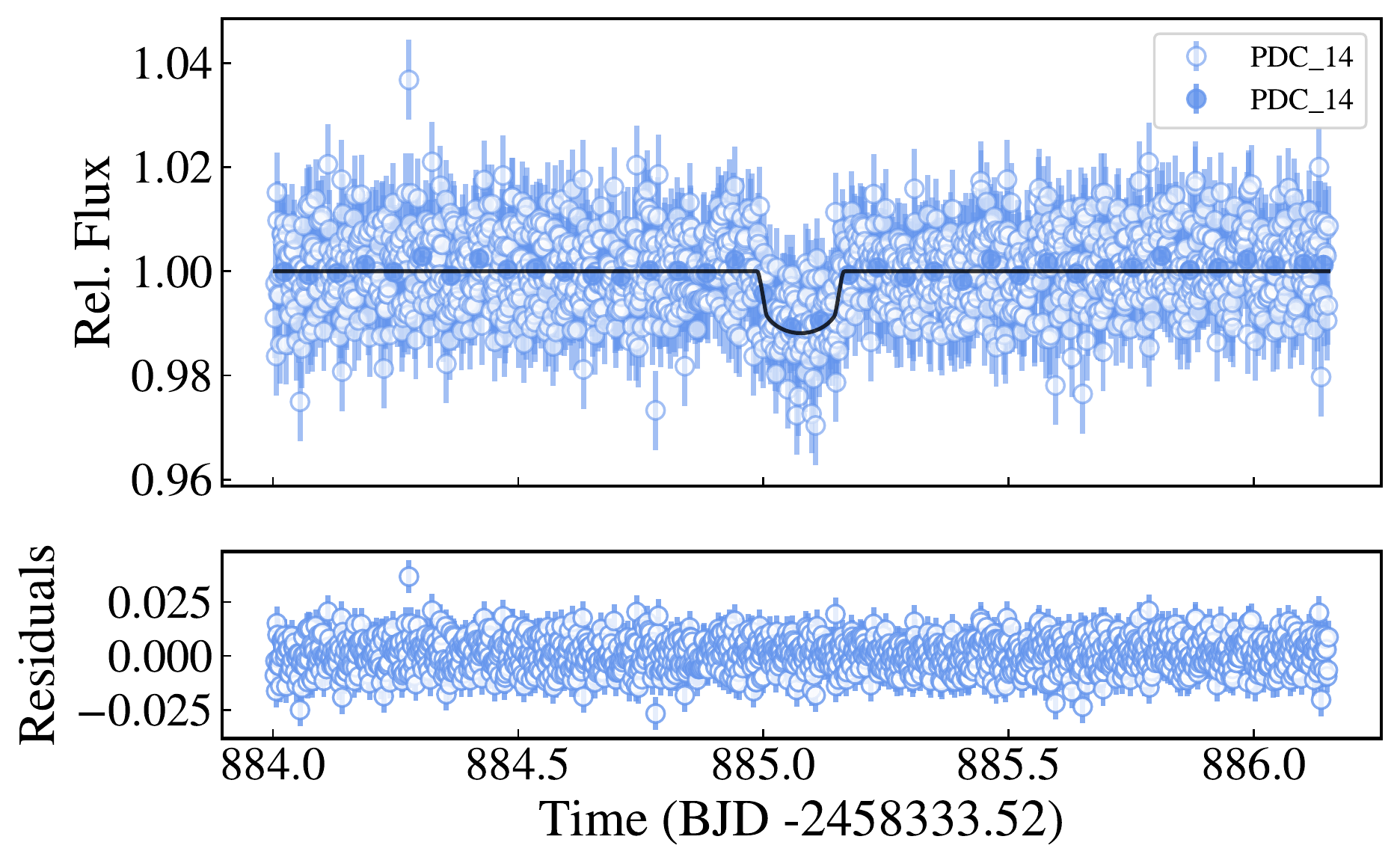}  &
    \includegraphics[width=.23\textwidth]{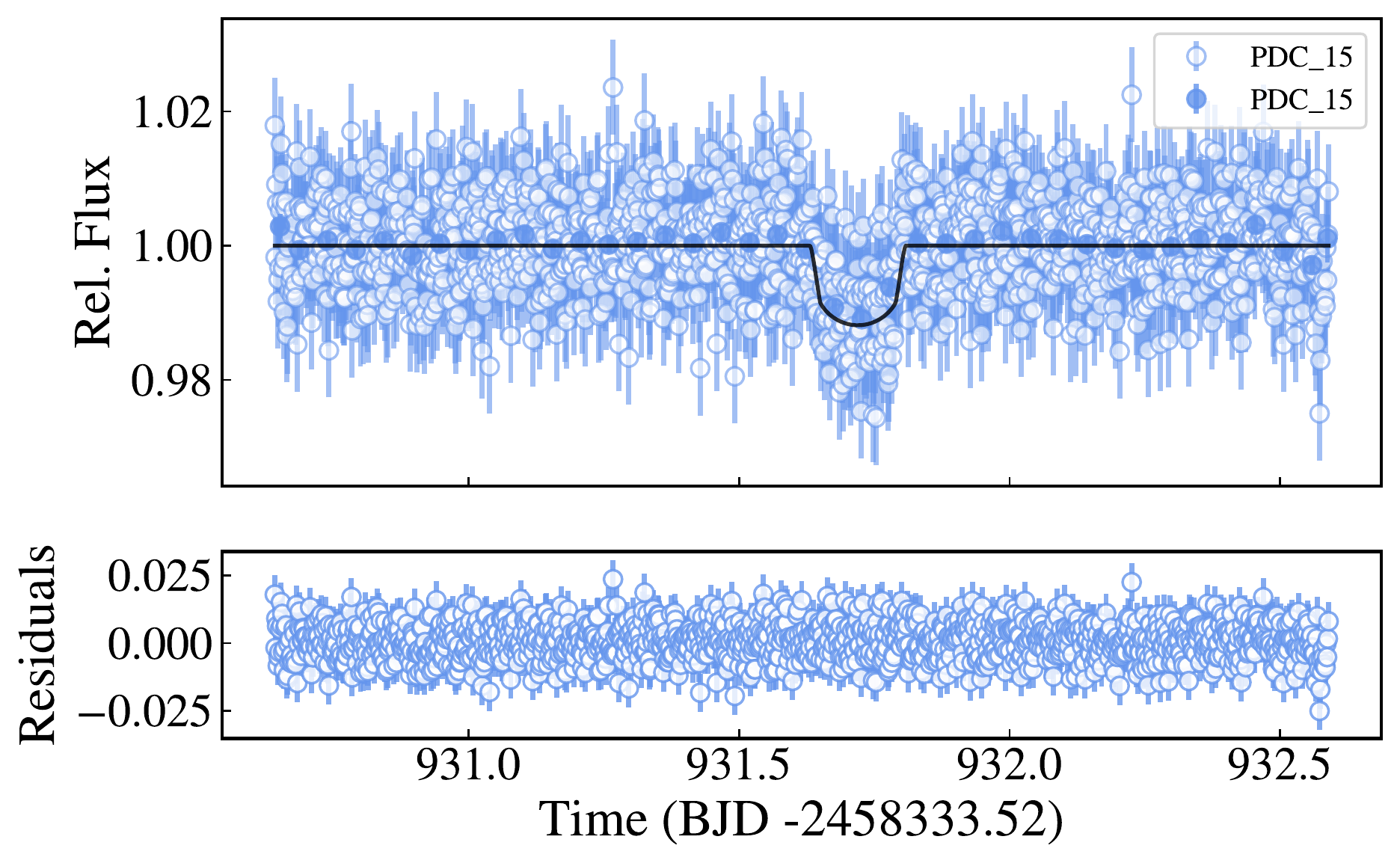} \\
    \includegraphics[width=.23\textwidth]{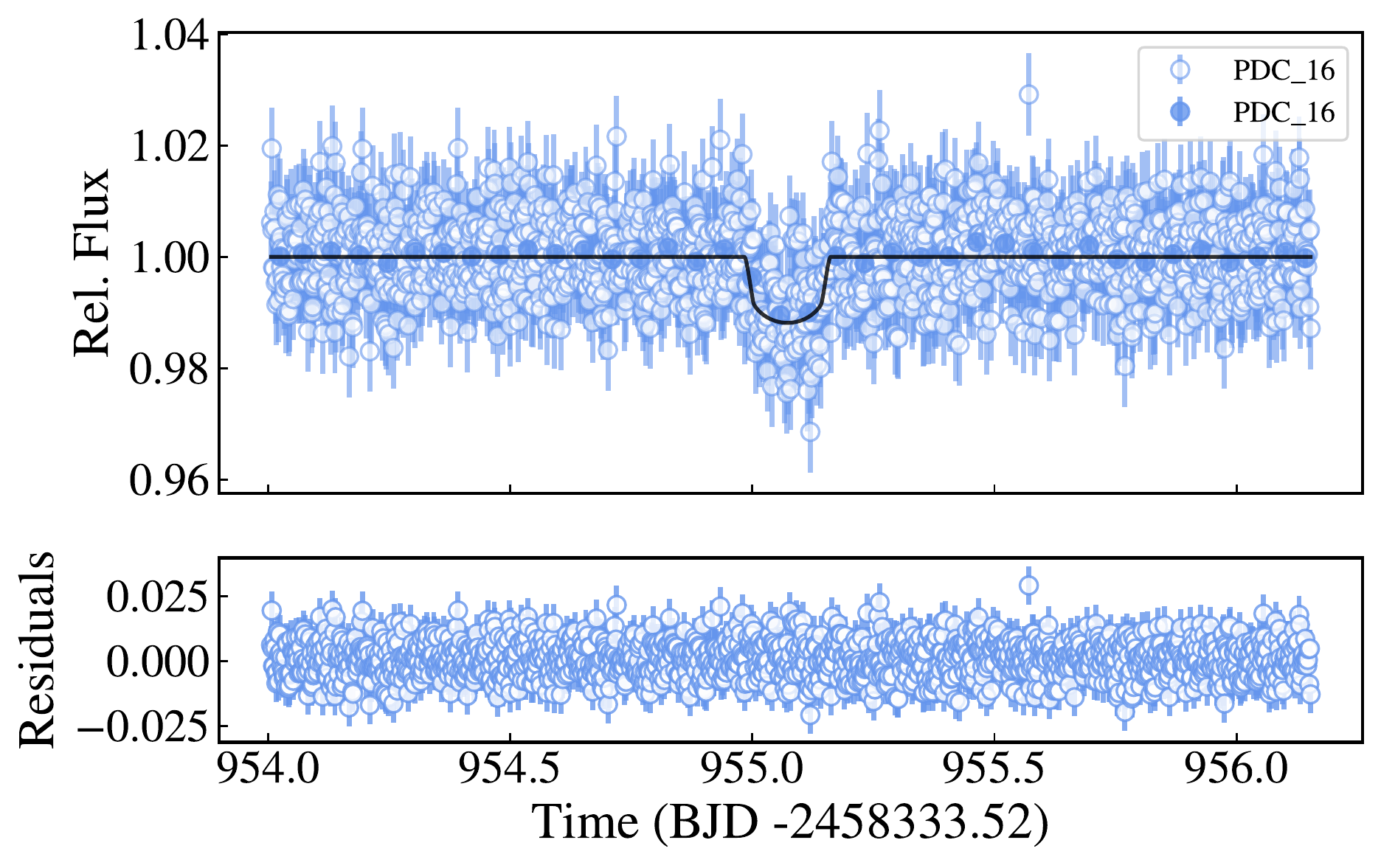} &
    \includegraphics[width=.23\textwidth]{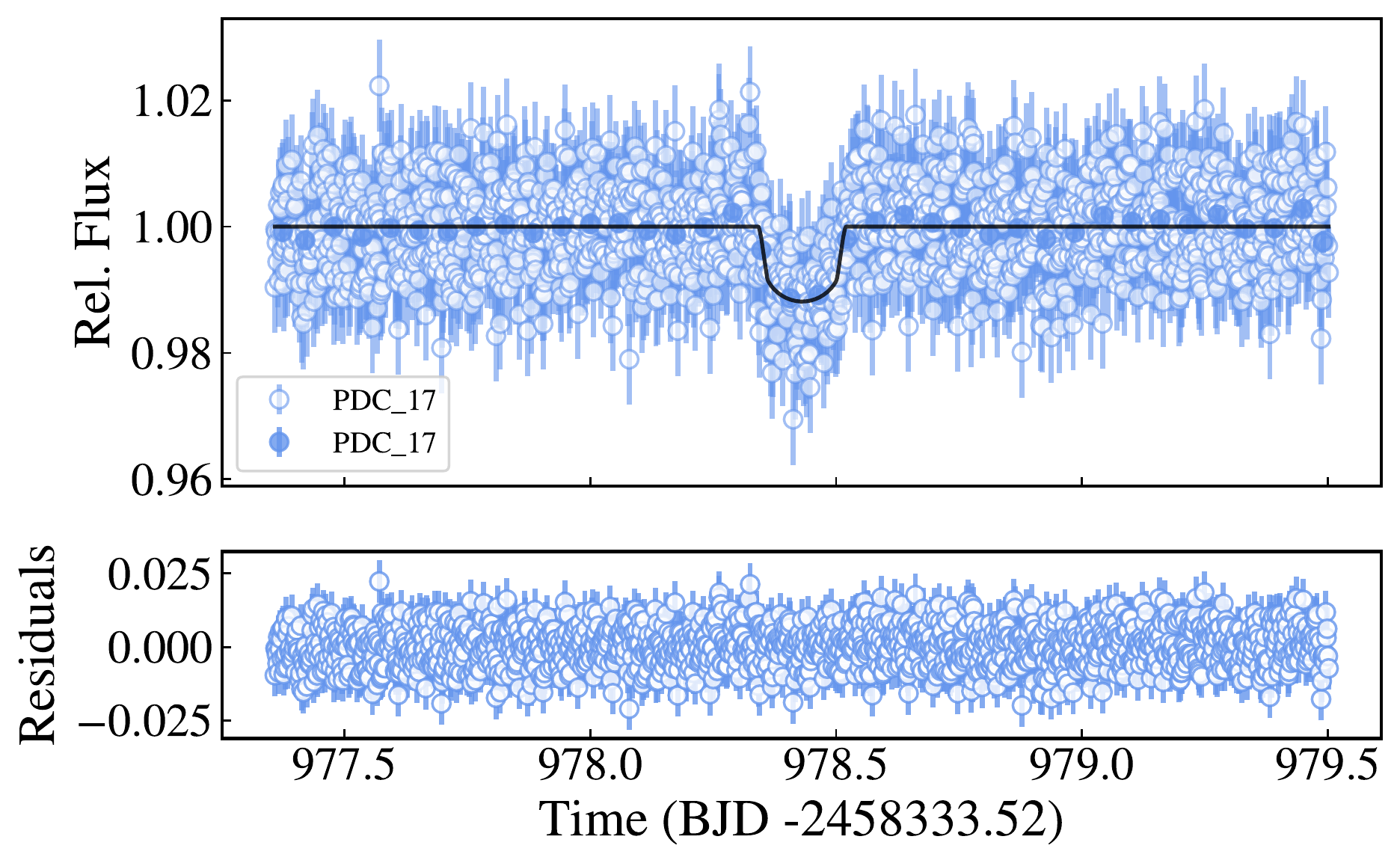}   &
    \includegraphics[width=.23\textwidth]{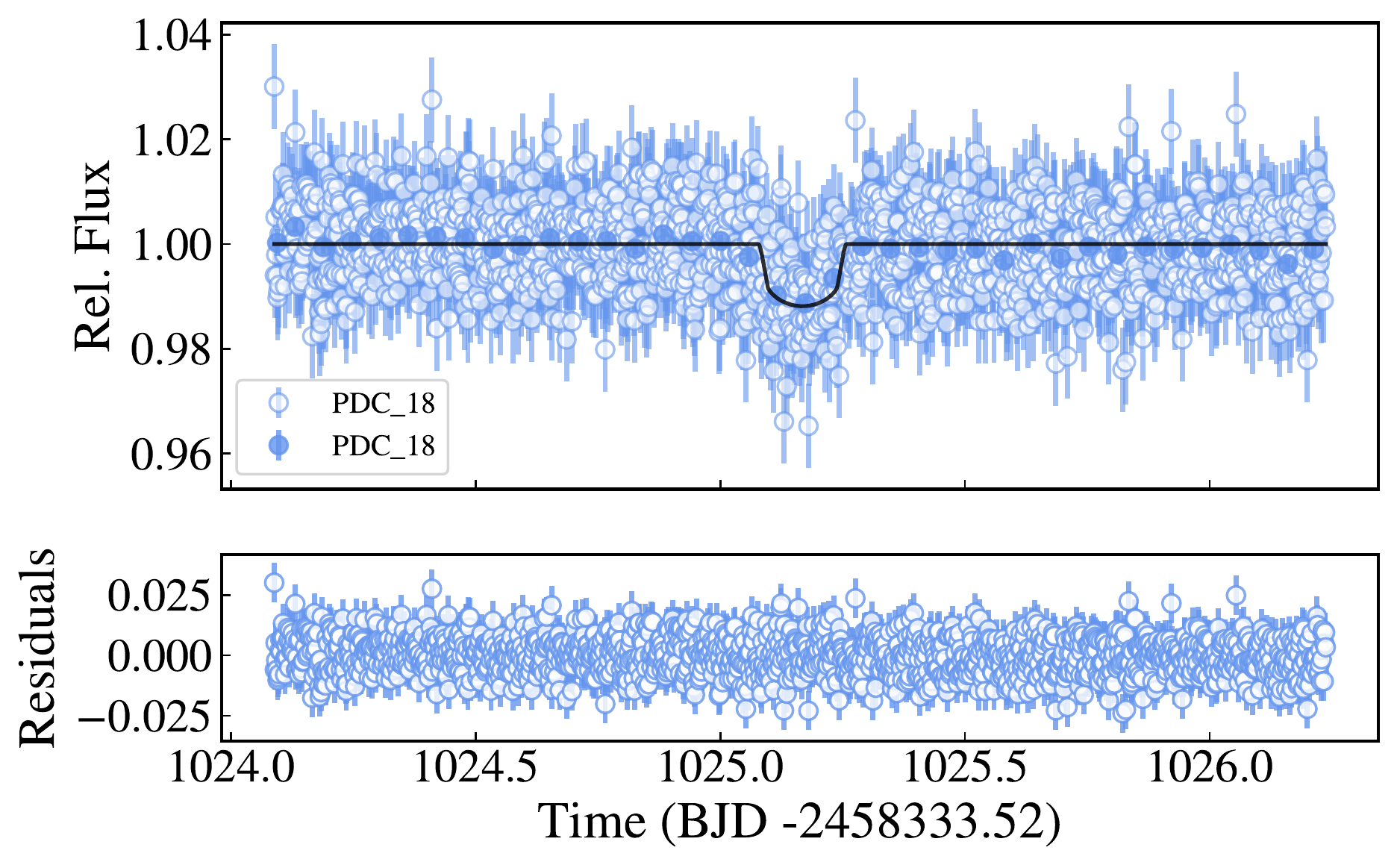}   &
    \includegraphics[width=.23\textwidth]{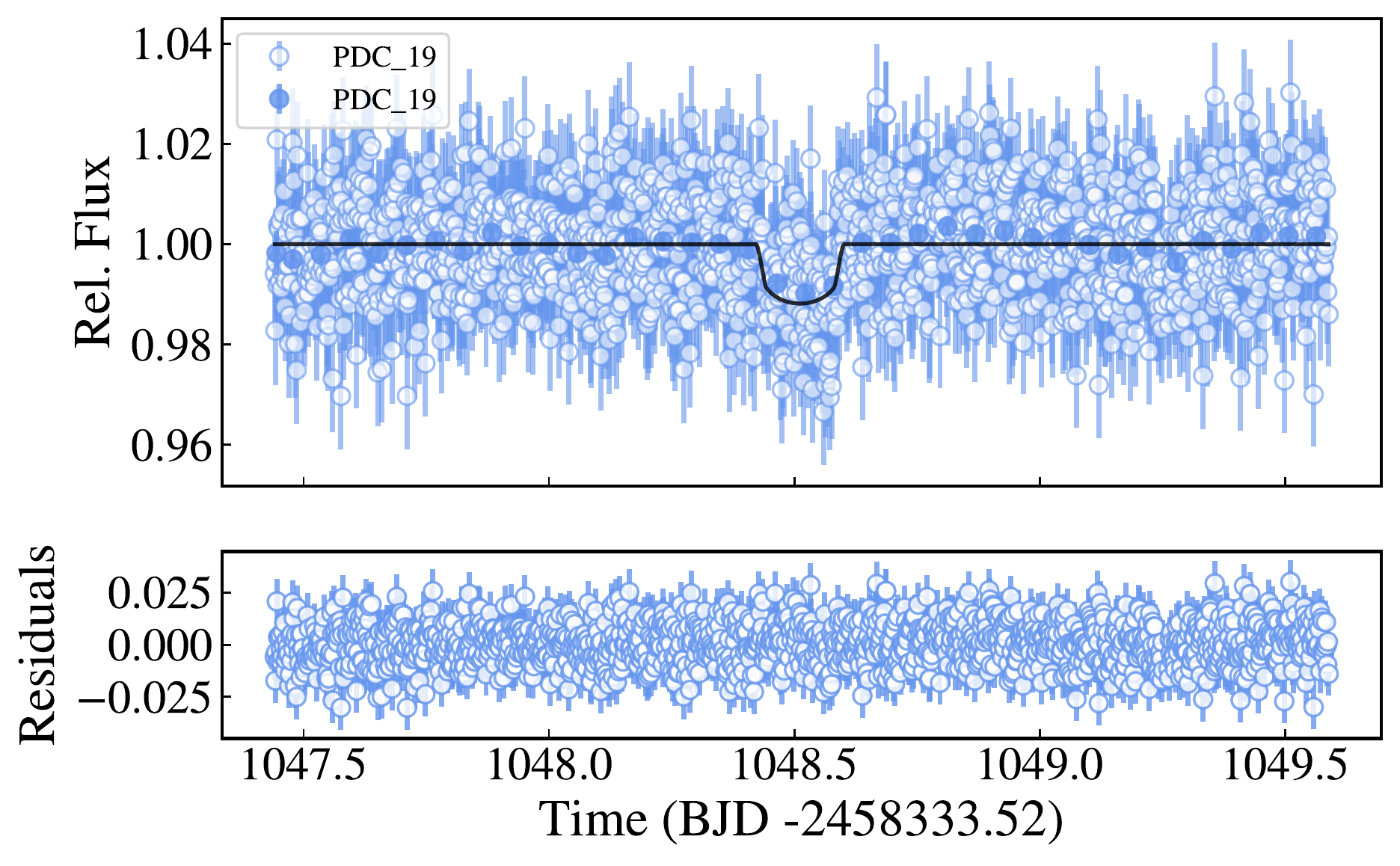} \\
    \includegraphics[width=.23\textwidth]{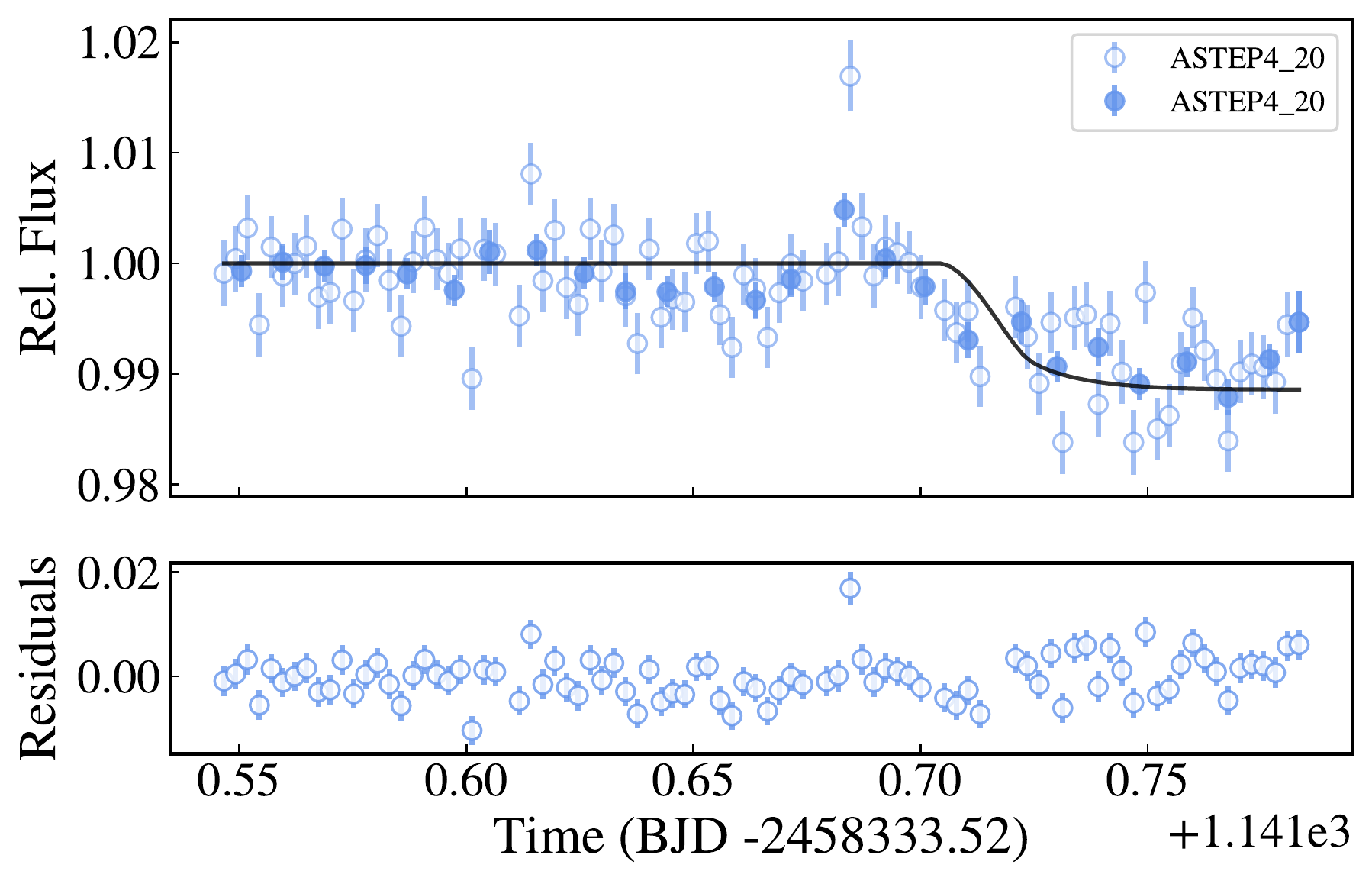}&
    \includegraphics[width=.23\textwidth]{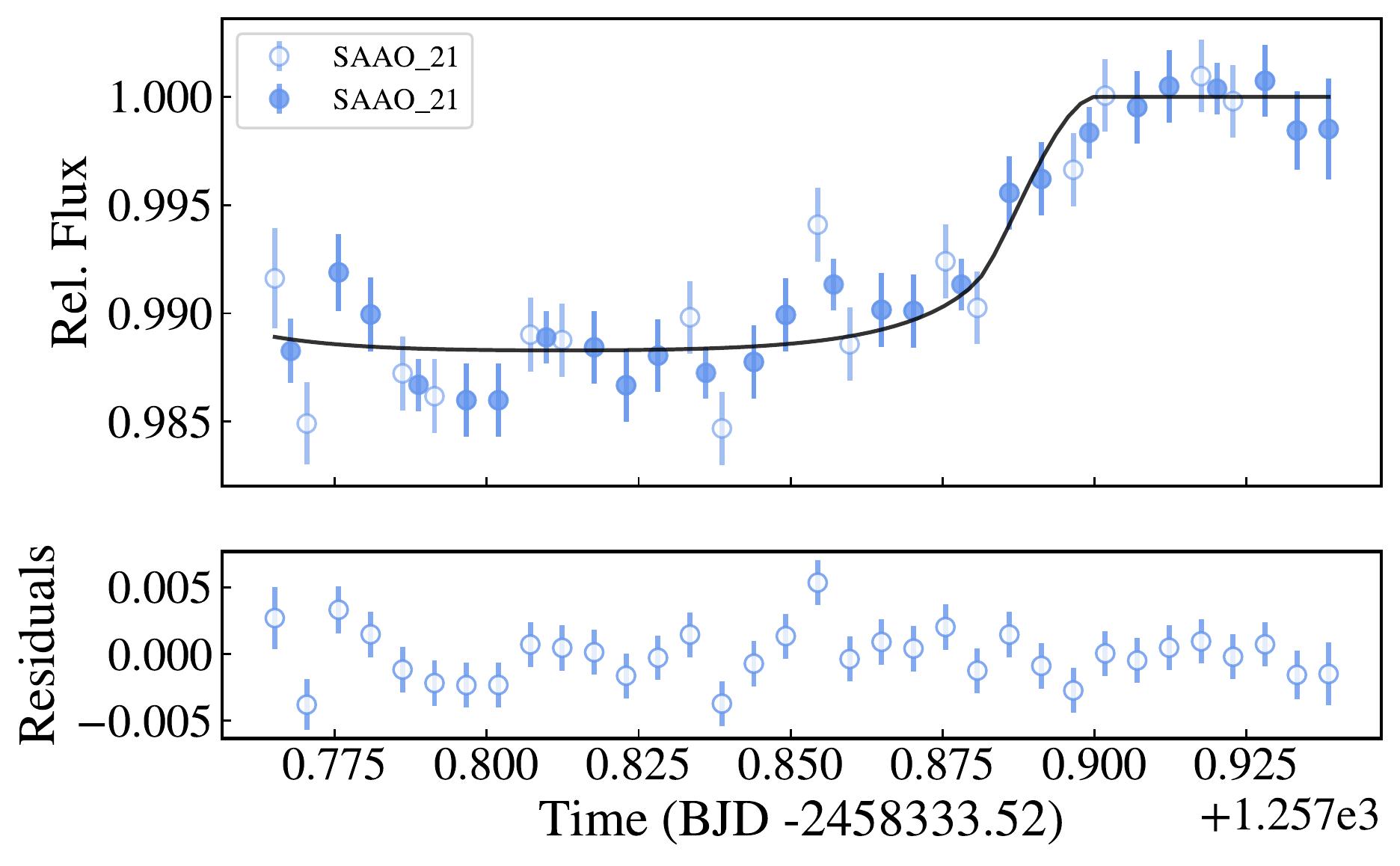}   &
    \includegraphics[width=.23\textwidth]{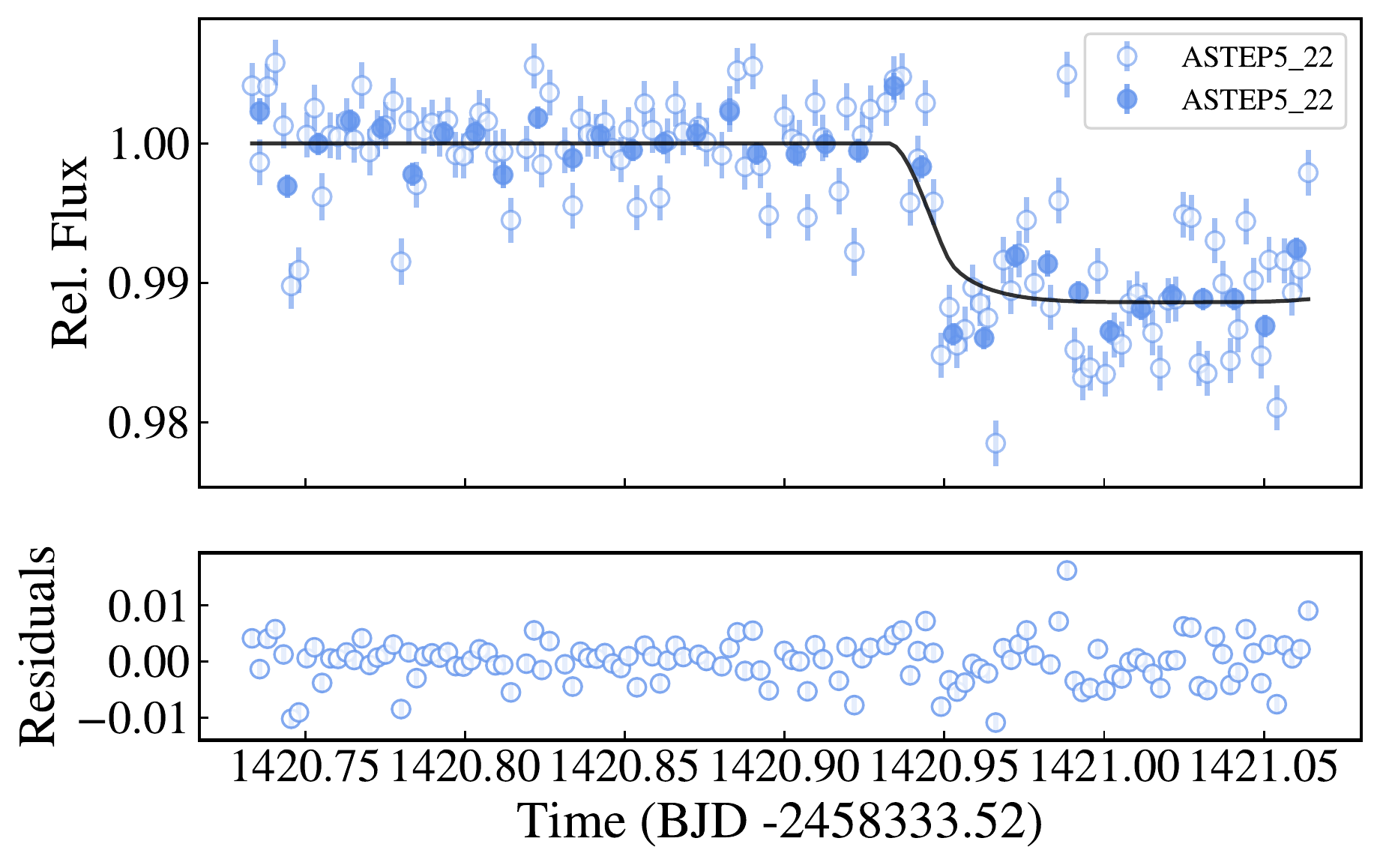} &

  \end{tabular}
  \caption{Shown are all Transit lightcurves of TOI-2525b from the photodynamic model observed by different telescopes. Nine of them are TESS observed (FFI), eleven are TESS observed transits (PDCSAP), two transits are observed by ASTEP, and one by SAAO. A 30-minute binning was introduced after the fit with errors calculated by standard deviation (filled circles).
  }
  \label{Transits2525b}
\end{figure}


\begin{figure}[]
\centering
  \begin{tabular}{@{}cccc@{}}
    \includegraphics[width=.23\textwidth]{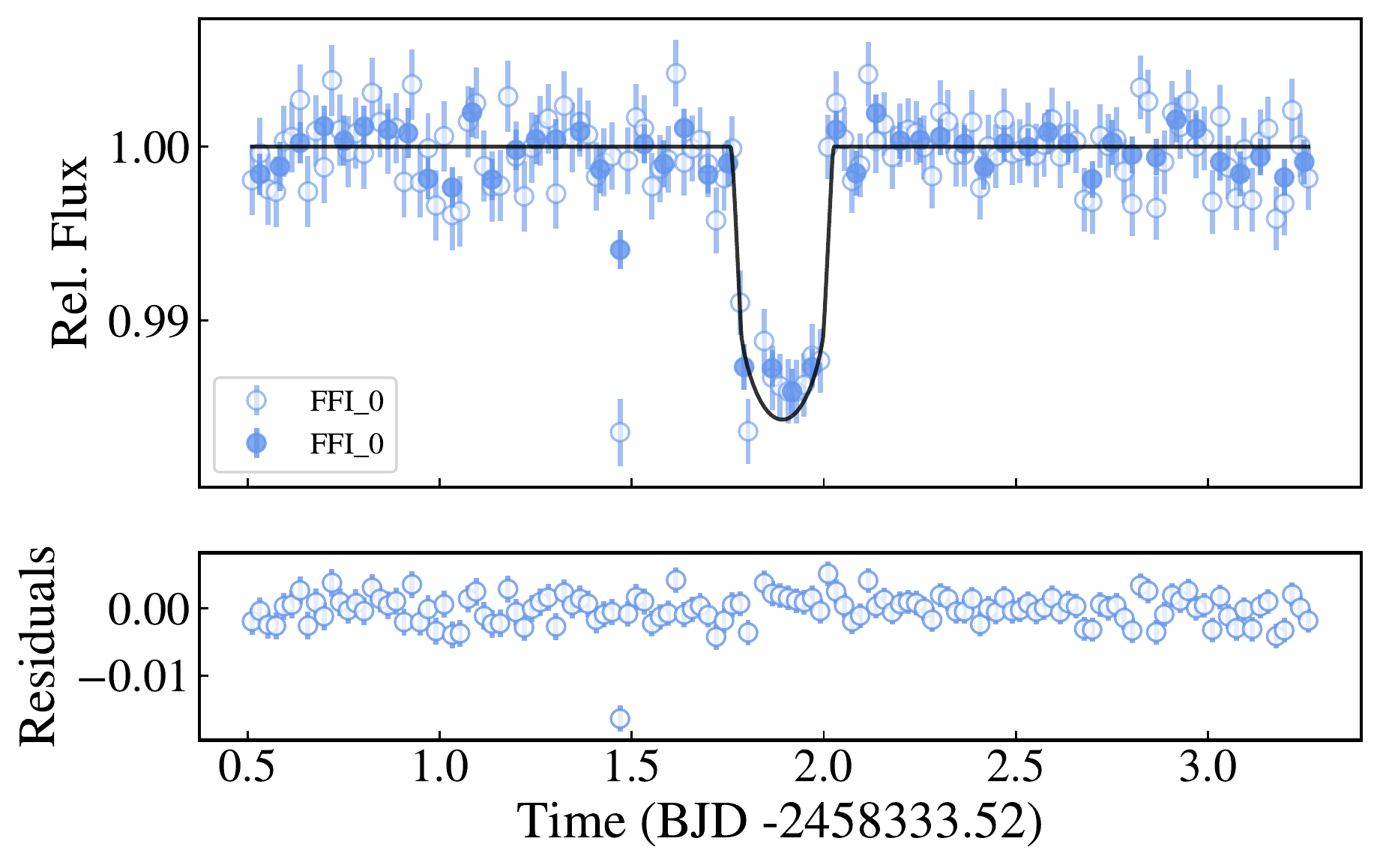} &
    \includegraphics[width=.23\textwidth]{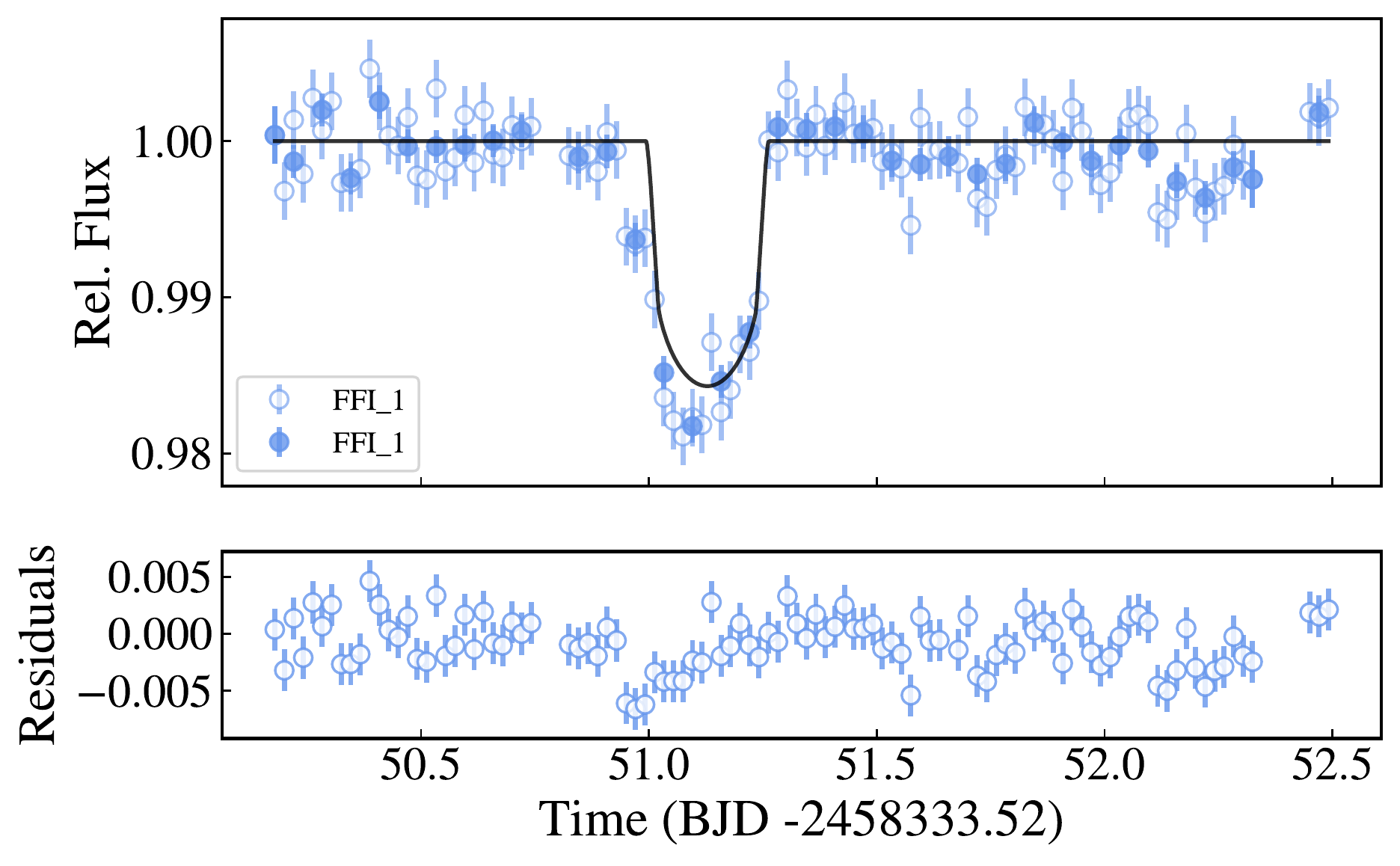} &
    \includegraphics[width=.23\textwidth]{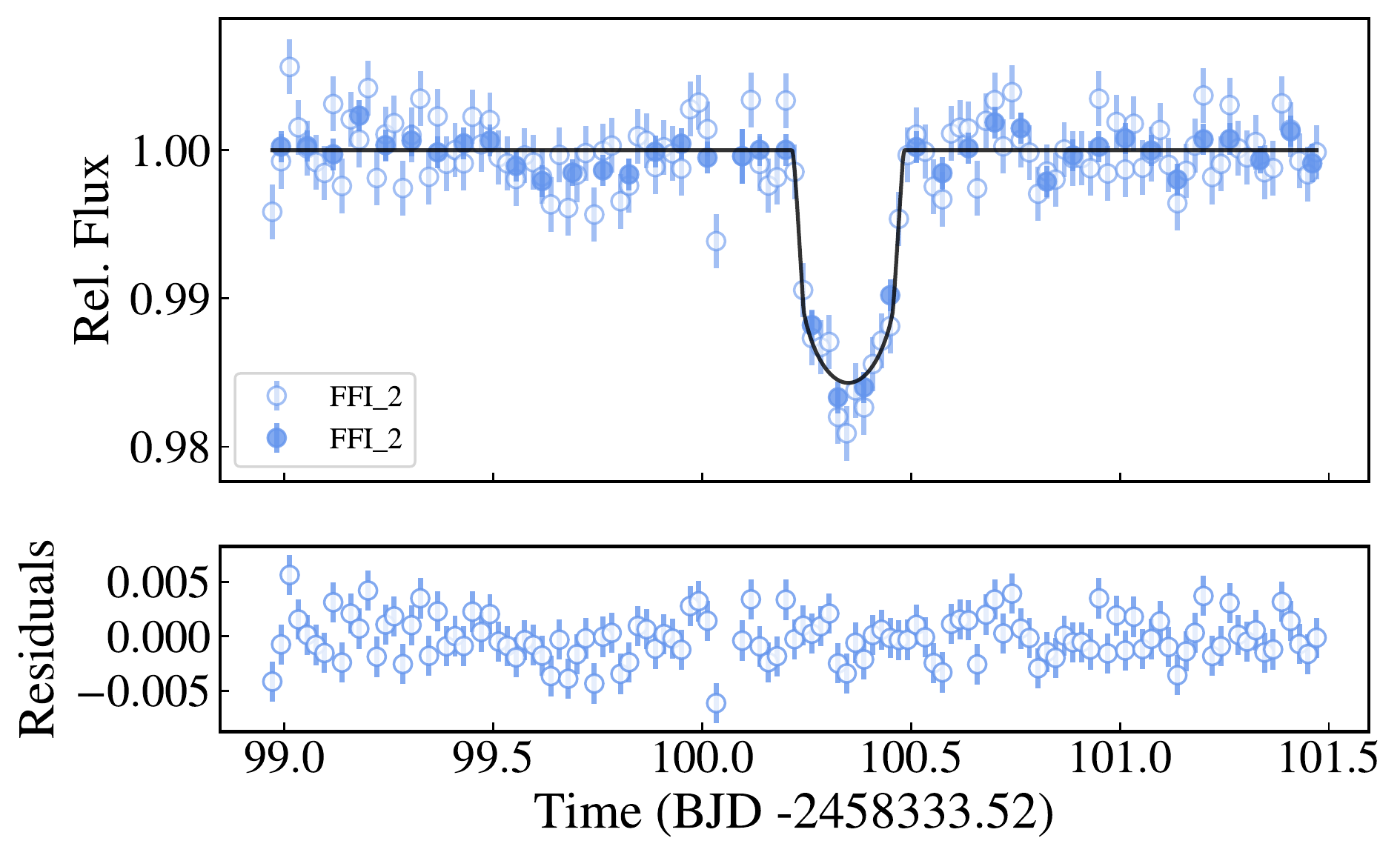}  &
    \includegraphics[width=.23\textwidth]{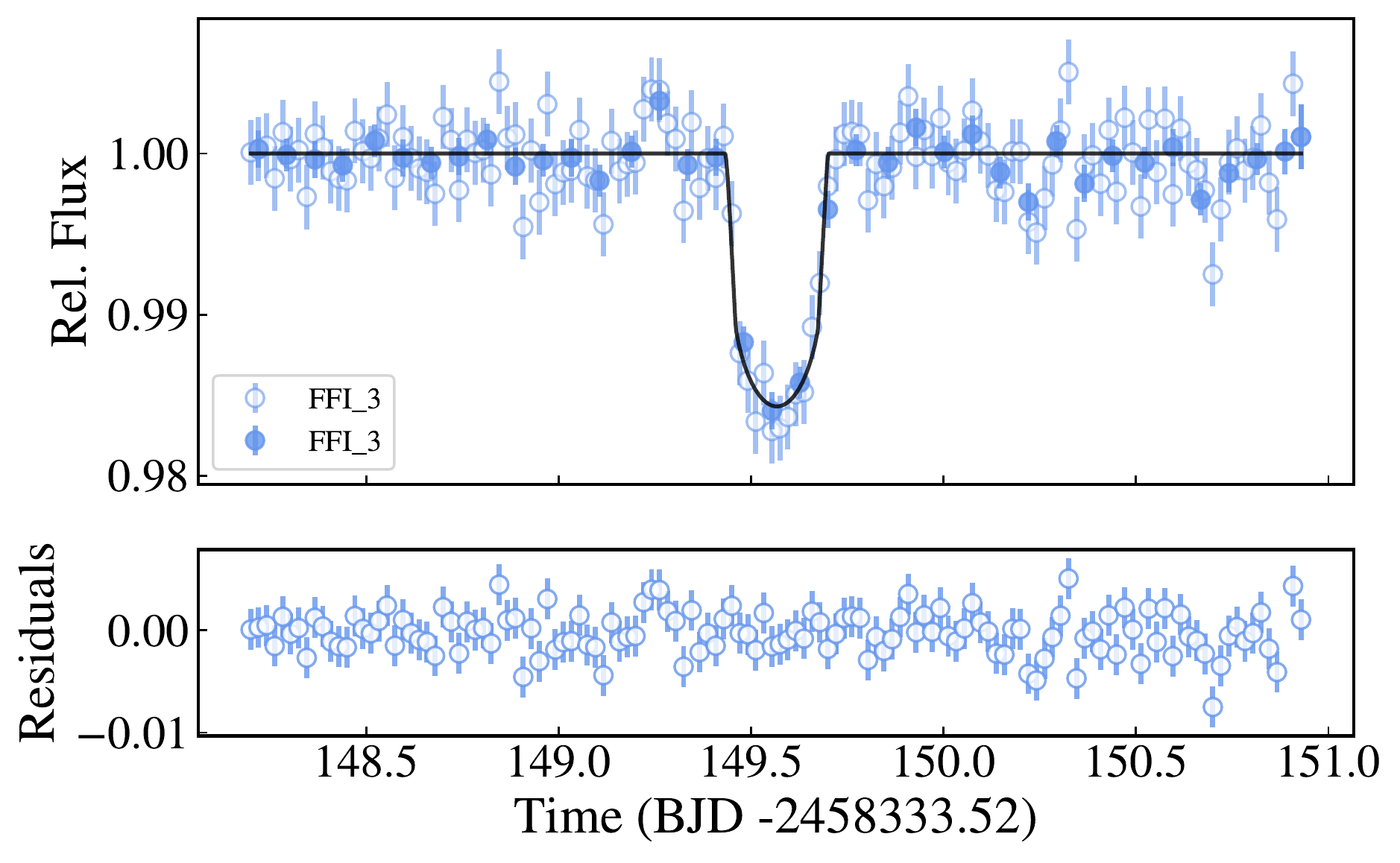} \\
    \includegraphics[width=.23\textwidth]{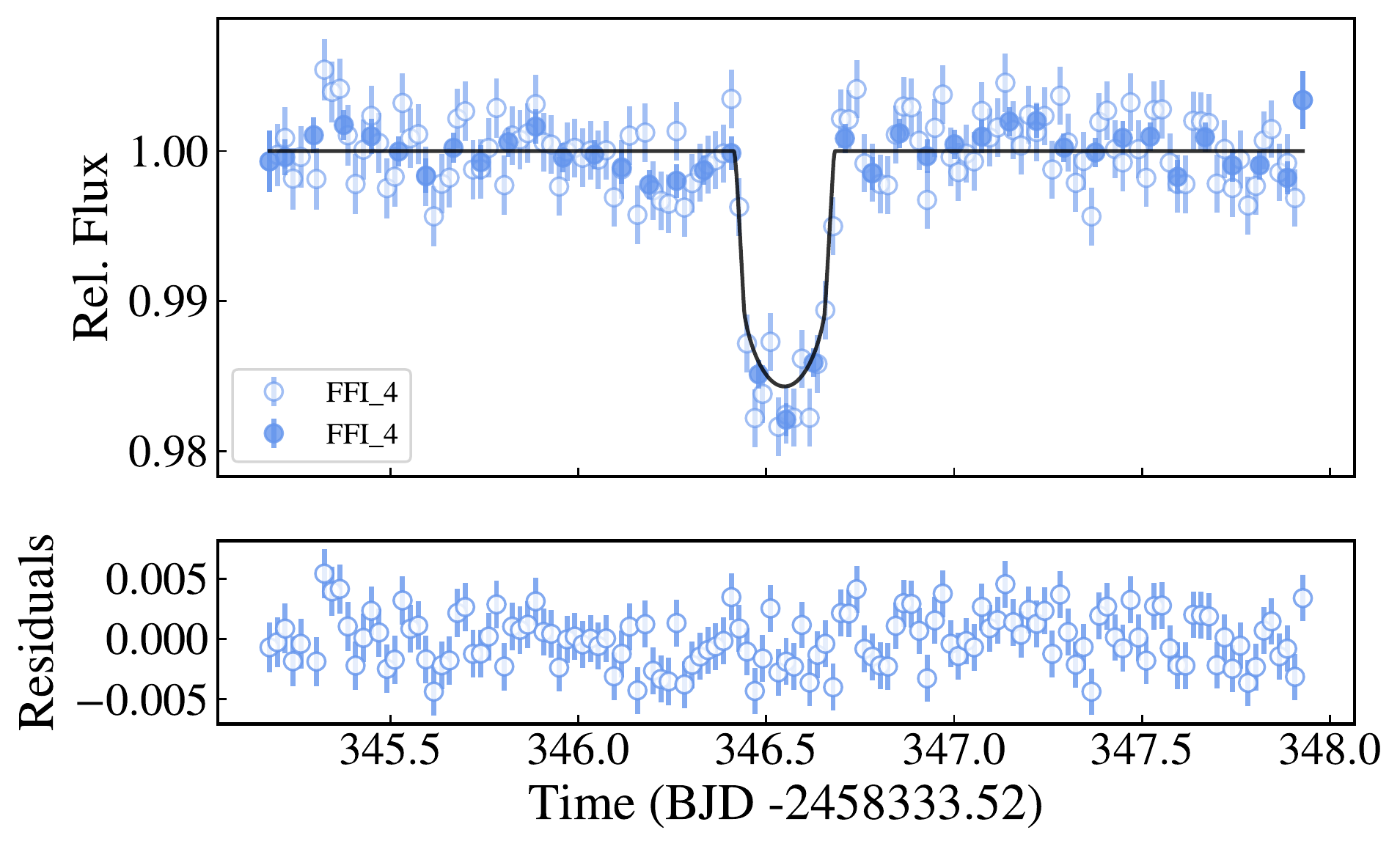} &
    \includegraphics[width=.23\textwidth]{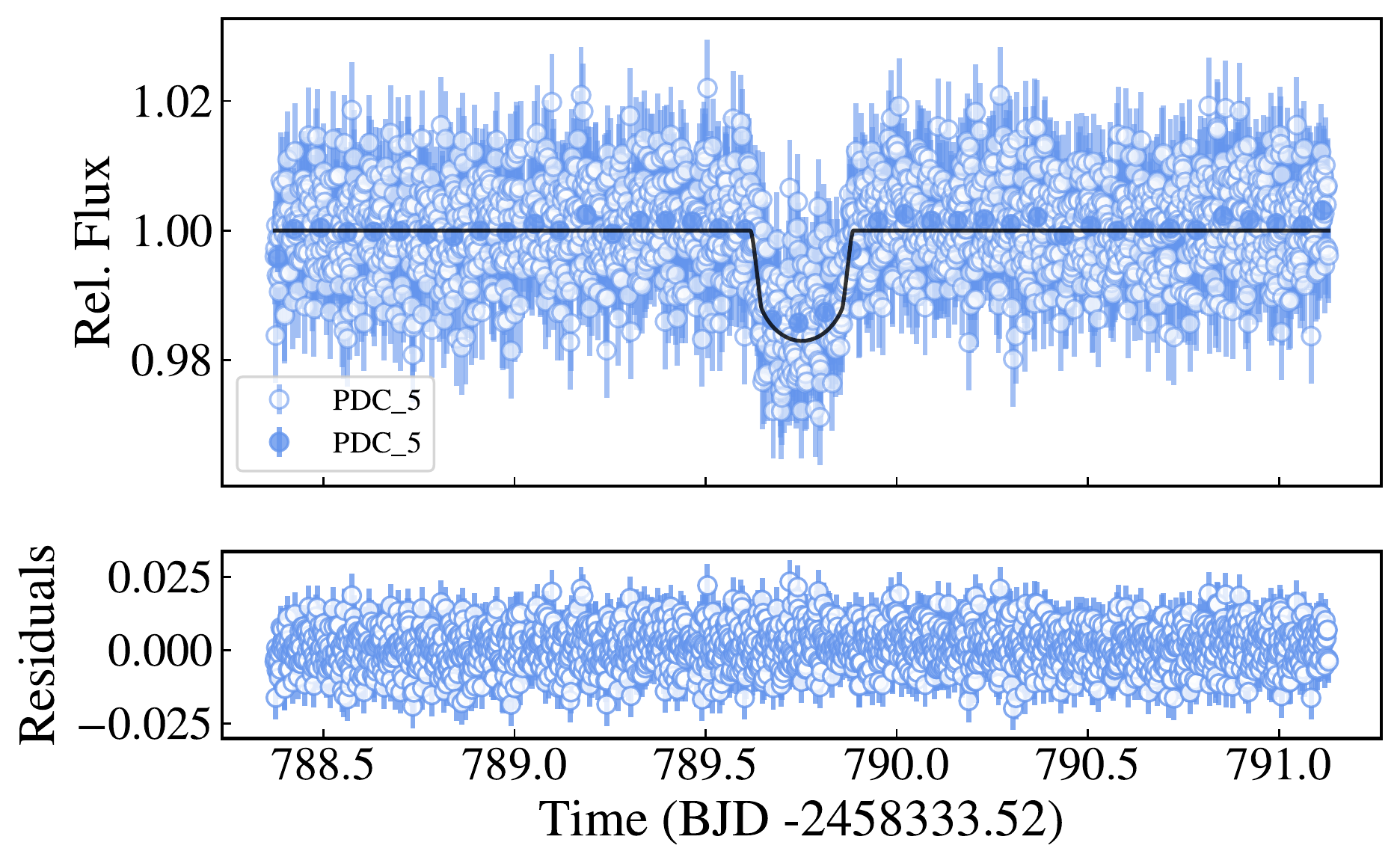}   &
    \includegraphics[width=.23\textwidth]{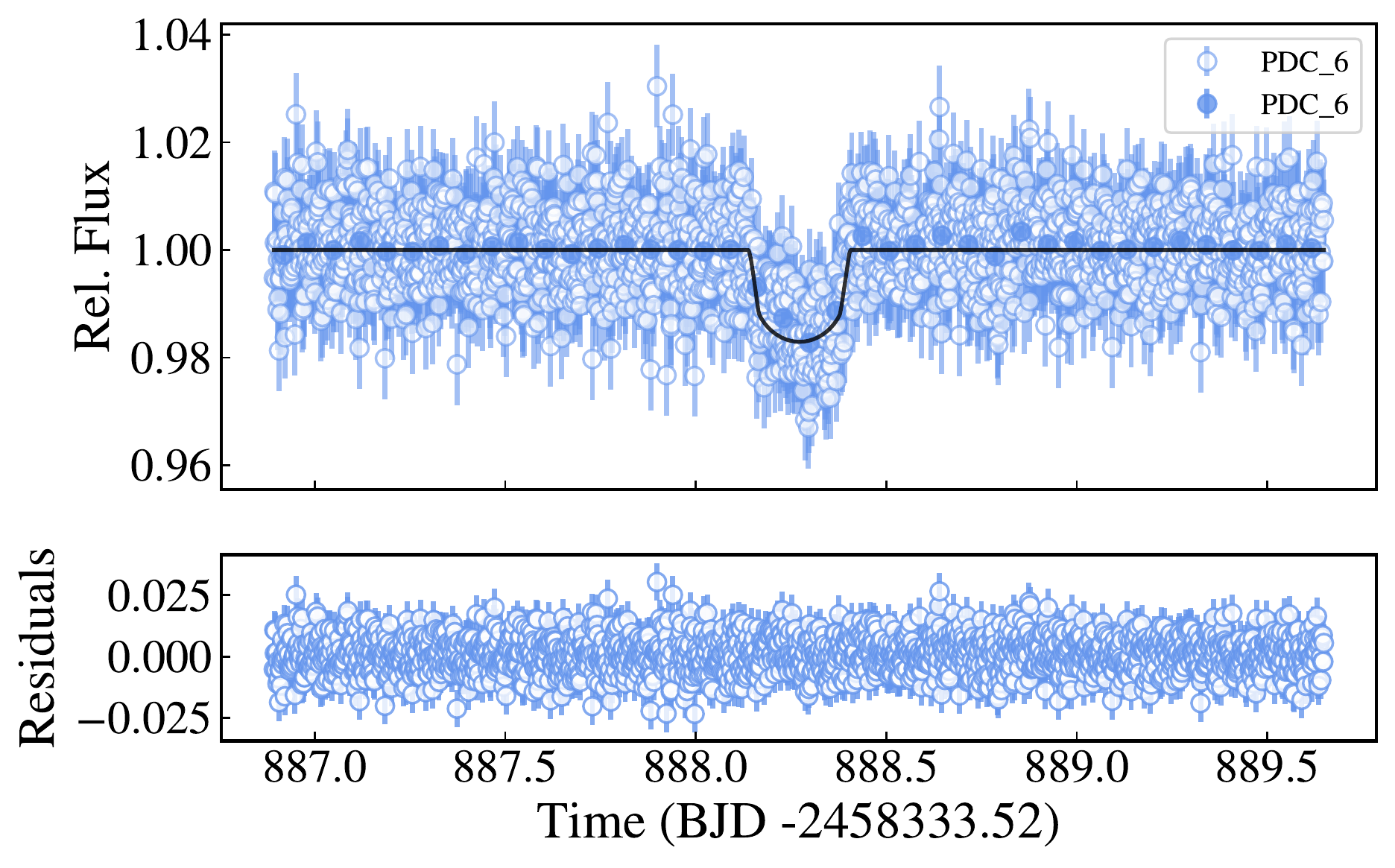} &
    \includegraphics[width=.23\textwidth]{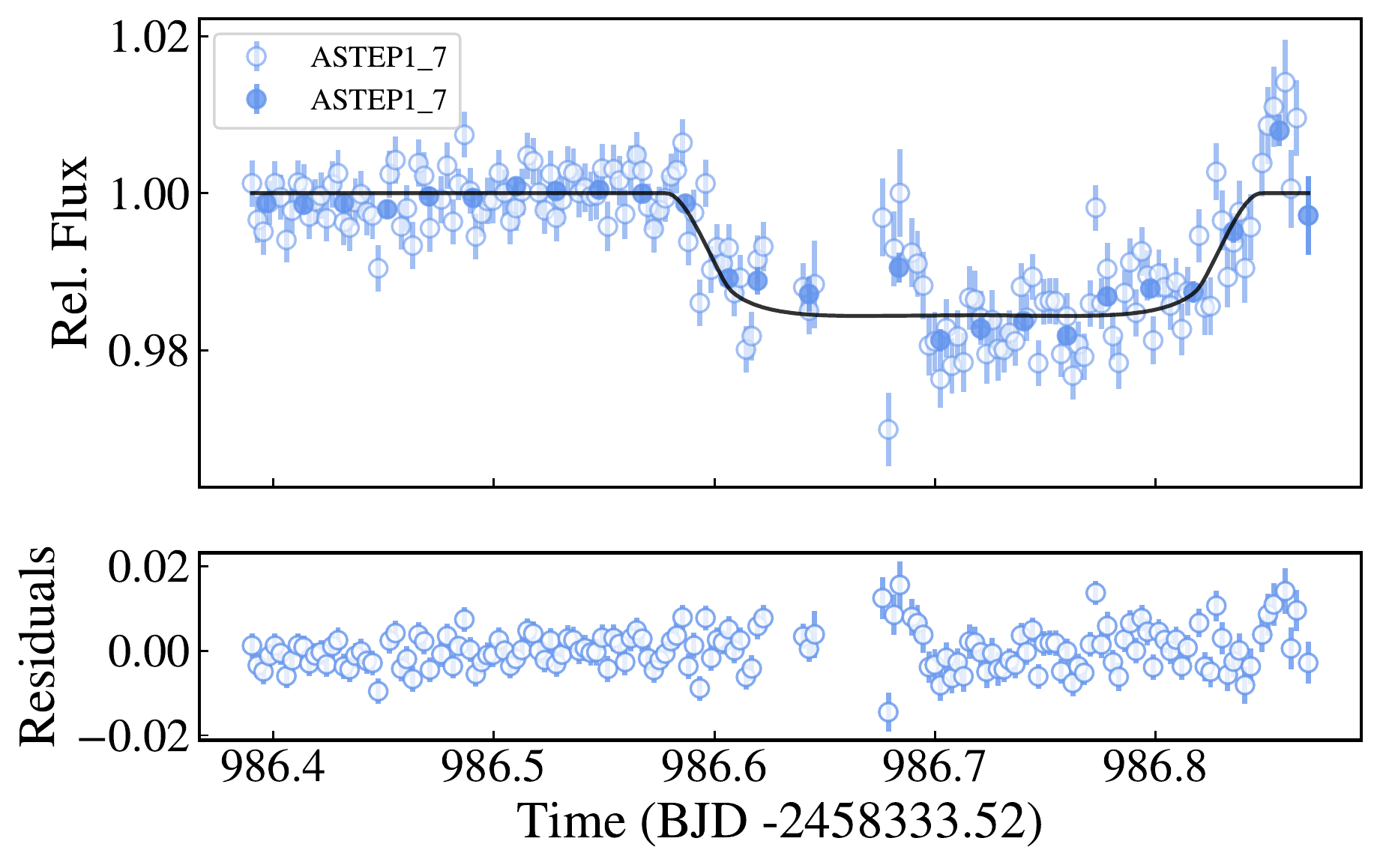} \\
    \includegraphics[width=.23\textwidth]{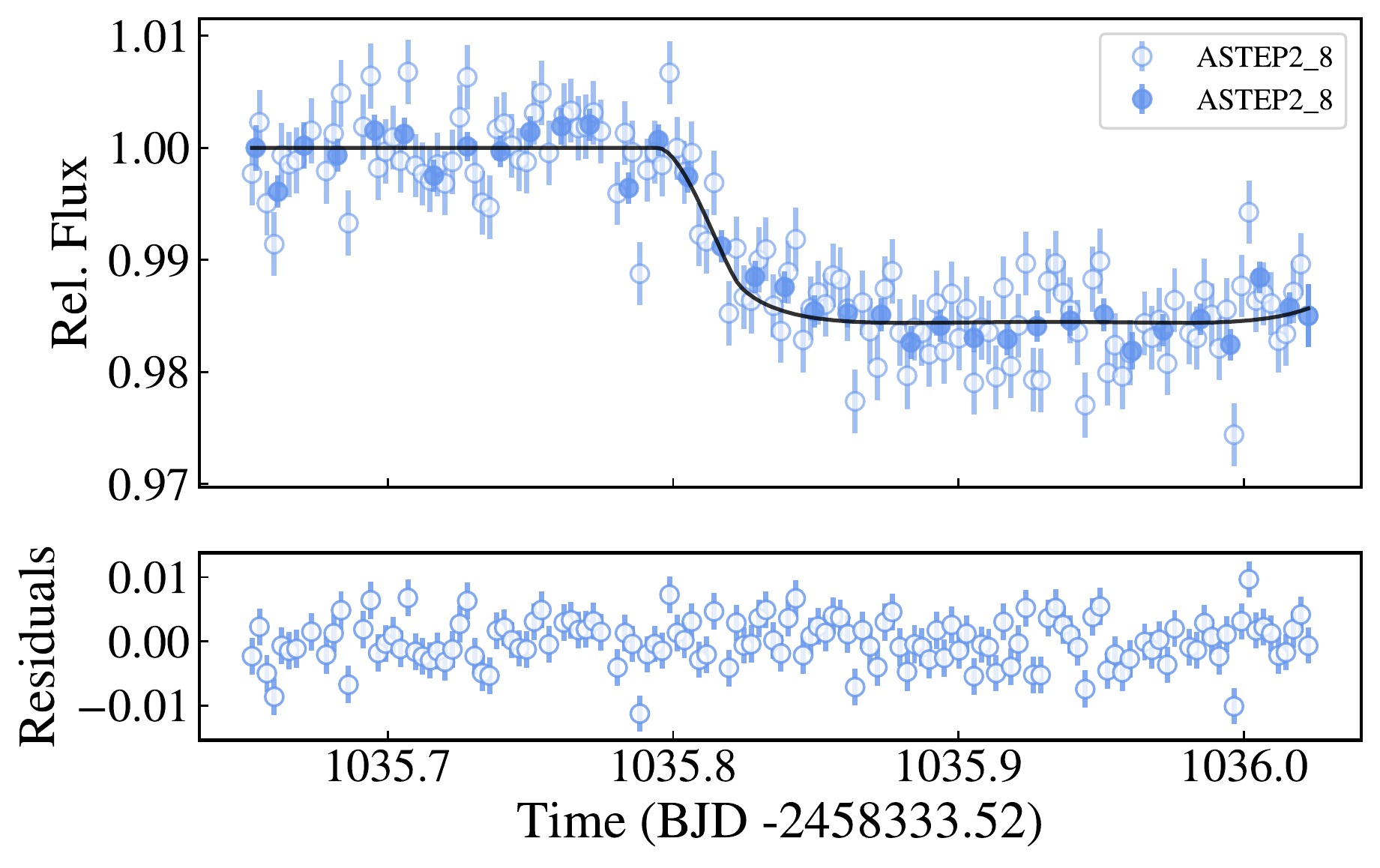}  &
    \includegraphics[width=.23\textwidth]{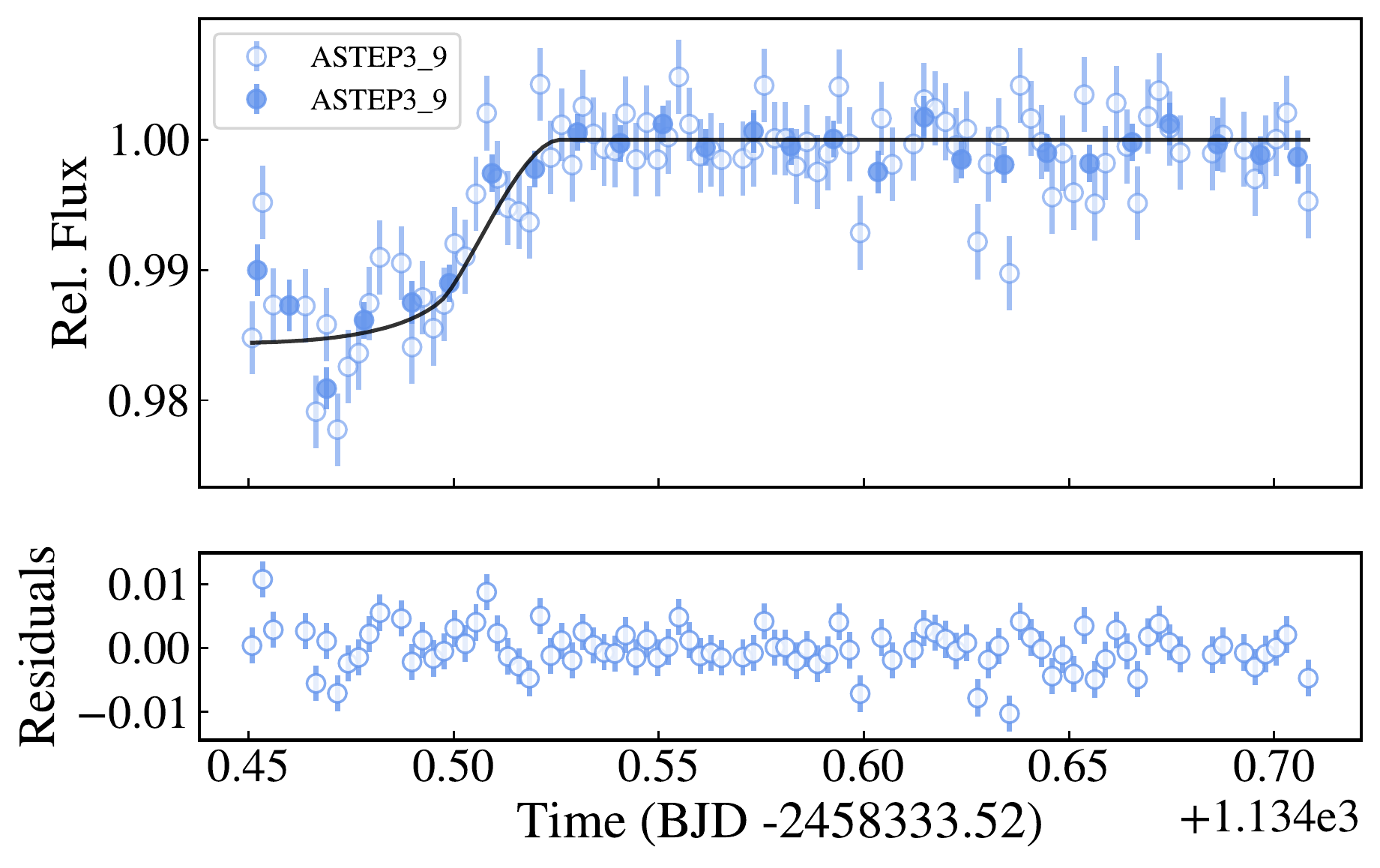} &
    \includegraphics[width=.23\textwidth]{TOI2525_PDC_10_15_model.pdf} &
    \includegraphics[width=.23\textwidth]{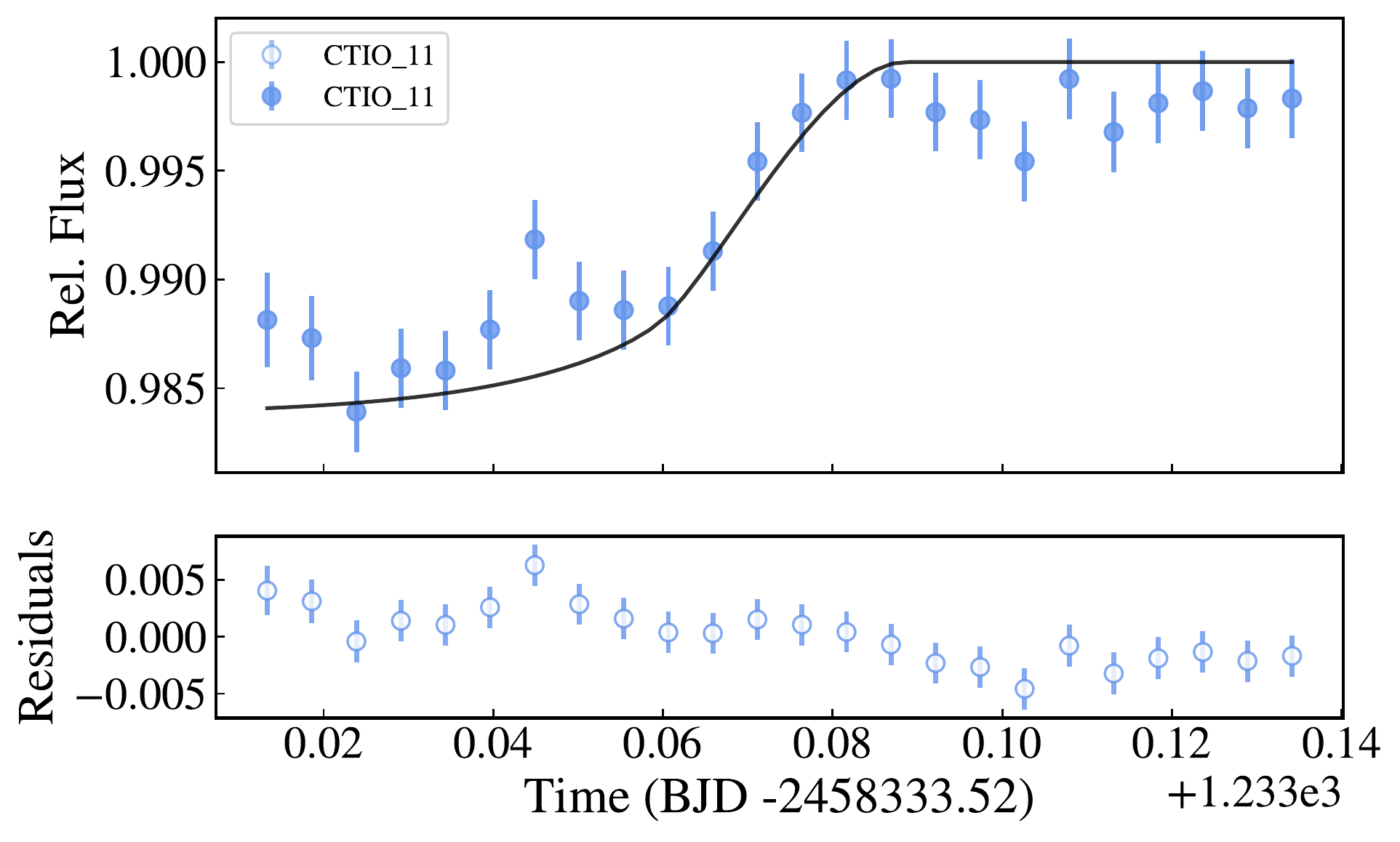}   \\
    \includegraphics[width=.23\textwidth]{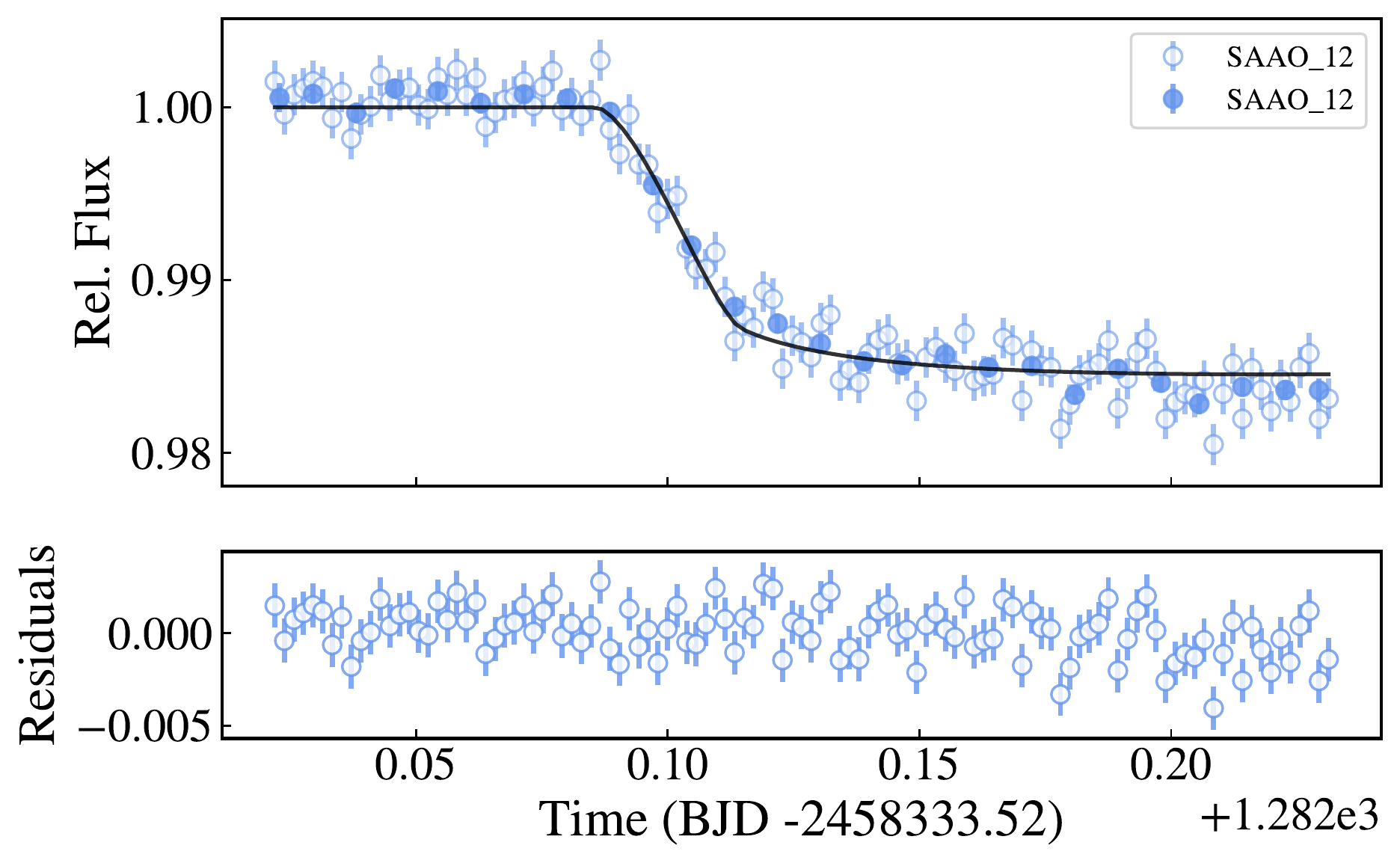} &
    \includegraphics[width=.23\textwidth]{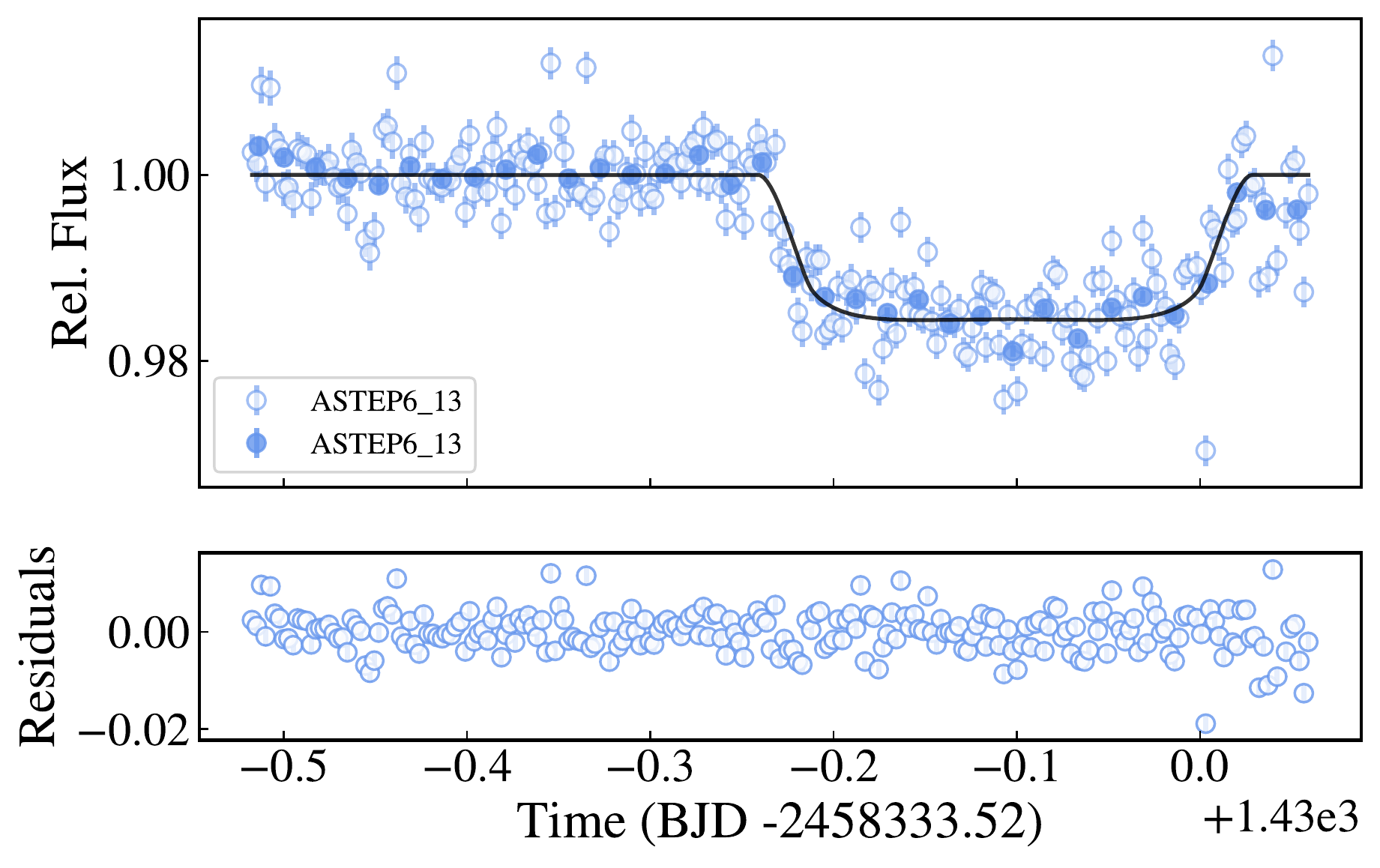} &

  \end{tabular}
  \caption{Shown are all Transit lightcurves of TOI-2525c from the photodynamic model observed by different telescopes. Five of them are TESS observed (FFI), three are TESS observed transits (PDCSAP), four transits are observed by ASTEP, one by CTIO and one by SAAO. A 30-minute binning was introduced after the fit with errors calculated by standard deviation (filled circles).
  }
  \label{Transits2525c}
\end{figure}

\begin{table}
 \caption{MCMC posterior medians, best-fits and adopted priors from orbital elements, offset and jitter terms from \texttt{flexi-fit}.}

    \begin{tabular}{   l   c   c  c   }

    \hline Parameter & median and $1\sigma$ & best-fit & adopted prior  \\
    \hline\hline
    \textit{TOI-2525b}                            & & &
                           
    \\ period $P$ [d]          &$23.2856_{-0.0017}^{+0.0017}$ &$23.2865$&$\mathcal{U}$(23.1,23.4)
    \\ mass $m$ [$\mathrm{M_{Jup}}$] & $0.084_{-0.005}^{+0.005}$     & $0.086$ &$\mathcal{U}$(10,100 M$_{\oplus}$)
    \\ eccentricity $e$ &$0.170_{-0.010}^{+0.011}$&$0.167$&$\mathcal{U}$(0,0.45)
    \\ longitude of periastron $\omega$[$^{\circ}$]&$345.9_{-0.8}^{+0.8}$ &$345.7$ &$\mathcal{U}$(0,360)
    \\ time of conjunction $t_{conj}$ [BJD-2457000d]&$1333.5289_{-0.0020}^{+0.0020}$&$1333.5293$&$\mathcal{U}$(1333.52,1356.82)
    \\ inclination $i$ [$^{\circ}$] & $89.50_{-0.07}^{+0.07}$&$89.47$&$\mathcal{U}$(80,100)
    \\ planet to star radius ratio $R_p/R_s$ & $0.1013_{-0.0008}^{+0.0008}$&$0.1009$ &$\mathcal{U}$(0.001,0.2)
    \\ longitude of ascending node $\Omega$[$^{\circ}$]& $0$ (fixed) &$0$ (fixed) & (fixed)
    \\ radius $R_p$ [$R_{Jup}$] & $0.774_{-0.010}^{+0.010}$&$0.771$&(derived)
    \\ density $\rho$ [$\mathrm{g/cm^3}$]        & $0.225_{-0.014}^{+0.015}$  &$0.235$&(derived) 
    \\ semi-major axis $a$ [au]& $0.1511_{-0.0020}^{+0.0020}$&$0.1511$&(derived)
    \\ transit duration $t_{dur}$[h]& $4.05_{-0.07}^{+0.07}$&$4.00$&(derived)
    \\ RV semi-amplitude $K$ [m/s]&$6.7_{-0.4}^{+0.4}$ &$6.9$&(derived)
    \\ mean longitude $\lambda$ [$^{\circ}$]&$107.8_{-0.8}^{+0.8}$&$107.5$&(derived)

    \\\hline \textit{TOI-2525c}&&&

    \\ period $P$ [d]  &$49.2519_{-0.0004}^{+0.0004}$ &$49.2518$&$\mathcal{U}$(49.1,49.4)
    \\ mass $m$ [$\mathrm{M_{Jup}}$] & $0.657_{-0.032}^{+0.031}$     & $0.675$ &$\mathcal{U}$(100,400 M$_{\oplus}$)
    \\ eccentricity $e$ &$0.157_{-0.007}^{+0.008}$&$0.153$&$\mathcal{U}$(0,0.45)
    \\ longitude of periastron $\omega$[$^{\circ}$]&$21.5_{-1.1}^{+1.1}$ &$21.5$ &$\mathcal{U}$(0,360)
    \\ time of conjunction $t_{conj}$ [BJD-2457000d]&$1335.4118_{-0.0014}^{+0.0014}$&$1335.4121$&$\mathcal{U}$(1333.52,1382.77)
    \\ inclination $i$ [$^{\circ}$] & $89.97_{-0.07}^{+0.09}$&$90.03$&$\mathcal{U}$(80,100)
    \\ planet to star radius ratio $R_p/R_s$ & $0.1183_{-0.0005}^{+0.0005}$&$0.1182$ &$\mathcal{U}$(0.001,0.25)
    \\ longitude of ascending node $\Omega$[$^{\circ}$]& $1.1_{-0.7}^{+1.0}$ &$0.4$ &$\mathcal{U}$(-10,10)
    \\ radius $R_p$ [$R_{Jup}$] & $0.904_{-0.010}^{+0.010}$                               &$0.903$&(derived)
    \\ density $\rho$ [$\mathrm{g/cm^3}$]  & $1.11_{-0.07}^{+0.07}$  &$1.14$&(derived) 

    \\ semi-major axis $a$ [au]& $0.249_{-0.004}^{+0.004}$&$0.249$&(derived)
    \\ transit duration $t_{dur}$[h]& $5.502_{-0.027}^{+0.013}$&$5.514$&(derived)
    \\ RV semi-amplitude $K$ [m/s]&$41.2_{-2.2}^{+2.2}$ &$42.2$&(derived)
    \\ mean longitude $\lambda$ [$^{\circ}$]&$93.2_{-1.1}^{+1.1}$&$93.2$&(derived)
   
    \\\hline &&&
    
    \\ $\mathrm{RV_{off}}$ PFS [m/s]&$-7_{-11}^{+11}$&$-4$&$\mathcal{U}(-\infty,\infty)$
    \\ $\mathrm{RV_{jit}}$ PFS [m/s]&$31_{-7}^{+10}$&$29$&$\mathcal{U}(exp(-5),exp(5))$
   
    \\ $\mathrm{TR_{off}}$ TESS FFI [ppm]&$-310_{-60}^{+60}$&$-300$&$\mathcal{U}(-\infty,\infty)$
    \\ $\mathrm{TR_{off}}$ TESS PDC (year 3) [ppm] &$380_{-60}^{+60}$&$330$&$\mathcal{U}(-\infty,\infty)$

    \\ $\mathrm{TR_{off}}$ ASTEP 1 [ppm]&$390_{-260}^{+250}$&$320$&$\mathcal{U}(-\infty,\infty)$
    \\ $\mathrm{TR_{off}}$ ASTEP 2 [ppm]&$350_{-260}^{+260}$&$340$&$\mathcal{U}(-\infty,\infty)$
    \\ $\mathrm{TR_{off}}$ ASTEP 3 [ppm]&$210_{-320}^{+310}$&$340$&$\mathcal{U}(-\infty,\infty)$
    \\ $\mathrm{TR_{off}}$ ASTEP 4 [ppm]&$650_{-330}^{+320}$&$940$&$\mathcal{U}(-\infty,\infty)$
    \\ $\mathrm{TR_{off}}$ SSO [ppm]&$-1870_{-310}^{+310}$&$-1850$&$\mathcal{U}(-\infty,\infty)$
    \\ $\mathrm{TR_{off}}$ CTIO [ppm]&$6690_{-410}^{+410}$&$6620$&$\mathcal{U}(-\infty,\infty)$
    \\ $\mathrm{TR_{off}}$ SAAO1 [ppm] &$7880_{-350}^{+350}$&$8000$&$\mathcal{U}(-\infty,\infty)$
    \\ $\mathrm{TR_{off}}$ SAAO2 [ppm] &$9140_{-150}^{+150}$&$9000$&$\mathcal{U}(-\infty,\infty)$
    \\ $\mathrm{TR_{off}}$ ASTEP5 [ppm] &$-500_{-170}^{+160}$&$-580$&$\mathcal{U}(-\infty,\infty)$
    \\ $\mathrm{TR_{off}}$ ASTEP6 [ppm] &$-280_{-130}^{+130}$&$-270$&$\mathcal{U}(-\infty,\infty)$

    \\\hline\end{tabular}

    \label{photodynparameters}
\end{table}

\begin{figure}
    
    \includegraphics[width=16cm, height=18cm]{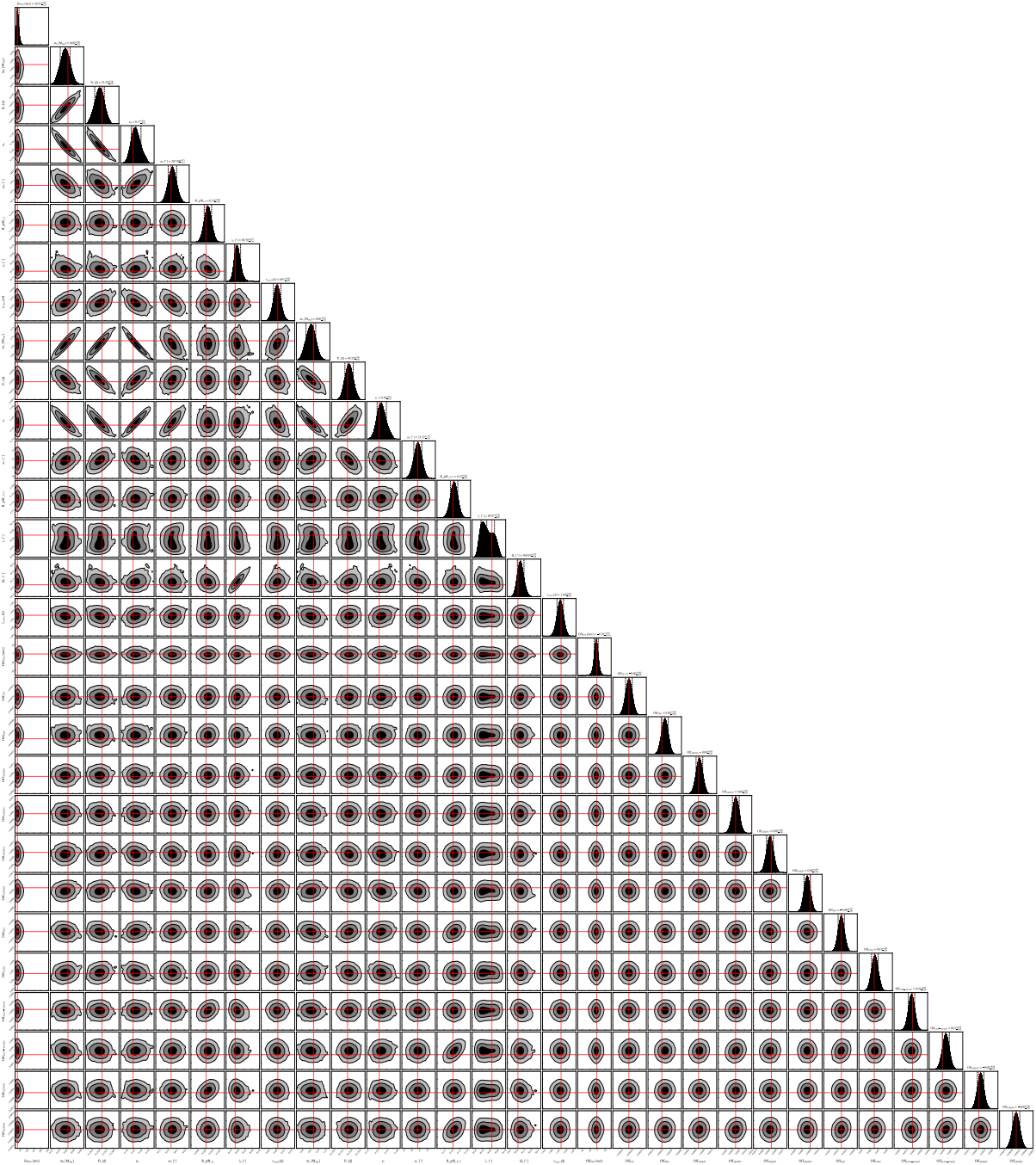}
    \caption{Posterior distribution for the joint photodynamical analysis. The orbital parameters of both planets and jitter and offset terms are obtained by an MCMC analysis in \texttt{flexi-fit} \url{https://gitlab.gwdg.de/sdreizl/exoplanet-flexi-fit}. The best-fit posteriors are marked in red. The scattered black lines represent the median and the $1\sigma$ intervals of the distribution.}
    
    \label{photodyposterior}
    
\end{figure} 
    
    
    
    
    
    

\end{appendix}

\end{document}